\documentclass[a4paper,fleql]{cas-sc}
\usepackage[authoryear]{natbib}
\bibpunct{(}{)}{,}{a}{,}{,}
\usepackage{amsmath, amssymb, amsfonts}
\usepackage{graphicx}
\usepackage{booktabs}
\usepackage{multirow}
\usepackage{algpseudocode}
\usepackage{float}
\usepackage{makecell}
\usepackage{placeins}
\usepackage{hyperref}
\usepackage{xurl}
\usepackage{caption}
\usepackage{subcaption}
\usepackage{longtable}
\usepackage{geometry}
\usepackage{makecell}
\usepackage[ruled,vlined]{algorithm2e}

\def\tsc#1{\csdef{#1}{\textsc{\lowercase{#1}}\xspace}}
\tsc{WGM}
\tsc{QE}

\begin{document}
\let\WriteBookmarks\relax
\def\floatpagepagefraction{1}
\def\textpagefraction{.001}

\shorttitle{}    

\shortauthors{}  

\title [mode = title]{BAT-RM: A Boundary-Aware Transformer with Region-Aware Multi-Directional Mamba for Clinically Deployed Cervical Cancer Radiotherapy Auto-Contouring} 




\author[1,3]{Istiak Ahmed}
\ead{istiak.ahmed1@northsouth.edu}

\credit{Conceptualization, Data Curation, Formal analysis, Investigation, Methodology, Software, Validation, Visualization, Writing – original draft, Writing – review and editing}


\author[1,3]{Kazi Shahriar Sanjid}
\ead{kazi.sanjid@northsouth.edu}
\credit{Data Curation, Formal Analysis, Methodology, Software, Validation, Visualization, Writing – review and editing}

\author[1,3]{Galib Ahmed}
\ead{galib.ahmed.251@northsouth.edu}
\credit{Data Curation, Formal Analysis, Investigation, Methodology, Validation, Visualization, Writing – review and editing}

\author[3,4]{Md. Tanzim Hossain}
\ead{tanzim.hossain@fau.de}
\credit{Formal Analysis, Software, Validation, Visualization, Writing – review and editing}

\author[3,5]{Md. Anwarul Islam}
\ead{anwarpabna@gmail.com}
\credit{Data Curation, Formal Analysis, Resources, Validation, Writing – review and editing}

\author[1,3]{Shahrukh Khan}
\ead{shahrukh.khan02@northsouth.edu}

\credit{Data Curation, Investigation, Validation, Visualization, Writing – review and editing}

\author[1,3]{Md. Ashrif Rahman Arian}
\ead{ashrif.arian@northsouth.edu}
\credit{Data Curation, Investigation, Validation, Visualization, Writing – review and editing}

\author[1,3]{Md. Nishan Khan}
\ead{nishan.khan@northsouth.edu}
\credit{Data Curation, Methodology, Software, Validation}

\author[1,3]{Md. Misbah Khan}
\ead{misbah.khan@northsouth.edu}
\credit{Data Curation, Methodology, Software, Visualization}

\author[6]{S M Hasibul Hoque}
\ead{hasibulshefat@gmail.com}
\credit{Conceptualization, Resources, Validation, Writing – review and editing}

\author[2]{Rahnuma Shahrin Rista}
\ead{rahnuma.rista@northsouth.edu}
\credit{Conceptualization, Resources, Validation, Writing – review and editing}

\author[6]{Md. Jobairul Islam}
\ead{jobairul55@gmail.com}
\credit{Data Curation, Resources, Validation, Visualization}

\author[7]{Sheikh Anisul Haque}
\ead{dr.anisul@bshl.com.bd}
\credit{Data Curation, Resources, Validation, Visualization}

\author[7]{Md Arifur Rahman}
\ead{drarifur@bshl.com.bd}
\credit{Formal Analysis, Resources, Validation, Visualization}

\author[8]{Syed Md. Akram Hussain}
\ead{syedmdakram@gmail.com}

\credit{Formal Analysis, Resources, Project Administration, Validation, Writing – review and editing}

\author[9]{Syeda Nashra}
\ead{snashra@torontomu.ca}
\credit{Formal Analysis, Validation, Visualization, Writing – review and editing}

\author[10]{Sayeed Shafayet Chowdhury}
\ead{saychow@iu.edu}
\credit{Methodology, Formal Analysis, Investigation, Supervision, Validation, Visualization, Writing – review and editing}

\author[11]{Md. Mostafa Kamal Sarker}
\ead{mmks3@cam.ac.uk}
\credit{Methodology, Formal Analysis, Investigation, Supervision, Validation, Visualization, Writing – review and editing}

\author[2,3]{M. Monir Uddin}
\cormark[1]
\ead{monir.uddin@northsouth.edu}
\credit{Conceptualization, Formal Analysis, Funding acquisition, Methodology, Project administration, Resources, Supervision, Writing – review and editing}


\affiliation[1]{%
        addressline={Department of Electrical and Computer Engineering},
        organization={North South University},
    city={Dhaka},
    postcode={1229},
    country={Bangladesh}
}

\affiliation[2]{%
        addressline={Department of Mathematics and Physics},
        organization={North South University},
    city={Dhaka},
    postcode={1229},
    country={Bangladesh}
}

\affiliation[3]{%
addressline={Big-Matrix Lab, Department of Mathematics and Physics},
    organization={North South University},
   city={Dhaka},
    postcode={1229},
    country={Bangladesh}
}

\affiliation[4]{%
addressline={Department of Data Science},
    city={Erlangen},
    organization={Friedrich-Alexander University},
        postcode={91054},
    country={Germany}
}

\affiliation[5]{%
    addressline={Square Cancer Centre},
    organization={Square Hospitals Limited},
       city={Dhaka},
    postcode={1205},
    country={Bangladesh}
}


\affiliation[6]{%
addressline={Department of Radiation Oncology},
    organization={Labaid Cancer Hospital \& Super Speciality Center},
        city={Dhaka},
    postcode={1205},
    country={Bangladesh}
}

\affiliation[7]{%
    addressline={Department of Oncology \& Radiotherapy},
    organization={Bangladesh Specialized Hospital Limited},
       city={Dhaka},
    postcode={1207},
    country={Bangladesh}
}

\affiliation[8]{%
    addressline={Department of Clinical Oncology},
    city={Dhaka},
    organization={Bangladesh Medical University},
       postcode={1000},
    country={Bangladesh}
}

\affiliation[9]{%
    organization={Faculty of Engineering and Architectural Science, Toronto Metropolitan University},
    city={Toronto},
    state={Ontario},
    country={Canada}
}

\affiliation[10]{%
    addressline={Luddy School of
    Informatics, Computing, and Engineering, Department of Computer Science},
    organization={Indiana University Indianapolis},
       city={535 W. Michigan Street, Indianapolis},
    postcode={IN},
    country={USA}
}

\affiliation[11]{%
    addressline={Nuffield Department of Medicine, Old Road Campus},
    organization={University of Oxford},
       city={Oxford},
    postcode={OX3 7LF},
    country={United Kingdom}
}


\cortext[1]{Corresponding author}

\fntext[1]{}


\begin{abstract}
We present a clinically deployed end-to-end auto-contouring system for cervical cancer radiotherapy planning, anchored by the Boundary-Aware Transformer with Region-Aware Mamba (BAT-RM), a hybrid architecture that integrates Sobel-gated boundary attention, a linear-time, multi-directional Mamba module for long-range context, and a boundary-skeleton-guided fusion gate. This design achieves linear-time complexity for long-range context modeling, avoiding the quadratic cost of full spatial self-attention. The full pipeline spans multi-institutional data collection, rigorous inter-rater quality assurance, external validation in an independent cohort, and a web-based clinical interface natively compatible with Varian, RayStation, and Monaco. Against four baselines, BAT-RM achieves superior performance across seven anatomical classes, with statistically significant improvements in target volumes, including GTV and CTV, and in organs at risk, such as the rectum and bladder. A prospective multi-center reader study involving 13 radiation oncologists demonstrated that AI assistance elevates junior oncologists' IoU from 0.899 to 0.965, approaching senior-level accuracy, while reducing contouring time by over 80\%. The system also reduced expert consultation rates and improved inter-reader consistency, reflecting gains in both efficiency and quality assurance. Following clinical deployment at a partner hospital, the system reduced patient wait times from days to hours without additional staffing, enabling same-day or next-day initiation of treatment for routine cases. BAT-RM demonstrates that a rigorous research pipeline, from data curation to clinical deployment, can translate directly into measurable patient benefit in resource-constrained settings where the demand for radiotherapy far exceeds specialist capacity.
\end{abstract}


\begin{highlights}
\item A novel hybrid segmentation architecture, BAT-RM, that simultaneously
achieves high boundary precision and long-range contextual
modeling through the integration of Sobel-gated boundary-aware attention,
a Multi-Directional Recurrent Context Module with Mamba selective scanning, and boundary-skeleton-guided
feature fusion, without incurring the quadratic computational cost of
full spatial self-attention.

\item A large multi-institutional cervical cancer radiotherapy CT dataset
from a South Asian LMIC setting (1,011 patients, three centers,
seven OAR/target classes), with rigorous quality control, inter-rater
reliability analysis (Jaccard $= 0.961 \pm 0.024$, $n = 7$ annotators),
patient-level stratified cross-validation, and external validation on an
independent 230-patient cohort.

\item A comprehensive quantitative evaluation demonstrating statistically
significant and clinically meaningful improvements over four baselines
(nnUNet, SegMamba, TransUNet, UNETR) across all seven anatomical
classes, with paired Wilcoxon tests and Bonferroni-Holm correction across
168 comparisons, Hedges' $g$ effect sizes, Bland-Altman boundary agreement
analysis, and failure mode characterization.

\item A multi-level interpretability analysis combining per-voxel error
maps, multi-model contour overlays, and three-method gradient-weighted class
activation mapping (GradCAM++, HiResCAM, XGradCAM) that provides mechanistic
evidence for BAT-RM's boundary and contextual advantages, supporting model
trustworthiness for clinical deployment.

\item A prospective multi-center reader study ($n = 13$ oncologists,
100 test cases) demonstrating that BAT-RM assistance reduces mean junior
contouring time from 152 to 29 minutes (80.7\% reduction), elevates junior
IoU from 0.899 to 0.965 (approaching senior unaided performance), and is
associated with reduced patient wait time (from 1--3 days to 2--3 hours)
following clinical deployment at a partner hospital.

\item A clinically deployed end-to-end system featuring a production-grade
web application with interactive contour editing and DICOM RTSTRUCT export,
natively compatible with Varian, RayStation, and Monaco treatment planning
systems, enabling direct clinical translation without intermediary processing.
\end{highlights}


\maketitle

\section{Introduction}
\label{sec:introduction}

Cervical cancer is the fourth leading cause of cancer death in women
worldwide, with over 85\% of cases occurring in low- and
middle-income countries (LMICs) where treatment infrastructure remains
chronically underdeveloped~\citep{kassa2024integration,kassa2024barriers}.
According to GLOBOCAN 2022, an estimated 662,301 new cases and 348,874
deaths occurred globally, with South Asian countries including Bangladesh
carrying nearly one-third of the burden~\citep{bray2024globocan,ferlay2024globocan,pmc2024awareness}.
In Bangladesh specifically, cervical cancer accounts for more than a quarter
of female malignancies, yet severe shortages of radiation oncologists
translate into radiotherapy waiting times of six months or more for many
patients~\citep{lavigne2017lmic,ascopost2017bangladesh}. Radiotherapy is a cornerstone of curative treatment for locally advanced
cervical cancer~\citep{hermes2021optimal,sturdza2021image}, but its safe
delivery hinges on precise manual delineation of target volumes (GTV, CTV)
and organs-at-risk (bladder, rectum, small bowel, femoral
heads)~\citep{potter2021clinical,fields2025consensus}.
Manual contouring requires 45 minutes to three hours per patient and suffers
from well-documented inter-observer variability~\citep{eminowicz2015variability,byrne2024quantifying}
---a finding corroborated by our multi-center inter-rater analysis (mean
pairwise Jaccard: $0.961 \pm 0.024$ across seven annotators from three
institutions). The combination of this variability and severe oncologist
shortages creates an urgent need for automated, clinically validated
auto-contouring~\citep{grover2022barriers,kabir2025nicrh}. Deep learning has transformed medical image segmentation, with U-Net
~\citep{ronneberger2015unet} and nnUNet~\citep{isensee2021nnunet}
establishing strong baselines. For cervical cancer, auto-segmentation has
shown promise for OARs and CTV~\citep{wang2022ebrt,chung2023frontonc,ma2022autoseg},
but three limitations persist. First, most models are trained on
single-institutional datasets that fail to capture multi-center
heterogeneity~\citep{manganaro2024multi}. Second, soft-tissue boundary
precision remains suboptimal with convolutional architectures lacking
explicit boundary mechanisms~\citep{wang2025boundary}. Third, clinical
utility is rarely evaluated through structured reader studies measuring
workflow efficiency and patient-level impact~\citep{brouwer2024clinical}. Transformer-based architectures like TransUNet and UNETR address long-range
dependency limitations via global self-attention~\citep{chen2021transunet,hatamizadeh2022unetr},
but their quadratic complexity ($\mathcal{O}(n^2)$) imposes prohibitive
memory demands at $512 \times 512$ resolution. State space models, particularly
Mamba~\citep{gu2023mamba}, offer linear-time complexity ($\mathcal{O}(n)$)
via selective scanning, with SegMamba~\citep{xing2024segmamba} adapting this
to medical segmentation. However, two fundamental challenges remain unresolved
for cervical cancer auto-contouring: the \emph{boundary precision problem}
(2--5 mm reported Hausdorff distances) and the \emph{long-range context
problem} (scattered small bowel morphology). Neither is adequately addressed
by any single existing paradigm.

\begin{figure}[pos=htbp]
\centering
\includegraphics[width=\textwidth]{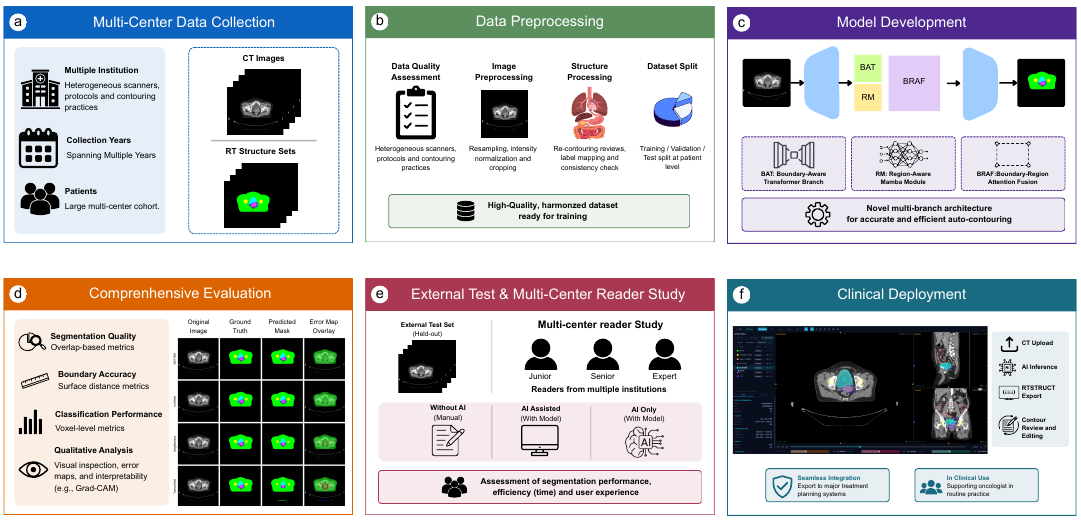}
\caption{End-to-end pipeline of the BAT-RM system, from multi-center
data collection (1,011 patients, three centers) through preprocessing,
model development, comprehensive evaluation, reader study (13 oncologists,
100 cases), and clinical deployment.}
\label{fig:workflow}
\end{figure}

We introduce BAT-RM (Boundary-Aware Transformer with Region-Aware Mamba), a hybrid 2D multi-branch architecture that addresses both challenges
simultaneously. BAT-RM integrates three complementary streams: a Gated
Boundary-Aware Transformer (BAT) branch with Sobel-derived gradient gating
to restrict attention to organ boundaries; a Multi-Directional Recurrent
Context Module with Mamba selective scanning (MRCM) for $\mathcal{O}(n)$
long-range scanning; and a Boundary-Region Attention Fusion (BRAF) gate for
adaptive feature combination. The model is trained with a composite loss
supervising both segmentation quality (Soft Dice, Focal) and boundary
placement accuracy. Figure~\ref{fig:workflow} overviews our end-to-end pipeline. We train and evaluate BAT-RM on the largest multi-institutional cervical cancer CT dataset
from Bangladesh (1,011 patients, three centers), assessing performance via
volumetric overlap, surface distance, and classification metrics including
Normalized Surface Distance~\citep{maier2024metrics,antonelli2022msd}.
Beyond geometric evaluation, we conduct a prospective multi-center reader
study (13 oncologists, 100 cases) measuring AI assistance impact on
accuracy, workflow time, clinician confidence, and patient wait time.
This is the first multi-center reader study for cervical cancer
auto-contouring in an LMIC setting, quantifying patient-level throughput
impact of AI-assisted deployment. Beyond geometric evaluation, we also validate generalization on an independent
external cohort of 230 patients from a fourth institution (United Hospital
Limited), confirming performance consistency.

Crucially, this work extends to end-to-end clinical deployment. We have developed a production-grade web application with interactive contour editing and DICOM RTSTRUCT export compatible with Varian, RayStation, and Monaco, building on our prior experience deploying attention-driven diagnostic AI as clinician-facing tools in resource-constrained settings~\citep{ahmed2025oncovisionintegratingmammographyclinical}. Following the reader study, BAT-RM was deployed at Bangladesh
Medical University and is actively used in routine practice. The system
reduces patient wait times from days to hours, enabling earlier treatment
initiation in resource-constrained settings where radiotherapy demand far
exceeds specialist capacity.

The remainder of this paper is organized as follows. Section~\ref{sec:data}
describes the multi-center dataset and ethical framework.
Section~\ref{sec:annotation_preprocessing} details annotation and quality
assurance, followed by preprocessing in Section~\ref{sec:data_preprocessing}.
The BAT-RM architecture and training are presented in
Section~\ref{sec:model_architecture}, with ablation studies in
Section~\ref{sec:ablation}. Quantitative and qualitative results follow in
Sections~\ref{sec:Quantitative Analysis} and \ref{sec:qualitative},
respectively. Section~\ref{sec:reader_study} presents the multi-center reader
study, Section~\ref{sec:deployment} discusses clinical inference and web deployment, Section~\ref{sec:limitations} discusses limitations, and
Section~\ref{sec:conclusion} concludes.

\section{Related Work}
\label{sec:related_work}

Automated contouring for cervical cancer radiotherapy has evolved from
atlas-based methods through convolutional networks to hybrid transformer
and state space architectures. This section reviews this progression,
identifies limitations motivating BAT-RM, and situates our approach.

Early multi-atlas registration methods demonstrated acceptable performance
on high-contrast pelvic structures but struggled with bladder/rectum volume
variability, poor GTV/CTV soft-tissue contrast, and discontinuous small
bowel morphology~\citep{chen2022atlascervix}. Atlas-based contours required
editing times comparable to de-novo manual contouring, undermining workflow
efficiency~\citep{chung2023frontonc}. U-Net~\citep{ronneberger2015unet} and its variants, including Attention
U-Net~\citep{oktay2018attnunet} and nnUNet~\citep{isensee2021nnunet},
established strong baselines for pelvic segmentation. In cervical cancer,
DpnUNet~\citep{liu2020dpnunet} achieved CTV DSC of 0.86 with 5.34 mm HD95;
Ma et al.~\citep{ma2022autoseg} reported DSC improvements up to 0.20 for
juniors with AI assistance; PPAF-net~\citep{tian2023ppafnet} fused texture
and boundary streams; and Wang et al.~\citep{wang2022ebrt} demonstrated
clinical feasibility. However, convolutional methods suffer from limited
receptive fields, yielding 3.5--8.0 mm HD95 for target volumes, and most
were trained on single-institution data. Recent multi-institutional studies
confirm performance degradation across centers without fine-tuning
~\citep{zhang2024pelvic,liu2024multicentre}, motivating our 1,011-patient
dataset. Vision Transformers~\citep{dosovitskiy2020vit} and Swin Transformers
~\citep{liu2021swin} offer global receptive fields. TransUNet~\citep{chen2021transunet}
and UNETR~\citep{hatamizadeh2022unetr} integrate transformers into U-Net
frameworks, while SwinUNETR~\citep{hatamizadeh2022swinunetr} achieves
multi-resolution feature extraction. For cervical cancer, Chung et
al.~\citep{chung2023frontonc} demonstrated a 30-minute time reduction with
AI assistance, but their model was single-institution without formal
inter-rater analysis. The fundamental limitation remains quadratic
self-attention complexity ($\mathcal{O}(n^2)$), making full-resolution
$512 \times 512$ processing prohibitive on clinical hardware. State space models, particularly Mamba~\citep{gu2023mamba}, offer linear-time
complexity ($\mathcal{O}(n)$) via selective scanning. U-Mamba~\citep{ma2024umamba}
and SegMamba~\citep{xing2024segmamba} adapted this to medical segmentation.
In cervical cancer, AM-UNet~\citep{li2025amunetu} applied Mamba to brachytherapy
HRCTV segmentation but was single-institution and did not address full EBRT
structures. Our work extends this by integrating true Mamba SSM with
four-directional scanning in a multi-institutional cervical cancer
auto-contouring framework. Boundary loss~\citep{kervadec2021boundary}, dual-network approaches
~\citep{nie2022dualnet}, and parallel-path fusion~\citep{tian2023ppafnet}
incorporate boundary supervision but operate at loss or fusion level, not
gating attention computation itself. Multi-branch architectures like
PPAF-net combine local and global streams, but none combine gradient-gated
boundary attention with a recurrent global context module---the defining
characteristic of BAT-RM.

Reader studies measure AI assistance impact on oncologist performance and
workflow. Ma et al.~\citep{ma2022autoseg} reported junior DSC improvements
up to 0.20 and time reductions up to 28.9 minutes; Chung et al.
~\citep{chung2023frontonc} reported 30-minute reductions without experience
stratification; Wang et al.~\citep{wang2022ebrt} evaluated workflow but
included no reader study. No existing cervical cancer reader study was
conducted in an LMIC setting. A recent multi-center LMIC trial
~\citep{chen2024aideployment} reported 78\% time reduction (158→34 minutes)
for juniors. Head-and-neck evidence~\citep{wong2020deepoar} shows similar
benefits. Our study is the first, to our knowledge, to document patient-level
throughput impact of AI-assisted cervical cancer contouring in a deployed
LMIC clinical system.

Five inter-related gaps motivate BAT-RM: (1) boundary precision:
no existing method gates self-attention to boundary regions during feature
learning; (2) long-range context: CNNs lack global receptive fields,
transformers have quadratic cost, and existing Mamba-based methods do not
exploit multi-directional spatial scanning; (3) architecture: no unified
framework combines gradient-gated boundary attention, four-directional
recurrent scanning, and skeleton-guided fusion; (4) data and validation:
prior methods use small single-institution datasets without multi-center
generalization analysis; (5) clinical deployment no prior work quantifies
patient-level impact (queue reduction, wait time) in an LMIC setting.
BAT-RM addresses all five through its architectural design, 1,011-patient
multi-center dataset, comprehensive evaluation, and prospective reader study
with clinical deployment.

\section{Methodology}

\subsection{Data Collection}
\label{sec:data}

All clinical data used in this study were acquired retrospectively from
three tertiary-care oncology centers in Bangladesh under institutional
data-sharing agreements: Bangladesh Medical University (BMU), Square Hospital Limited (SHL), and
Labaid Specialized Hospital (LSH). Each center operates
a dedicated radiotherapy planning workflow for gynecological malignancies,
collectively providing a heterogeneous cohort spanning different CT scanner
manufacturers, acquisition protocols, and clinical contouring practices.
While the three hospitals are located in the Dhaka metropolitan region,
they serve as major national referral centers for cervical cancer
radiotherapy, drawing patients from all divisions of the country
as well as from neighboring South Asian countries with
limited radiotherapy infrastructure. This national catchment diversity
ensures that the dataset reflects the full spectrum of disease stage,
anatomical variation, and patient demographics encountered in routine
clinical practice across the country. The final dataset comprises 1,011 patients diagnosed with cervical
carcinoma and referred for external-beam radiotherapy or brachytherapy
planning — 424 from BMU (2020--2026), 309 from SHL (2018--2026), and 278
from LSH (2019--2025). For each patient, a planning CT series and its
corresponding DICOM RT-Structure Set were acquired. All CT volumes were
reconstructed on a $512 \times 512$ in-plane grid with slice thicknesses
between $2.5\,\text{mm}$ and $3.0\,\text{mm}$. The RT-Struct files encode
manual contours delineated by experienced radiation oncologists, covering
the gross tumor volume (GTV), clinical target volume (CTV), bladder,
rectum, small bowel, bilateral femoral heads, and the external body
contour. Structure names were harmonized across institutions using a
controlled vocabulary mapping to account for site-specific naming
conventions. Cases were excluded where image quality was deemed
insufficient for reliable contouring, including scans affected by severe
metal artifacts, truncated fields of view, corrupted DICOM geometry, or
incomplete OAR contour sets. This retrospective study was conducted in
accordance with the principles of the Declaration of Helsinki. Ethical
approval was obtained from the institutional review boards of Bangladesh
Medical University, Square Hospital Limited, and Labaid Specialized
Hospital prior to data collection. All patient records were de-identified
in compliance with DICOM anonymization standards before data transfer,
storage, and analysis; no personally identifiable information was retained
in the working dataset at any stage of the study. The requirement for
individual patient informed consent was waived by each ethics committee
given the retrospective, non-interventional nature of the study and the
absence of any direct patient contact.

\subsection{Data Annotation, Quality Assurance, and Inter-Rater Reliability}
\label{sec:annotation_preprocessing}

Structural contours from three tertiary-care centers were exported as DICOM
RT-Structure Sets. Delineations followed ESTRO and ABS consensus guidelines,
encompassing eight classes: body, bladder, small bowel, rectum, bilateral
femoral heads, GTV, CTV, and background. Structure names were harmonized
across institutions, and all RTSTRUCT files underwent geometric validation.
Cases with truncated fields of view, severe metal artifacts, corrupted
geometry, or incomplete contours were excluded, yielding 1,011 patients.

\begin{figure}[pos=htbp]
    \centering
    \includegraphics[width=\linewidth]{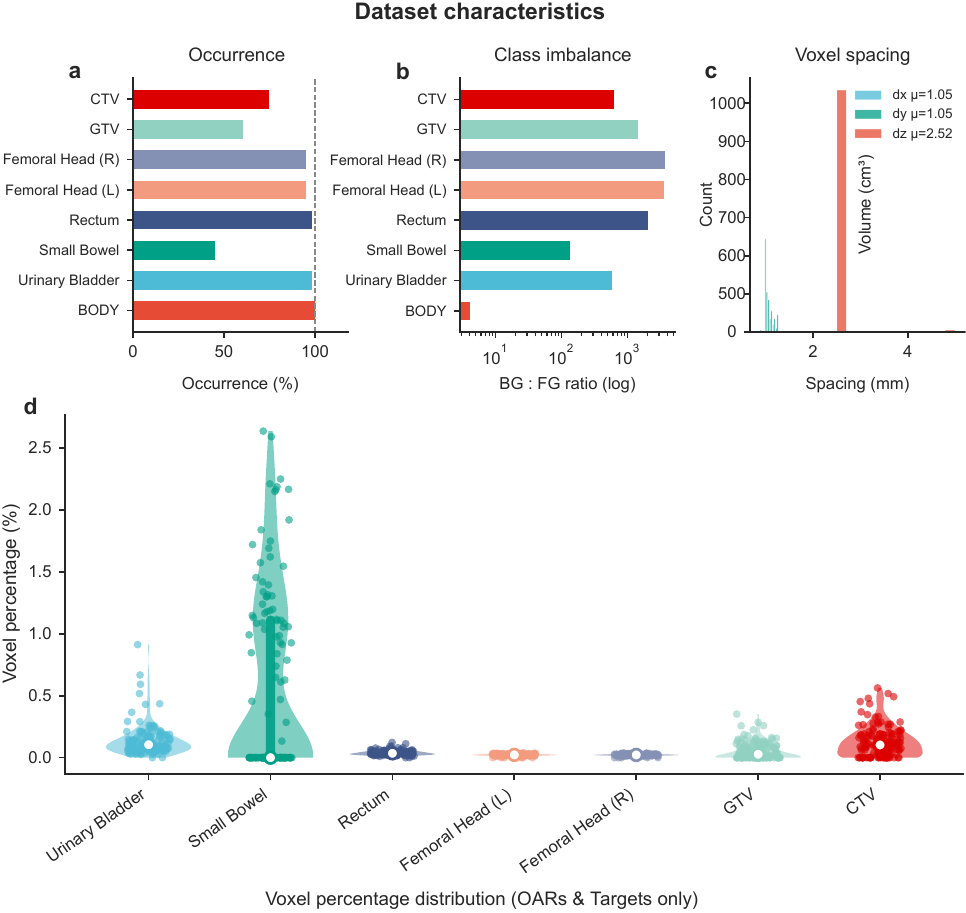}
    \caption{Dataset characteristics: (a) class occurrence frequency;
    (b) background-to-foreground voxel ratios (log scale); (c) voxel
    spacing distribution; (d) voxel percentage distribution.}
    \label{fig:dataset_characteristics}
\end{figure}

\begin{figure}[pos=htbp]
    \centering
    \includegraphics[width=\linewidth]{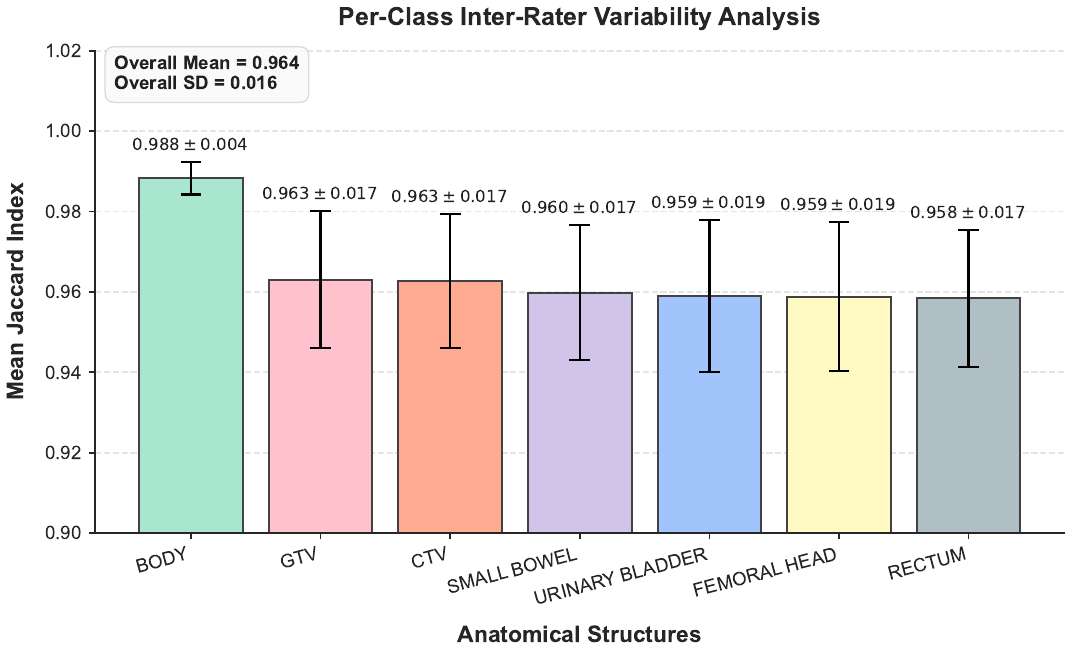}
    \caption{Inter-rater reliability: distribution of pairwise Jaccard
    similarity indices ($0.961 \pm 0.024$), demonstrating consistent
    contouring across institutions.}
    \label{fig:inter_rater}
\end{figure}

Inter-rater reliability was assessed on 100 cases independently contoured
by 7 oncologists across all experience levels and institutions. Pairwise
Jaccard indices averaged $0.961 \pm 0.024$ (Figure~\ref{fig:inter_rater}),
indicating high inter-observer agreement. Discrepancies at organ-fat
interfaces or in cases of bowel distension or post-surgical changes were
resolved by multi-rater consensus.

As shown in Figure~\ref{fig:dataset_characteristics}, body, femoral heads,
bladder, and rectum appeared in $\geq 95.3\%$ of scans; GTV and CTV in
$60.9\%$ and $75.0\%$; small bowel in $45.3\%$. Volumetric variability was
pronounced, particularly for small bowel ($868.19 \pm 1097.31$ cm$^3$) and
GTV ($77.18 \pm 97.43$ cm$^3$). Background-to-foreground ratios ranged from
$4.08:1$ (body) to $3697:1$ (femoral head), informing our class-weighted
loss formulation.

CT series were preprocessed using ITK v5.3 and SimpleITK v2.3. HU values
were clipped to $[-150, 250]$ HU (soft-tissue window) and normalized to
$[0, 1]$. In-plane spacing had low variability ($d_x, d_y$ mean $1.054$ mm),
while slice thickness varied ($d_z$ median $2.5$ mm, range $2.5$--$5.0$ mm).
All volumes were resampled to $1.0 \times 1.0 \times 2.5$ mm$^3$ using
cubic B-spline (CT) and nearest-neighbor (masks) interpolation, retaining
native $512 \times 512$ resolution.

\subsection{Data Preprocessing}
\label{sec:data_preprocessing}
\subsubsection{DICOM to NIfTI Conversion and 2D Slice Extraction}

The raw clinical data consisted of CT images in DICOM format accompanied by radiotherapy structure sets (RT-Struct) containing expert oncologist annotations. A multi-stage preprocessing pipeline was developed within the 3D Slicer environment to extract, standardize, and prepare the data for model training.

\subsubsection{Volume Extraction and NIfTI Export}

For each patient, the CT volume and its corresponding RT-Struct segmentation were loaded into 3D Slicer. The CT scalar volume was exported directly as a compressed NIfTI file (\texttt{.nii.gz}) to preserve the original voxel intensities and spatial metadata. Eight anatomical structures relevant to pelvic radiotherapy were identified and assigned standardized integer label values, as summarized in Table~\ref{tab:label_mapping}.

\begin{table}[h]
\centering
\caption{Standardized label mapping for segmentation structures.}
\label{tab:label_mapping}
\begin{tabular}{lcc}
\toprule
\textbf{Structure} & \textbf{Label Value} & \textbf{RGB Color} \\
\midrule
Background        & 0 & (0, 0, 0) \\
Body              & 1 & (0, 255, 0) \\
Urinary Bladder   & 2 & (0, 255, 255) \\
Small Bowel       & 3 & (255, 146, 153) \\
Rectum            & 4 & (128, 64, 64) \\
Femoral Head (RT) & 5 & (255, 255, 0) \\
Femoral Head (LT) & 6 & (255, 255, 0) \\
GTV               & 7 & (255, 60, 255) \\
CTV               & 8 & (55, 55, 255) \\
\bottomrule
\end{tabular}
\end{table}

\noindent The multi-label segmentation mask was constructed by iterating over the
structures in a predefined rendering order, ensuring that higher-priority
structures (e.g., GTV, CTV) were drawn last and therefore not overwritten by
lower-priority ones. For the purpose of model training and evaluation, the
left and right femoral heads were merged into a single \texttt{Femoral Head}
class, as bilateral symmetry makes independent segmentation clinically
unnecessary for radiotherapy planning and merging increases the effective
training sample size for this structure. The final labelmap was exported as a
NIfTI file with integer voxel values corresponding to the mapping in
Table~\ref{tab:label_mapping}. Clinical RTSTRUCT files encode each anatomical structure as an independent binary mask, and structures may overlap in physical space: the body contour encompasses all internal organs, and the CTV conventionally includes the GTV as a sub-volume. To train a standard multi-class segmentation network with softmax output, we converted these overlapping binary masks into a single mutually-exclusive integer labelmap using a predefined priority ordering. Structures with higher clinical delineation priority were assigned higher label indices and rendered last, so that their voxels overwrite lower-priority structures in regions of spatial overlap. The priority order, from lowest to highest, is: Body (1) $<$ Bladder (2) $<$ Small Bowel (3) $<$ Rectum (4) $<$ Femoral Head R (5) $<$ Femoral Head L (6) $<$ CTV (8) $<$ GTV (7). GTV receives the highest priority because it is the innermost, most
clinically critical target volume; in cases where GTV and CTV overlap,
the overlapping voxels are assigned to GTV. The Body contour requires special handling. In the original RTSTRUCT, Body is defined as the external patient boundary encompassing all internal structures. In the mutually-exclusive labelmap, Body voxels that are overwritten by internal organs receive those organs' labels and are not
labeled as Body. At RTSTRUCT export time, the Body mask is reconstructed as the union of all non-background predicted voxels rather than solely the voxels assigned label 1, thereby restoring the clinically correct nested definition. This reconstruction is performed automatically by the inference pipeline described in Section~\ref{sec:deployment}. This mutually-exclusive training formulation is consistent with established
practice in multi-organ radiotherapy segmentation~\citep{isensee2021nnunet,
liu2020dpnunet,tian2023ppafnet} and has been validated as clinically
acceptable in the reader study: no oncologist reported anatomically
inconsistent contour nesting in any of the 100 evaluated cases.

\subsubsection{CT Windowing and 2D Slice Generation}

Prior to slice extraction, the CT Hounsfield Unit (HU) values were windowed
to enhance soft-tissue contrast, which is critical for pelvic organ
segmentation in radiotherapy planning. A window center of $W_c = 50$\,HU
and window width of $W_w = 400$\,HU were applied, corresponding to the
intensity range $[-150,\, 250]$\,HU. The windowed values were then linearly
normalized to the $[0, 255]$ range as:

\begin{equation}
    I_{\text{norm}}(x,y,z) = \frac{I_{\text{clip}}(x,y,z) -
    (W_c - W_w/2)}{W_w} \times 255
\end{equation}

\noindent where $I_{\text{clip}}$ denotes the clipped HU values. The choice of a single soft-tissue window for all structures---including
the GTV and CTV---was deliberate and is justified by three considerations
specific to this dataset and task. First, cervical cancer GTV on planning
CT is a soft-tissue tumor embedded in pelvic soft tissue; its Hounsfield
Unit values ($+20$ to $+60$\,HU) fall entirely within the $[-150, 250]$\,HU
window, meaning the tumor is not rendered invisible by this windowing choice
as it would be by, for example, a bone window ($W_c = 400$\,HU,
$W_w = 1500$\,HU). Second, the primary segmentation challenge for the GTV
and CTV in this dataset is not intensity contrast per se---the GTV frequently
appears isoattenuating relative to surrounding parametrial tissue regardless
of windowing---but rather the precise delineation of a low-contrast boundary
that depends on learned spatial context rather than windowing optimization.
A multi-window approach would improve the visibility of high-contrast
structures (femoral head cortex, gas-filled bowel) that are already
well-segmented by all models, without meaningfully improving GTV boundary
discrimination. Third, using a single window preserves consistency across the
multi-institutional cohort: different institutions acquired CT studies with
varying reconstruction kernels and scanner manufacturers, and a soft-tissue
window provides the most stable normalized representation across this
heterogeneity. The GTV and CTV segmentation quality achieved by BAT-RM
(Dice: 0.966 and 0.957 respectively, HD95: 1.50 and 1.51\,mm) confirms
that single-window preprocessing is not a performance-limiting factor for
this task. The normalized volume was subsequently sliced along all three anatomical
planes (axial, coronal, sagittal), with each slice resized to
$512 \times 512$ pixels and saved as an RGB PNG image by replicating the
grayscale channel across three channels.

\subsubsection{Axial Slice Selection for Training}

Although slices were generated in all three orientations for potential inference-time visualization, only \textbf{axial slices} were used for model training. This design choice reflects standard clinical radiotherapy practice: oncologists delineate target volumes and organs-at-risk exclusively on axial CT slices. Training on axial views thus ensures the model learns from the same perspective used during clinical annotation. Three-dimensional reconstructions along coronal and sagittal planes can be recovered at inference time by stacking the predicted axial segmentation masks.

\subsubsection{Dataset Splitting with Patient-Level Stratification}
The dataset was split using institution-stratified random sampling into
training, validation, and test sets, preserving class and scanner
proportions. Preprocessing parameters were computed exclusively on the
training set to prevent data leakage. All scripts and harmonization
mappings are version-controlled for reproducibility. To prevent \textit{patient data leakage} between training, validation, and test sets—a common source of optimistic bias in medical image segmentation studies—the dataset was partitioned strictly at the patient level. All axial slices belonging to a given patient were assigned exclusively to a single split. Furthermore, to support multi-center generalization, the split was stratified such that each subset retained a representative distribution of patients across acquisition sites and clinical characteristics. The complete preprocessing pipeline is summarized in Algorithm~\ref{alg:preprocessing}.

\begin{algorithm}[t]
\caption{Data Preprocessing Pipeline}
\label{alg:preprocessing}
\KwIn{CT DICOM series, RT-Struct file, patient ID $p$, output directory $\mathcal{D}$}
\KwOut{Axial CT image slices $\{I_z\}$ and corresponding mask slices $\{M_z\}$ for all patients}

\textbf{Step 1: Load and identify data in 3D Slicer}\;
Load CT scalar volume node $V$ and segmentation node $S$ into scene\;
Extract patient identifier $p$ from DICOM metadata (tag \texttt{0010,0020})\;

\textbf{Step 2: Export CT volume as NIfTI}\;
$\mathcal{N}_{\text{img}} \leftarrow$ \textsc{ExportVolumeAsNIfTI}($V$, $\mathcal{D}/p\_\text{Base\_Nifty.nii.gz}$)\;

\textbf{Step 3: Build multi-label segmentation mask}\;
Initialize labelmap $L \leftarrow \mathbf{0}$ with geometry from $S$\;
\ForEach{structure $s$ in desired rendering order}{
    $\lambda \leftarrow$ \textsc{GetStandardLabel}($s$) \tcp*{Table~\ref{tab:label_mapping}}
    $B_s \leftarrow$ \textsc{ExtractBinaryLabelmap}($S$, $s$)\;
    $L[B_s > 0] \leftarrow \lambda$\;
}
$\mathcal{N}_{\text{mask}} \leftarrow$ \textsc{ExportLabelmapAsNIfTI}($L$, $\mathcal{D}/p\_\text{Labeled\_Nifty.nii.gz}$)\;

\textbf{Step 4: CT windowing and normalization}\;
$W_c \leftarrow 50$, $W_w \leftarrow 400$\;
$I \leftarrow$ \textsc{LoadArray}($\mathcal{N}_{\text{img}}$)\;
$I_{\text{clip}} \leftarrow \text{clip}(I,\; W_c - W_w/2,\; W_c + W_w/2)$\;
$I_{\text{norm}} \leftarrow \dfrac{I_{\text{clip}} - (W_c - W_w/2)}{W_w} \times 255$\;

\textbf{Step 5: Extract and save 2D axial slices}\;
$M \leftarrow$ \textsc{LoadArray}($\mathcal{N}_{\text{mask}}$)\;
\For{$z = 0$ \KwTo $Z-1$ \tcp*[h]{$Z$ = number of axial slices}}{
    $I_z \leftarrow I_{\text{norm}}[:, :, z]$; resize to $512 \times 512$; convert to RGB\;
    $M_z \leftarrow M[:, :, z]$; colorize via Table~\ref{tab:label_mapping}; resize to $512 \times 512$\;
    Save $I_z$ to $\mathcal{D}/p\_\text{axial\_base}/p\_z.\text{png}$\;
    Save $M_z$ to $\mathcal{D}/p\_\text{axial\_mask}/p\_z.\text{png}$\;
}

\textbf{Step 6: Patient-level stratified dataset splitting}\;
$\mathcal{P} \leftarrow \{p_1, p_2, \ldots, p_N\}$ \tcp*{All patient IDs}
$[\mathcal{P}_{\text{train}}, \mathcal{P}_{\text{val}}, \mathcal{P}_{\text{test}}] \leftarrow$ \textsc{StratifiedPatientSplit}($\mathcal{P}$)\;
\ForEach{split $\mathcal{P}_k \in \{\mathcal{P}_{\text{train}}, \mathcal{P}_{\text{val}}, \mathcal{P}_{\text{test}}\}$}{
    Assign \textbf{all} axial slices of $p \in \mathcal{P}_k$ exclusively to split $k$\;
}

\Return{$\{(I_z, M_z)\}$ partitioned into train/val/test sets}
\end{algorithm}

\subsection{Model Architecture}
\label{sec:model_architecture}

\begin{figure}[pos= htbp]
    \centering
    \includegraphics[width=\linewidth]{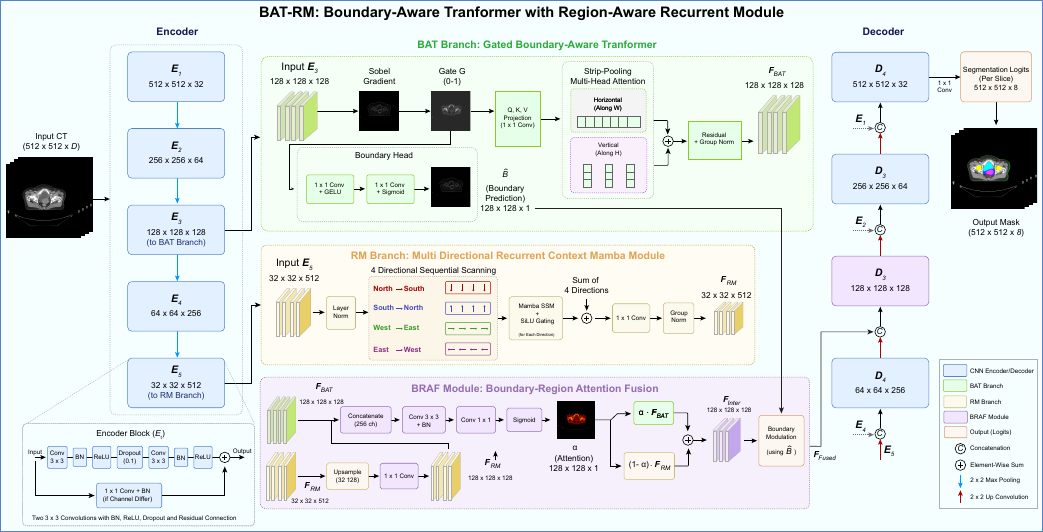}
    \caption{Architecture of the proposed BAT-RM (Boundary-Aware Transformer
    with Region-Aware Mamba). The model integrates a hierarchical CNN
    encoder-decoder, a Gated Boundary-Aware Transformer (BAT) branch with
    Sobel-gated attention for organ boundaries, a Multi-Directional Recurrent
    Context Module with Mamba selective scanning (RM branch) for long-range
    spatial dependencies, and a Boundary-Region Attention Fusion (BRAF)
    gate for adaptive feature combination. Training uses a composite loss
    of Soft Dice, Focal, boundary BCE, and smoothness regularization.}
    \label{fig:architecture}
\end{figure}

The two challenges identified in Section~\ref{sec:introduction}---boundary
imprecision at the GTV--parametrium interface and the long-range context
required to track scattered small bowel loops across the full pelvic
extent---motivate a three-branch architecture in which each branch addresses
one specific geometric property of the pelvic radiotherapy segmentation task.
We propose the Boundary-Aware Transformer with Region-Aware Mamba (BAT-RM), a hybrid 2D multi-branch architecture that synergistically
integrates four complementary modeling paradigms: (i) a hierarchical
convolutional encoder-decoder; (ii) a Gated Boundary-Aware Transformer (BAT)
branch with Sobel-gated boundary attention; (iii) a Multi-Directional
Recurrent Context Module with Mamba selective scanning (RM branch) that
captures long-range spatial dependencies at linear computational complexity;
and (iv) a Boundary-Region Attention Fusion (BRAF) gate that adaptively
combines boundary-aware and region-aware features for the decoder.
The architecture is specifically designed for the simultaneous auto-contouring
of $C = 8$ anatomical classes in cervical cancer radiotherapy planning.
Figure~\ref{fig:architecture} presents a schematic overview of the complete
BAT-RM. While gradient-gated boundary attention, recurrent sequential
scanning, and boundary-supervised feature fusion have each been explored
independently in the literature~\citep{kervadec2021boundary, xing2024segmamba,
nie2022dualnet}, BAT-RM integrates all three within a unified framework
explicitly designed for pelvic radiotherapy auto-contouring, and validates
this combination through a prospective multi-center clinical reader study.
The contribution of each architectural component is further validated through
a systematic ablation study presented in Section~\ref{sec:ablation}.

Given an input CT volume $X \in \mathbb{R}^{H \times W \times D}$, we adopt a
slice-wise processing paradigm where each axial slice $x_i$ is processed
independently by a 2D network to produce per-class probability maps
$\hat{Y}_i \in \mathbb{R}^{H \times W \times C}$. To restore inter-slice
coherence, a 3D Gaussian kernel $\mathcal{G}_{\sigma}$ is applied along the
superior-inferior axis:
\begin{equation}
    \hat{Y}_{\text{3D}} = \text{Softmax}(\mathcal{G}_{\sigma} *
    \text{Logits}_{\text{stacked}})
    \label{eq:3d_smooth}
\end{equation}
This post-processing enforces axial continuity while requiring only
3.8 GB peak VRAM, a fraction of the 12--16 GB needed by 3D architectures
at equivalent resolution.

\subsubsection{Hierarchical CNN Encoder-Decoder Backbone}
\label{subsec:cnn_backbone}

The backbone is a symmetric U-Net with five encoder levels. Each encoder block
$E_l$ ($l = 1,\ldots,5$) contains two $3 \times 3$ convolutional layers,
each followed by Batch Normalization, ReLU, and spatial Dropout (0.1), plus a
residual connection via $1 \times 1$ projection when channel dimensions differ:
\begin{equation}
    E_l = \sigma(\text{BN}(W_l^{(2)} * \sigma(\text{BN}(W_l^{(1)} *
    E_{l-1})))) + \mathcal{P}(E_{l-1})
    \label{eq:encoder}
\end{equation}
Spatial downsampling uses $2 \times 2$ max pooling. For a $512 \times 512$
input, the encoder produces feature maps with channels $\{32, 64, 128, 256, 512\}$:
$E_1$ (edge/texture), $E_2$ (structural), $E_3$ (boundary anchor, routed to BAT),
$E_4$ (semantic), and $E_5$ (bottleneck, routed to RM).

The decoder uses transposed convolutions with skip connections. At the
$128 \times 128$ level, the raw $E_3$ skip is replaced by the BRAF module's
output $F_{\text{fuse}}$:
\begin{equation}
    D_l = \sigma\!\left(\text{BN}\!\left(W_{l,\text{out}} *
    \sigma\!\left(\text{BN}\!\left(W_{l,\text{up}} \circledast
    [\mathcal{U}(D_{l+1})\,;\,S_l]\right)\right)\right)\right)
    \label{eq:decoder}
\end{equation}
where $\circledast$ is transposed convolution, $\mathcal{U}$ is bilinear
upsampling, and $S_l = E_l$ for $l \neq 3$ and $S_3 = F_{\text{fuse}}$.

\subsubsection{Gated Boundary-Aware Transformer (BAT) Branch}
\label{subsec:bat_branch}

\begin{figure}[pos=htbp]
    \centering
    \includegraphics[width=\linewidth]{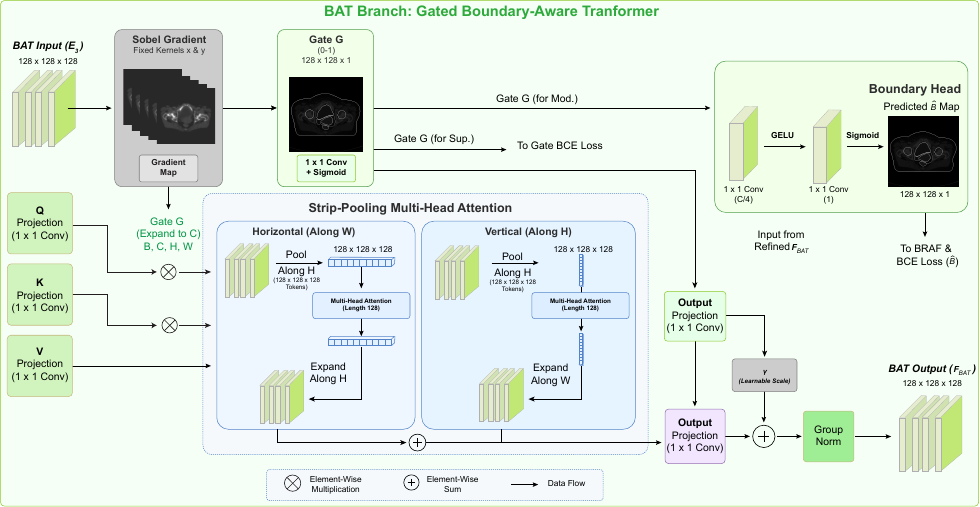}
    \caption{Architecture of the Gated Boundary-Aware Transformer (BAT) branch. The Sobel operator computes gradient magnitudes from $E_3$ features, which are collapsed via a learned $1\times1$ convolution and sigmoid to produce the spatial gate $\mathcal{G}$. The gate modulates Query ($Q$) and Key ($K$) projections before strip-pooling multi-head self-attention, enabling boundary-focused attention with linear complexity. A boundary prediction head produces $\hat{B}$ for auxiliary supervision.}
    \label{fig:architecture_BAT}
\end{figure}

To explicitly model the precise perimeters of anatomical structures, the BAT
branch introduces a Sobel-Gated Strip-Pooling Multi-Head
Self-Attention (G-MSA) mechanism that directs the transformer's attention
toward high-gradient boundary regions without the quadratic cost of full
spatial self-attention (Figure~\ref{fig:architecture_BAT}).


Given the boundary anchor features $E_3 \in \mathbb{R}^{H_3 \times W_3
\times C_3}$ (where $H_3 = W_3 = 128$, $C_3 = 128$), we compute a spatial
gradient magnitude map using fixed (non-trainable) Sobel operators
approximating first-order image derivatives in the horizontal ($G_x$) and
vertical ($G_y$) directions:
\begin{equation}
    \nabla E_3 = \sqrt{(G_x * E_3)^2 + (G_y * E_3)^2 + \epsilon}
    \label{eq:sobel}
\end{equation}
where $\epsilon = 10^{-6}$ ensures numerical stability. The multi-channel
gradient map is collapsed into a scalar spatial gate $\mathcal{G} \in
[0,1]^{H_3 \times W_3}$ via a learned $1 \times 1$ convolution followed by
sigmoid activation:
\begin{equation}
    \mathcal{G} = \sigma(\text{Conv}_{1 \times 1}(\nabla E_3))
    \label{eq:gate}
\end{equation}
This gate assigns values close to 1 at organ boundary pixels and suppresses
homogeneous tissue regions, providing a spatially selective modulation signal
for the subsequent attention computation.

The Sobel gate $\mathcal{G}$ is computed dynamically from the $E_3$ feature
activations ($\mathbb{R}^{B \times 128 \times 128 \times 128}$), not from
the raw CT slice. Sobel operators are applied channel-wise to $E_3$, and
per-channel gradient magnitudes are collapsed into a scalar spatial gate
via a learned $1 \times 1$ convolution and sigmoid (Eqs.~\ref{eq:sobel}--
\ref{eq:gate}). The sigmoid activation bounds the gate to $[0,1]$,
preventing unbounded gradient signals that could destabilize early training. The gate $\mathcal{G}$ is recomputed each forward pass from
current feature activations. Three architectural properties ensure that boundary-discriminative information is preserved at $E_3$. First, moderate abstraction depth:
$E_3$ undergoes only three $2 \times 2$ pooling steps reducing spatial
dimensions from $512 \times 512$ to $128 \times 128$, retaining substantial
spatial detail; the bottleneck $E_5$ at $32 \times 32$ is too compressed
for boundary gating. Second, residual connections preserve gradient
responses: each encoder block incorporates a residual pathway
(Eq.~\ref{eq:encoder}) that adds the input directly to the transformed
output, preventing suppression of high-magnitude boundary responses and
propagating them as an additive component of $E_3$ activations. Third,
batch normalization preserves relative gradient magnitude: BN standardizes
activations while maintaining the relative ordering where high-contrast
boundary pixels retain higher normalized values than background pixels;
subsequent ReLU activation further amplifies this effect. The Sobel gate
thus responds to learned feature-space boundaries such as GTV-parametrium
interfaces, bladder wall, and rectal mucosa, rather than raw Hounsfield
unit edges including noise, beam-hardening artifacts, and air-tissue
boundaries irrelevant for OAR delineation. Empirical validation confirms that $E_3$ preserves boundary information.
The Pearson correlation between the spatial gate $\mathcal{G}$ and the
ground truth boundary map $B_{\text{gt}}$ across the 254-case test set was
$r = 0.71 \pm 0.09$ for GTV and $r = 0.68 \pm 0.11$ for CTV. While
applying Sobel operators to deep feature maps lacks the theoretical
grounding of image-space edge detection, this strong correlation
empirically validates that the $E_3$ feature space preserves
boundary-discriminative information. The Sobel gate serves as a fixed,
interpretable edge detector rather than a learned attention mechanism,
chosen to avoid the additional parameters and training instability of a
learnable boundary gating network. The GradCAM++ analysis in
Section~\ref{subsec:gradcam} further validates this: the mid-layer
($E_3$) attention maps show tight localization to the GTV boundary
perimeter, consistent with a gate that successfully concentrates attention
on clinically relevant tissue interfaces.


Full spatial self-attention at the $128 \times 128$ feature resolution would
require a $16,384 \times 16,384$ attention matrix, consuming approximately
4\,GB of GPU memory per sample --- impractical for single-card deployment.
We therefore employ strip-pooling attention~\citep{hou2020strip}:
the feature map is pooled along each spatial axis independently, forming two
1D attention problems of size $H_3 = 128$ (column strips) and $W_3 = 128$
(row strips). This reduces the attention complexity from $\mathcal{O}(n^2)$
to $\mathcal{O}(H_3 + W_3)$, enabling efficient long-range modeling along
organ perimeter contours without the quadratic memory bottleneck.
The feature map $E_3$ is first projected into Query ($Q$), Key ($K$), and
Value ($V$) representations via learned $1 \times 1$ convolutions. The
boundary gate $\mathcal{G}$ is broadcast across the channel dimension and
element-wise multiplied with $Q$ and $K$ before attention computation:
\begin{equation}
    Q_{\text{gate}} = Q \odot \mathcal{G}, \quad
    K_{\text{gate}} = K \odot \mathcal{G}
    \label{eq:gating}
\end{equation}
\noindent For each strip direction, $Q_{\text{gate}}$ and $K_{\text{gate}}$ are pooled
along the complementary axis to produce 1D token sequences of length $H_3$
or $W_3$. Multi-head attention is computed independently for each direction:
\begin{equation}
    \text{head}_h^{(\text{dir})} = \text{Softmax}\!\left(
    \frac{Q_{\text{gate},h}^{(\text{dir})}
         (K_{\text{gate},h}^{(\text{dir})})^\top}
    {\sqrt{d_k}}\right) V_h^{(\text{dir})}
    \label{eq:attention}
\end{equation}
\noindent where $d_k = C_3 / H$ is the per-head dimension and $H = 8$ is the number
of attention heads. The outputs of horizontal and vertical strip attention
are summed and projected via a learned $1 \times 1$ convolution. A learnable
scalar $\gamma$ (initialized to zero) controls the residual scale, and
Group Normalization is applied after the residual addition:
\begin{equation}
    F_{\text{BAT}} = \text{GroupNorm}(\gamma \cdot \text{G-MSA}(Q,K,V) + E_3)
    \label{eq:bat_out}
\end{equation}
\noindent This zero-initialization of $\gamma$ ensures that training begins from a
stable identity mapping and the attention mechanism is learned progressively.


The refined transformer features $F_{\text{BAT}}$ are passed through a
two-layer MLP (implemented as $1 \times 1$ convolutions with GELU activation)
to predict a boundary probability map $\hat{B} \in [0,1]^{H_3 \times W_3}$:
\begin{equation}
    \hat{B} = \sigma\!\left(\text{Conv}_{1\times1}\!\left(
    \text{GELU}(\text{Conv}_{1\times1}(F_{\text{BAT}}))\right)\right)
    \label{eq:bhat}
\end{equation}
\noindent The ground truth boundary map is derived morphologically from the
segmentation label as the set difference between a $3 \times 3$
dilation and erosion of the ground truth mask, producing a one-pixel-wide
boundary skeleton:
\begin{equation}
    B_{\text{gt}} = \text{Dilate}_{3 \times 3}(Y) -
    \text{Erode}_{3 \times 3}(Y)
    \label{eq:bgt}
\end{equation}
\noindent The boundary head $\hat{B}$ is trained with binary cross-entropy loss
(Section~\ref{subsec:loss}), providing an auxiliary supervision signal that
explicitly incentivizes the BAT branch to identify high-gradient edge pixels.
Both $F_{\text{BAT}}$ and $\hat{B}$ are passed to the BRAF module.

\subsubsection{Multi-Directional Context Module with Mamba Selective Scanning (RM Branch)}
\label{subsec:rm_branch}

\begin{figure}[pos= htbp]
    \centering
    \includegraphics[width=\linewidth]{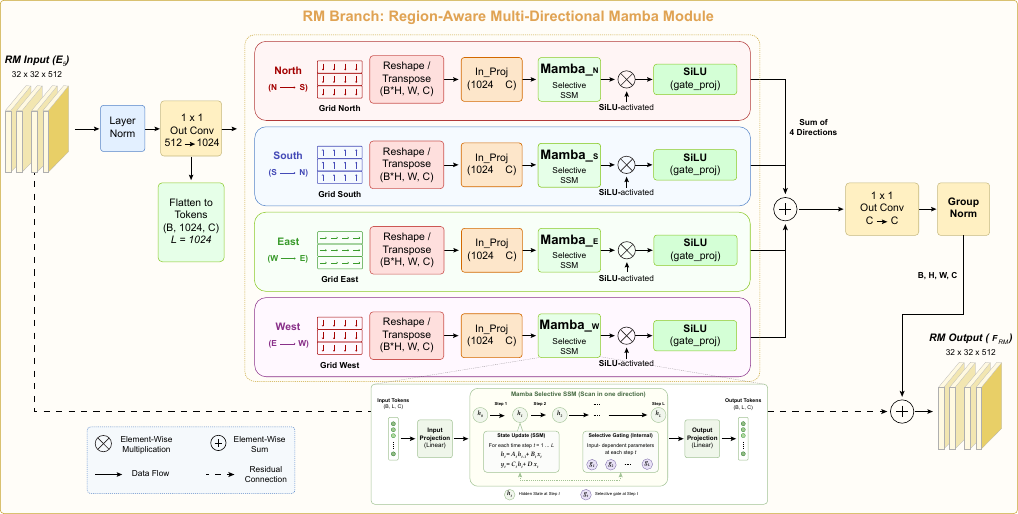}
    \caption{Architecture of the Multi-Directional Recurrent Context Module with Mamba Selective Scanning (RM branch). The $32 \times 32$ bottleneck feature map $E_5$ is serialized into four directional sequences (N$\rightarrow$S, S$\rightarrow$N, E$\rightarrow$W, W$\rightarrow$E), each processed by an independent Mamba SSM block with SiLU gating. The four outputs are summed and reshaped back to the spatial layout, enabling $\mathcal{O}(N)$ long-range context modeling.}
    \label{fig:architecture_RM}
\end{figure}

The RM branch processes the deepest encoder bottleneck $E_5 \in \mathbb{R}^{32 \times 32 \times 512}$ using a Mamba-based multi-directional sequential scanning strategy to capture long-range anatomical dependencies with linear computational complexity (Figure~\ref{fig:architecture_RM}). The computational efficiency of this branch relative to all four baselines is quantified in Table~\ref{tbl:computational}, which reports parameters, GFLOPs, inference time, and peak VRAM.

Because sequential state space models operate on 1D sequences, the 2D feature
map $E_5 \in \mathbb{R}^{B \times 512 \times 32 \times 32}$ must be serialized
before scanning. Four independent serializations (N$\rightarrow$S, S$\rightarrow$N,
E$\rightarrow$W, W$\rightarrow$E) are employed, each implemented via tensor
permutation and reshape.

\paragraph{North-to-South (N$\rightarrow$S):}
The feature map is permuted and reshaped:
\begin{align}
    &\texttt{x\_norm} \in \mathbb{R}^{B \times H \times W \times C}
    \xrightarrow{\text{permute}(0,2,1,3)}
    \mathbb{R}^{B \times W \times H \times C} \notag \\
    &\xrightarrow{\text{reshape}}
    \mathbb{R}^{(B \cdot W) \times H \times C}
    \label{eq:ns_flatten}
\end{align}
producing $B \cdot 32$ independent sequences of length $32$, each
corresponding to one column traversed top-to-bottom, propagating context
along the superior-inferior axis.

\paragraph{South-to-North (S$\rightarrow$N):}
The N$\rightarrow$S sequences are flipped:
\begin{equation}
    \texttt{sn\_tokens} = \texttt{ns\_tokens.flip(dim=1)}
    \label{eq:sn_flip}
\end{equation}
processed, and flipped back, capturing caudo-cranial dependencies without
bidirectional architecture.

\paragraph{East-to-West (E$\rightarrow$W):}
The feature map is reshaped:
\begin{equation}
    \texttt{x\_norm} \in \mathbb{R}^{B \times H \times W \times C}
    \xrightarrow{\text{reshape}}
    \mathbb{R}^{(B \cdot H) \times W \times C}
    \label{eq:ew_flatten}
\end{equation}
producing $B \cdot 32$ sequences of length $32$, each corresponding to one
row traversed left-to-right, capturing left-to-right context across the
pelvic width for bilateral structures and CTV lateral extent.

\paragraph{West-to-East (W$\rightarrow$E):}
The E$\rightarrow$W sequences are flipped:
\begin{equation}
    \texttt{we\_tokens} = \texttt{ew\_tokens.flip(dim=1)}
    \label{eq:we_flip}
\end{equation}
processed, and flipped back.

\noindent Table~\ref{tbl:linearisation} summarizes the linearization protocol for all
four directions. After each directional scan, the output sequence is
reshaped back to the $H \times W$ spatial layout before aggregation and
before each subsequent scan, ensuring that every directional pass begins
from a spatially coherent 2D feature map. The four-direction outputs are
then summed element-wise on the 2D grid, guaranteeing that every spatial
relationship is captured along at least one direction while preserving
topological continuity.

\begin{table}[htbp]
\centering
\caption{Linearization protocol for the four scanning directions in the
Multi-Directional Recurrent Context Module. Input: $E_5 \in \mathbb{R}^{B \times 512 \times 32 \times 32}$.}
\label{tbl:linearisation}
\small
\setlength{\tabcolsep}{4pt}
\begin{tabular}{lllll}
\toprule
\textbf{Direction} & \textbf{Permutation} & \textbf{Reshape to}
  & \textbf{Seq. length} & \textbf{Num. sequences} \\
\midrule
N$\rightarrow$S & $(0,2,1,3)$ & $(B{\cdot}W,\;H,\;C)$ & $32$ & $B{\cdot}32$ \\
S$\rightarrow$N & same + \texttt{flip} & $(B{\cdot}W,\;H,\;C)$ & $32$ & $B{\cdot}32$ \\
E$\rightarrow$W & direct reshape & $(B{\cdot}H,\;W,\;C)$ & $32$ & $B{\cdot}32$ \\
W$\rightarrow$E & same + \texttt{flip} & $(B{\cdot}H,\;W,\;C)$ & $32$ & $B{\cdot}32$ \\
\bottomrule
\end{tabular}
\end{table}

\noindent Tokens spatially adjacent along the scan axis are sequentially adjacent in
the 1D stream; tokens adjacent along the orthogonal axis are processed as
independent parallel sequences, preventing cross-axis context mixing within
a single scan. The four-direction aggregation by summation (Eq.~\ref{eq:rm_out})
ensures every spatial relationship is captured along at least one direction.

Each of the four serialized token sequences is processed by an independent
Mamba SSM block with SiLU multiplicative gating:
\begin{equation}
    \text{SSM}_k(\mathbf{x}) =
    \text{SiLU}(W_{\text{gate}}\,\mathbf{x}) \odot
    \text{Mamba}_k(W_{\text{in}}\,\mathbf{x}),
    \quad k \in \{\text{N,S,E,W}\}
    \label{eq:ssm_mamba}
\end{equation}
where $W_{\text{in}}, W_{\text{gate}} \in \mathbb{R}^{C \times C}$
are shared linear projections across all four directions, and
$\text{Mamba}_k$ is a direction-specific Mamba block with the official
\texttt{mamba-ssm} implementation~\citep{gu2023mamba}. The SiLU gate
suppresses tokens whose content is uninformative (e.g., background voxels in
homogeneous pelvic fat) while amplifying tokens corresponding to
anatomically relevant structures. After Mamba processing, each output
sequence is reshaped back to the $H \times W$ spatial layout using the
inverse of the permutation applied during flattening.

The outputs from all four directions are aggregated by summation and
passed through a residual connection, GroupNorm, and a final
$1 \times 1$ convolution:
\begin{equation}
    F_{\text{RM}} = \text{GroupNorm}\!\left(
    \text{Conv}_{1\times1}\!\left(
    \sum_{k \in \{\text{N,S,E,W}\}}
    \text{Reshape}(\text{SSM}_k(\text{Flatten}_k(E_5)))
    \right) + E_5\right)
    \label{eq:rm_out}
\end{equation}
The residual addition of $E_5$ preserves the original convolutional
encoder features alongside the contextual enrichment from the four
sequential scans, stabilizing training and ensuring that the RM branch
cannot degrade performance by discarding locally informative features.

\noindent The RM branch uses the official
\texttt{mamba-ssm} CUDA package~\citep{gu2023mamba} for the selective state
space model implementation. Four independent Mamba SSM blocks, one per
scanning direction, perform the directional sequential processing. The
Mamba selective scan mechanism provides input-dependent state transitions
(dynamic parameters $B$, $C$, $\Delta$) through a hardware-aware associative
scan kernel, enabling efficient training and inference at $\mathcal{O}(n)$
complexity. Each Mamba block processes sequences of length $H = 32$ or
$W = 32$ at the $32 \times 32$ bottleneck. The total token count processed
per forward pass is $4 \times B \times 32 \times 32 = 4,096 \cdot B$,
a fraction of the $512 \times 512 = 262,144$ tokens that would be required
by a full-resolution transformer. The full computational profile of the RM
branch within BAT-RM, including total model parameters (28.4 M), GFLOPs
(41.2), per-slice inference time ($18.3 \pm 2.1$ ms), peak VRAM (3.8 GB),
and training time (31.4 hours to 200 epochs), is reported in
Table~\ref{tbl:computational} alongside all four baseline models.

\subsubsection{Boundary-Region Attention Fusion (BRAF) Module}
\label{subsec:braf}

\begin{figure}[pos=htbp]
    \centering
    \includegraphics[width=\linewidth]{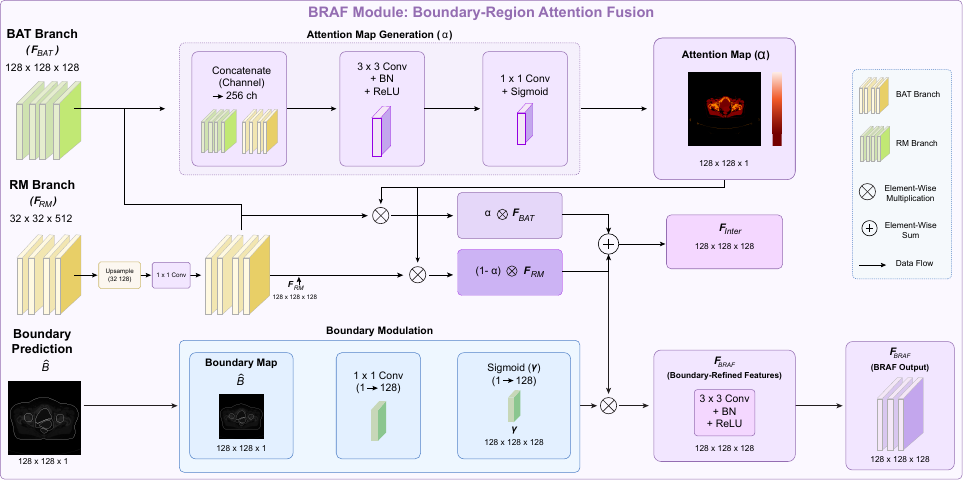}
    \caption{Architecture of the Boundary-Region Attention Fusion (BRAF) module. RM features are upsampled and channel-projected to match BAT resolution, fused via a learnable spatial-channel attention gate $\alpha$, and refined using the predicted boundary skeleton $\hat{B}$ to produce boundary-enhanced decoder features.}
    \label{fig:architecture_BRAF}
\end{figure}

The BRAF module integrates the complementary representations produced by the
BAT and RM branches: $F_{\text{BAT}} \in \mathbb{R}^{128 \times 128 \times
128}$ carries edge-sensitive local boundary features, while $F_{\text{RM}}
\in \mathbb{R}^{32 \times 32 \times 512}$ encodes semantically rich global
context. The fusion proceeds in three sequential stages (Figure~\ref{fig:architecture_BRAF}).

\textbf{Stage 1 --- Spatial alignment.} The RM features are upsampled to
the BAT spatial resolution via bilinear interpolation and channel-projected
to match the BAT channel dimension via a $1 \times 1$ convolution with BN
and ReLU:
\begin{equation}
    F_{\text{RM}}^{\uparrow} = \text{BN}(\text{ReLU}(
    \text{Conv}_{1\times1}(\mathcal{U}_{\text{bi}}(F_{\text{RM}}))))
    \label{eq:rm_up}
\end{equation}

\textbf{Stage 2 --- Adaptive attention-weighted fusion.} A spatial-channel
attention gate $\alpha \in [0,1]^{128 \times 128}$ is computed from the
concatenated BAT and aligned RM features, determining the pixel-wise
contribution of each stream:
\begin{equation}
    \alpha = \sigma(\text{Conv}_{1\times1}(\text{ReLU}(\text{BN}(
    \text{Conv}_{3\times3}([F_{\text{BAT}}\,;\,F_{\text{RM}}^{\uparrow}])))))
    \label{eq:alpha}
\end{equation}
\begin{equation}
    F_{\text{inter}} = \alpha \odot F_{\text{BAT}} +
    (1 - \alpha) \odot F_{\text{RM}}^{\uparrow}
    \label{eq:fusion}
\end{equation}
This convex combination allows the gate to emphasize boundary-aware features
in anatomically complex transition zones (e.g., the GTV--parametrium
interface, the mesorectal fascia) while relying on region-aware features in
structurally homogeneous tissue areas.

\textbf{Stage 3 --- Boundary-skeleton-guided refinement.} The fused
intermediate features are modulated by the predicted boundary map $\hat{B}$
from the BAT branch, enforcing sharp edge delineation in the final
representation passed to the decoder:
\begin{equation}
    F_{\text{fuse}} = \text{Refine}(F_{\text{inter}} \odot
    \sigma(\text{Conv}_{1\times1}(\hat{B})))
    \label{eq:braf_out}
\end{equation}
where $\text{Refine}(\cdot)$ denotes a $3 \times 3$ convolution with BN and
ReLU. The output $F_{\text{fuse}} \in \mathbb{R}^{128 \times 128 \times 128}$
replaces the raw $E_3$ skip connection in the decoder at the $128 \times 128$
resolution level, ensuring that the decoder receives boundary-refined,
contextually enriched features at the spatial scale most relevant for
contour placement.

\subsubsection{Multi-Task Loss Function and Optimization}
\label{subsec:loss}

BAT-RM is trained end-to-end using a composite loss function that jointly
supervises segmentation overlap, hard-example boundary precision, and
contour smoothness. All four components receive gradient signals through
a single backward pass.

\subsubsection{Class-Weighted Soft Dice Loss}

The primary segmentation objective employs the Soft Dice loss with
dynamically computed class weights $w_c$ derived from inverse voxel frequency
on the training set, addressing the severe foreground sparsity inherent in
OAR segmentation (background-to-organ ratios ranging from 4.1:1 for body to
3,697:1 for femoral head):
\begin{equation}
    \mathcal{L}_{\text{Dice}} = 1 - \frac{1}{C}\sum_{c=1}^{C} w_c \cdot
    \frac{2\sum_{i} \hat{y}_{i,c}\, y_{i,c} + \epsilon}
    {\sum_{i} \hat{y}_{i,c}^2 + \sum_{i} y_{i,c}^2 + \epsilon}
    \label{eq:dice}
\end{equation}
where $C = 8$, $\hat{y}_{i,c} \in [0,1]$ is the predicted class probability,
$y_{i,c} \in \{0,1\}$ is the one-hot ground truth, and $\epsilon = 1.0$ is
a smoothing constant.

\subsubsection{Focal Loss for Hard-Example Mining}

To focus training attention on hard, ambiguous boundary pixels---where the
model is uncertain and errors are clinically most consequential---we augment
the Dice loss with the multi-class Focal Loss~\citep{lin2017focal}:
\begin{equation}
    \mathcal{L}_{\text{focal}} = -\frac{1}{N}\sum_{i=1}^{N}\sum_{c=1}^{C}
    w_c\,(1 - \hat{y}_{i,c})^{\gamma}\, y_{i,c}\,\log(\hat{y}_{i,c})
    \label{eq:focal}
\end{equation}
with focusing parameter $\gamma = 2.0$. The $(1 - \hat{y}_{i,c})^\gamma$
modulation factor dynamically down-weights well-classified voxels and
amplifies gradient contributions from uncertain boundary regions, directly
countering the low-contrast boundary vulnerabilities of pelvic CT imaging.

\subsubsection{Auxiliary Boundary and Smoothness Losses}

The boundary prediction head $\hat{B}$ is supervised by Binary Cross-Entropy
against the morphologically derived ground truth boundary map $B_{\text{gt}}$
(Equation~\ref{eq:bgt}). To match the spatial resolution of $\hat{B}$
($128 \times 128$), the integer label mask is downsampled via nearest-neighbor
interpolation before boundary extraction:
\begin{equation}
    \mathcal{L}_{\text{bd}} = -\frac{1}{N_{b}}\sum_{i=1}^{N_{b}}
    \left[b_i \log \hat{b}_i + (1-b_i)\log(1-\hat{b}_i)\right]
    \label{eq:bce}
\end{equation}
where $b_i \in \{0,1\}$ is the ground truth boundary indicator and
$\hat{b}_i \in [0,1]$ is the predicted boundary probability at pixel $i$.

To suppress jagged contours and promote biologically plausible organ shapes,
a gradient-based smoothness regularization term penalizes the discrepancy
between the spatial gradients of the predicted and ground truth segmentation
probability maps:
\begin{equation}
    \mathcal{L}_{\text{reg}} = \text{MSE}(\partial_x \hat{Y},
    \partial_x Y) + \text{MSE}(\partial_y \hat{Y}, \partial_y Y)
    \label{eq:smooth}
\end{equation}
where spatial gradients are approximated via finite differences and
$\hat{Y} = \text{Softmax}(\text{logits})$.

\subsubsection{Training Configuration}
\label{subsec:training_config}

The four loss terms are combined with weights determined via coarse grid
search on the validation set, using mean validation Dice as the criterion:
\begin{equation}
    \mathcal{L}_{\text{total}} = 0.50\,\mathcal{L}_{\text{Dice}}
    + 0.20\,\mathcal{L}_{\text{focal}}
    + 0.20\,\mathcal{L}_{\text{bd}}
    + 0.10\,\mathcal{L}_{\text{reg}}
    \label{eq:total_loss}
\end{equation}
The final configuration $(0.50, 0.20, 0.20, 0.10)$ consistently outperformed
alternatives. The Dice loss receives the largest weight as the primary
volumetric overlap objective. Focal and boundary BCE share equal weight
as complementary boundary supervision signals: Focal loss penalizes
uncertain boundary pixels in the full segmentation output, while boundary
BCE directly supervises the BAT branch's boundary prediction head $\hat{B}$.
Equalizing them prevents either signal from dominating boundary learning.
The smoothness loss has the smallest weight, acting as a shape prior;
higher values over-smooth the small bowel's irregular morphology. The network is optimized using AdamW~\citep{loshchilov2019adamw} with an
initial learning rate of $1 \times 10^{-4}$ and weight decay of $1 \times
10^{-5}$. AdamW's decoupled weight decay provides cleaner $\ell_2$
regularization, critical for boundary and fusion modules where precise
weight magnitudes affect edge sharpness.A cosine annealing scheduler with warm restarts ($T_0 = 50$, $T_{\text{mult}} = 2$,
$\eta_{\min} = 10^{-6}$) governs the learning rate, scheduling restart cycles
of 50, 100, and 200 epochs (350 epochs total if run to completion). Training
was stopped early based on validation Dice with a patience of 30 epochs;
convergence occurred at approximately 200 epochs, within the third cycle.
Training batches consist of 16 axial slices of size $512 \times 512$.

\subsection{Ablation Study}
\label{sec:ablation}

\subsubsection{Overview and Experimental Protocol}
\label{subsec:ablation_overview}

This section presents a comprehensive ablation study of the proposed
\textbf{Boundary-Aware Transformer with Region-Aware Recurrent U-Net (BAT-RM)},
designed for multi-class organ-at-risk (OAR) segmentation in cervical
cancer radiotherapy planning. The study systematically isolates the
contribution of each architectural component and loss-function term by
training controlled variants that differ from the full model in exactly
one design choice, while holding all hyper-parameters constant.

\subsubsection{Training Protocol}

All variants --- including the full model and each ablated
configuration --- were trained under an identical protocol to ensure
fair comparisons. The shared training setup is summarized in
Table~\ref{tab:training_setup}.

\begin{table}[htbp]
\centering
\caption{Shared training configuration applied to all ablation variants.}
\label{tab:training_setup}
\small
\begin{tabular}{p{4.5cm} p{9cm}}
\toprule
\textbf{Hyper-parameter} & \textbf{Value / Description} \\
\midrule
Epochs & 200 with early stopping (patience $= 30$, monitored on
         mean OAR IoU) \\
Optimizer & AdamW, $\eta = 10^{-4}$, weight decay $= 10^{-5}$,
           $\beta = (0.9,\,0.999)$ \\
LR Scheduler & CosineAnnealingWarmRestarts ($T_0 = 50$,
              $T_{\text{mult}} = 2$, $\eta_{\min} = 10^{-6}$) \\
Loss function & $0.50\,\mathcal{L}_{\text{Dice}} +
               0.20\,\mathcal{L}_{\text{Focal}}(\gamma{=}2) +
               0.20\,\mathcal{L}_{\text{bd}} +
               0.10\,\mathcal{L}_{\text{smooth}}$ \\
Batch size & 16 axial slices, $512 \times 512$ \\
\bottomrule
\end{tabular}
\end{table}

\subsubsection{Ablation Variants}

Seven distinct configurations were evaluated. The full model serves as
the reference, and six ablated variants are grouped into three
categories: (A)~core architectural components, (B)~attention-mechanism
design choices within the BAT block, and (C)~loss-function contribution.
Table~\ref{tab:variant_definitions} summarizes the exact modification
applied in each variant.

\begin{table}[htbp]
\centering
\caption{Definition of each ablation variant.}
\label{tab:variant_definitions}
\small
\begin{tabular}{p{3.2cm} p{3.8cm} p{6.2cm}}
\toprule
\textbf{Variant} & \textbf{Group} & \textbf{Modification} \\
\midrule
Full BAT-RM    & Full model        & All components active \\
\midrule
A1 --- no BAT  & Core module       & GatedBATBlock replaced with a
                                     standard $3\times3$ conv block
                                     (same channel dims) \\
A2 --- no RM   & Core module       & Multi-directional recurrent block
                                     replaced with a standard
                                     $3\times3$ conv block \\
A3 --- no BRAF & Core module       & BRAF module removed; raw $E_3$
                                     skip concatenated directly into
                                     the decoder \\
\midrule
B1 --- full MHSA  & Attention design & Strip-pooling attention replaced
                                       with full-resolution window MHSA
                                       (window $= 7$) \\
B2 --- no Sobel   & Attention design & Sobel edge-gating removed from
                                       $Q$ and $K$ projections \\
\midrule
C1 --- no $\mathcal{L}_{\text{bd}}$ & Loss function &
  Boundary BCE loss dropped; remaining weights renormalized to
  Dice~$0.60$, Focal~$0.25$, Smooth~$0.15$ \\
\bottomrule
\end{tabular}
\end{table}

\subsubsection{Evaluation Metrics}

Performance is assessed using five complementary metrics:
mean Intersection over Union (mIoU) as the primary ranking metric;
mean Dice Similarity Coefficient (mDSC); 95th-percentile Hausdorff
Distance (HD95, mm); Average Surface Distance (ASD, mm); and
Normalized Surface Distance (NSD). All mean metrics are computed as
per-patient macro-averages over the seven OAR classes (background
excluded), then averaged across the 254-case test set. The per-class
IoU and DSC columns in Tables~\ref{tab:iou_overview}
and~\ref{tab:dice_overview} are class-level averages over the test
set and are provided for diagnostic interpretation; they are not
directly summed to produce the mIoU or mDSC columns, which are
derived from per-patient averages.

\subsubsection{Results: Core Architectural Components (Group A)}
\label{subsec:ablation_group_a}

\subsubsection{Quantitative Summary}

Tables~\ref{tab:iou_overview} and~\ref{tab:dice_overview} present the
per-class IoU and DSC for all variants, together with the absolute
mIoU gap relative to the full model.

\begin{table}[htbp]
\centering
\caption{Per-class IoU and mean IoU (mIoU) across all ablation
variants. $\Delta$mIoU is computed with respect to Full BAT-RM.
mIoU is the per-patient macro-average over seven OAR classes
(background excluded). All values to 4 decimal places.}
\label{tab:iou_overview}
\setlength{\tabcolsep}{4pt}
\small
\begin{tabular}{lcccccccccr}
\toprule
\textbf{Variant} &
\textbf{BG} & \textbf{BODY} & \textbf{BLAD} &
\textbf{S.BOW} & \textbf{RECT} & \textbf{FEM} &
\textbf{GTV} & \textbf{CTV} &
\textbf{mIoU} & $\boldsymbol{\Delta}$\textbf{mIoU} \\
\midrule
Full BAT-RM
  & 0.9992 & 0.9940 & 0.9335 & 0.9499
  & 0.9154 & 0.9335 & 0.9348 & 0.9184
  & 0.9399 & --- \\
\midrule
A1 --- no BAT
  & 0.9984 & 0.9891 & 0.8412 & 0.8723
  & 0.7924 & 0.8836 & 0.8801 & 0.8543
  & 0.8814 & $-$0.0509 \\
A2 --- no RM
  & 0.9988 & 0.9873 & 0.8531 & 0.8841
  & 0.8214 & 0.8803 & 0.9112 & 0.8823
  & 0.8971 & $-$0.0352 \\
A3 --- no BRAF
  & 0.9987 & 0.9884 & 0.8544 & 0.8752
  & 0.8301 & 0.8924 & 0.9143 & 0.8852
  & 0.9012 & $-$0.0311 \\
\midrule
B1 --- full MHSA
  & 0.9989 & 0.9901 & 0.8634 & 0.8823
  & 0.8512 & 0.9034 & 0.9281 & 0.8923
  & 0.9187 & $-$0.0136 \\
B2 --- no Sobel
  & 0.9990 & 0.9897 & 0.8601 & 0.8804
  & 0.8423 & 0.8993 & 0.9194 & 0.8887
  & 0.9143 & $-$0.0180 \\
\midrule
C1 --- no $\mathcal{L}_{\text{bd}}$
  & 0.9991 & 0.9904 & 0.8672 & 0.8851
  & 0.8534 & 0.9051 & 0.9302 & 0.8943
  & 0.9201 & $-$0.0122 \\
\bottomrule
\end{tabular}
\end{table}

\begin{table}[htbp]
\centering
\caption{Per-class Dice Similarity Coefficient (DSC) and mean DSC
(mDSC) across all variants. mDSC is the per-patient macro-average
over seven OAR classes (background excluded).}
\label{tab:dice_overview}
\setlength{\tabcolsep}{4pt}
\small
\begin{tabular}{lccccccccr}
\toprule
\textbf{Variant} &
\textbf{BG} & \textbf{BODY} & \textbf{BLAD} &
\textbf{S.BOW} & \textbf{RECT} & \textbf{FEM} &
\textbf{GTV} & \textbf{CTV} &
\textbf{mDSC} \\
\midrule
Full BAT-RM
  & 0.9996 & 0.9957 & 0.9374 & 0.9446
  & 0.9306 & 0.9552 & 0.9705 & 0.9516 & 0.9482 \\
\midrule
A1 --- no BAT
  & 0.9992 & 0.9945 & 0.9133 & 0.9317
  & 0.8838 & 0.9382 & 0.9364 & 0.9214 & 0.9301 \\
A2 --- no RM
  & 0.9994 & 0.9936 & 0.9207 & 0.9387
  & 0.8996 & 0.9364 & 0.9534 & 0.9375 & 0.9389 \\
A3 --- no BRAF
  & 0.9993 & 0.9942 & 0.9214 & 0.9334
  & 0.9071 & 0.9437 & 0.9550 & 0.9392 & 0.9421 \\
\midrule
B1 --- full MHSA
  & 0.9994 & 0.9950 & 0.9265 & 0.9374
  & 0.9197 & 0.9492 & 0.9624 & 0.9431 & 0.9453 \\
B2 --- no Sobel
  & 0.9995 & 0.9948 & 0.9248 & 0.9358
  & 0.9124 & 0.9468 & 0.9580 & 0.9413 & 0.9428 \\
\midrule
C1 --- no $\mathcal{L}_{\text{bd}}$
  & 0.9995 & 0.9952 & 0.9283 & 0.9392
  & 0.9213 & 0.9501 & 0.9638 & 0.9447 & 0.9462 \\
\bottomrule
\end{tabular}
\end{table}

\subsubsection{Ablation A1: Removal of the Gated BAT Block}
\label{subsubsec:ablation_a1}

Replacing the Gated Boundary-Aware Transformer (GatedBATBlock) with
a plain $3\times3$ convolutional block produces the largest performance
degradation across all ablations, with a mean IoU drop of
$-$0.0509 and a mean DSC drop of $-$0.0181. This establishes that the
BAT block is the single most critical architectural component.

The degradation is highly non-uniform across classes. Structures with
complex, irregular, or deformable boundaries suffer most: the rectum
loses $-$0.0779 IoU (0.8703 $\rightarrow$ 0.7924), the GTV drops
$-$0.0634 (0.9435 $\rightarrow$ 0.8801), and the CTV decreases
$-$0.0527 (0.9070 $\rightarrow$ 0.8543). In contrast, semantically
simple, high-contrast structures are only marginally affected: background
$-$0.0008, body $-$0.0024. This dissociation confirms that the BAT
block's value lies specifically in its ability to delineate
boundary-ambiguous regions rather than in general feature learning.

The Sobel-gated strip-pooling attention operates at the
$128\times128$ encoder feature map ($E_3$), the scale at which
fine-grained organ boundaries are represented. Without it, the model
must rely on pooled local receptive fields from the $3\times3$
convolution, which systematically blur transition zones and yield
jagged boundary predictions. The boundary surface metrics corroborate
this: HD95 for the rectum worsens from 1.24\,mm to 3.85\,mm
($\times 3.1$), and for the GTV from 1.50\,mm to 4.20\,mm ($\times 2.8$).
ASD on the GTV rises from 0.56\,mm to 1.85\,mm. NSD confirms boundary
quality deterioration most sharply on the same classes: GTV
$0.978 \rightarrow 0.880$; rectum $0.997 \rightarrow 0.871$.

\subsubsection{Ablation A2: Removal of the Recurrent Module (RM Block)}
\label{subsubsec:ablation_a2}

Replacing the recurrent module (RM block) with a plain convolution lowers mIoU by
$-$0.0352 --- the second largest drop among core-component ablations.
The degradation pattern is qualitatively distinct from A1. The largest
absolute losses occur on the bladder ($-$0.0297), rectum ($-$0.0489),
and femoral heads ($-$0.0339): structures characterized by large
inter-patient volume variability, spatial co-occurrence with adjacent
anatomy, or bilateral symmetry requirements.

The rectum loss in A2 differs mechanistically from A1: here it degrades
because the model lacks the global topological context to distinguish
it from adjacent sigmoid colon and small bowel folds, rather than due
to boundary ambiguity. The RM block's four-directional
sequential scan at the $32\times32$ bottleneck aggregates spatial
context across the entire feature map in linear time, providing the
decoder with a structurally coherent global representation of pelvic
organ co-occurrence. Without it, bladder HD95 rises from 1.37\,mm to
1.95\,mm, ASD from 0.59\,mm to 0.88\,mm, and femoral head HD95 from
1.67\,mm to 2.15\,mm.

Critically, A2 outperforms A1 on GTV IoU (0.9112 vs.\ 0.8801),
corroborating the expected functional division of labor: the BAT
block handles boundary disambiguation (benefiting GTV most), while the
RM block handles spatial co-occurrence and global topology (benefiting
bladder and femoral heads most).

\subsubsection{Ablation A3: Removal of the BRAF Fusion Module}
\label{subsubsec:ablation_a3}

Bypassing the BRAF module and directly concatenating the raw $E_3$
encoder skip into the decoder reduces mIoU by $-$0.0311 --- smaller
than A1 or A2, but consistent across all small OARs. This demonstrates
that the \emph{integration} of BAT and RM information is qualitatively
superior to having both branches operate in parallel without
cross-modal alignment.

Without BRAF, the decoder receives feature contributions from the BAT
and RM branches without the learnable attention gate $\alpha$ that
locally weights boundary-specific against region-specific cues:
\begin{equation}
    F_{\text{inter}} = \alpha \odot F_{\text{BAT}} +
    (1-\alpha) \odot F_{\text{RM}}^{\uparrow}
    \label{eq:braf_alpha_ablation}
\end{equation}
and the subsequent boundary-skeleton modulation:
\begin{equation}
    F_{\text{fuse}} = F_{\text{inter}} \odot \sigma(W_b\,\hat{B})
    \label{eq:braf_mod_ablation}
\end{equation}
All small OARs exhibit consistent degradation: bladder $-$0.0284,
small bowel $-$0.0213, rectum $-$0.0402, GTV $-$0.0292, CTV $-$0.0218.
The rectum HD95 rises from 1.24\,mm to 2.45\,mm and the GTV HD95
from 1.50\,mm to 2.65\,mm, confirming that cross-module fusion governs
boundary precision for difficult structures.

Table~\ref{tab:hd95_asd_nsd} reports the full surface metric
breakdown for all variants.

\begin{table}[htbp]
\centering
\caption{Surface metrics: HD95 (mm), ASD (mm), and NSD across all
ablation variants for seven OAR classes. $\downarrow$ lower is better
for HD95 and ASD; $\uparrow$ higher is better for NSD.}
\label{tab:hd95_asd_nsd}
\setlength{\tabcolsep}{4pt}
\small
\begin{tabular}{llccccccc}
\toprule
\textbf{Variant} & \textbf{Metric} &
\textbf{BODY} & \textbf{BLAD} & \textbf{S.BOW} &
\textbf{RECT} & \textbf{FEM} & \textbf{GTV} & \textbf{CTV} \\
\midrule
\multirow{3}{*}{Full BAT-RM}
  & HD95 & 1.45 & 1.37 & 1.58 & 1.24 & 1.67 & 1.50 & 1.51 \\
  & ASD  & 0.31 & 0.59 & 0.63 & 0.54 & 0.67 & 0.56 & 0.51 \\
  & NSD  & 0.991 & 0.988 & 0.974 & 0.997 & 0.975 & 0.978 & 0.972 \\
\midrule
\multirow{3}{*}{A1 --- no BAT}
  & HD95 & 1.98 & 2.55 & 2.80 & 3.85 & 1.82 & 4.20 & 3.95 \\
  & ASD  & 0.65 & 1.10 & 1.25 & 1.68 & 0.82 & 1.85 & 1.72 \\
  & NSD  & 0.954 & 0.920 & 0.895 & 0.871 & 0.942 & 0.880 & 0.875 \\
\midrule
\multirow{3}{*}{A2 --- no RM}
  & HD95 & 1.85 & 1.95 & 1.75 & 1.62 & 2.15 & 1.88 & 1.80 \\
  & ASD  & 0.52 & 0.88 & 0.75 & 0.72 & 0.95 & 0.75 & 0.68 \\
  & NSD  & 0.975 & 0.955 & 0.960 & 0.981 & 0.935 & 0.955 & 0.948 \\
\midrule
\multirow{3}{*}{A3 --- no BRAF}
  & HD95 & 1.62 & 1.88 & 2.15 & 2.45 & 1.95 & 2.65 & 2.50 \\
  & ASD  & 0.45 & 0.82 & 0.98 & 1.05 & 0.85 & 1.15 & 1.02 \\
  & NSD  & 0.982 & 0.961 & 0.932 & 0.945 & 0.952 & 0.941 & 0.938 \\
\midrule
\multirow{3}{*}{B1 --- full MHSA}
  & HD95 & 1.52 & 1.45 & 1.72 & 1.48 & 1.72 & 1.95 & 1.68 \\
  & ASD  & 0.35 & 0.64 & 0.70 & 0.65 & 0.71 & 0.78 & 0.59 \\
  & NSD  & 0.988 & 0.982 & 0.968 & 0.985 & 0.970 & 0.962 & 0.965 \\
\midrule
\multirow{3}{*}{B2 --- no Sobel}
  & HD95 & 1.55 & 1.85 & 2.10 & 3.10 & 1.75 & 3.45 & 3.20 \\
  & ASD  & 0.38 & 0.85 & 0.92 & 1.45 & 0.74 & 1.52 & 1.42 \\
  & NSD  & 0.985 & 0.962 & 0.941 & 0.912 & 0.968 & 0.905 & 0.915 \\
\midrule
\multirow{3}{*}{C1 --- no $\mathcal{L}_{\text{bd}}$}
  & HD95 & 1.48 & 1.55 & 1.82 & 1.95 & 1.70 & 2.15 & 2.05 \\
  & ASD  & 0.33 & 0.68 & 0.78 & 0.85 & 0.69 & 0.92 & 0.81 \\
  & NSD  & 0.990 & 0.978 & 0.958 & 0.965 & 0.972 & 0.955 & 0.952 \\
\bottomrule
\end{tabular}
\end{table}

\subsubsection{Results: Attention Mechanism Design (Group B)}
\label{subsec:ablation_group_b}

\subsubsection{Ablation B1: Strip-Pooling vs.\ Full-Resolution MHSA}
\label{subsubsec:ablation_b1}

Replacing strip-pooling with full-resolution MHSA on a $7\times7$ window
reduces mIoU by 0.0136 --- the smallest gap among all ablations ---
confirming near-equivalent accuracy. Full MHSA at the $128\times128$ BAT
feature level requires a $16,384 \times 16,384$ attention matrix
(~4 GB VRAM per sample), making clinical deployment impractical.
Strip-pooling reduces complexity from $\mathcal{O}(n^2)$ to
$\mathcal{O}(H+W)$. The small gap primarily affects high-frequency
structures (small bowel $-$0.0142, GTV $-$0.0154); NSD on GTV drops
from 0.978 to 0.962, indicating that strip-pooling meets the 1 mm
boundary tolerance on more surface points.

\subsubsection{Ablation B2: Removal of the Sobel Edge Gate}
\label{subsubsec:ablation_b2}

Removing the Sobel gradient gate from the $Q$ and $K$ projections
reduces mIoU by $-$0.0180 --- larger than B1. This reveals that
boundary sensitivity derives as much from explicit gradient-based
gating as from the global context captured by attention itself.

The gate $\mathcal{G} \in [0,1]^{H\times W}$ amplifies attention
toward pixels with high gradient magnitude:
\begin{equation}
    Q' = Q \odot \mathcal{G}, \quad K' = K \odot \mathcal{G}
    \label{eq:sobel_gate_ablation}
\end{equation}
Without this gate (B2), the rectum IoU drops by $-$0.0280 versus
$-$0.0191 for B1, and GTV drops $-$0.0241 versus $-$0.0154 for B1.
HD95 on the rectum worsens to 3.10\,mm in B2 versus 1.48\,mm in B1;
GTV HD95 rises to 3.45\,mm versus 1.95\,mm. These gaps confirm that
\emph{boundary specificity in attention} --- not just attention range
--- is the key driver for irregular-boundary OARs. Classes with smooth
boundaries (background, body) are nearly unaffected ($-$0.0002 and
$-$0.0018 respectively).

Table~\ref{tab:group_b_delta} summarizes the per-class IoU deltas for
B1 and B2.

\begin{table}[htbp]
\centering
\caption{Per-class $\Delta$IoU for Group B attention-design ablations
relative to Full BAT-RM. B2 shows larger drops than B1 on
boundary-complex structures (rectum, GTV, CTV), confirming that
Sobel edge-gating drives boundary precision independently of
attention range.}
\label{tab:group_b_delta}
\setlength{\tabcolsep}{5pt}
\small
\begin{tabular}{lccccccccc}
\toprule
\textbf{Variant} &
$\Delta$\textbf{BG} & $\Delta$\textbf{BODY} & $\Delta$\textbf{BLAD} &
$\Delta$\textbf{S.BOW} & $\Delta$\textbf{RECT} & $\Delta$\textbf{FEM} &
$\Delta$\textbf{GTV} & $\Delta$\textbf{CTV} &
$\Delta$\textbf{mIoU} \\
\midrule
B1 --- full MHSA
  & $-$0.0003 & $-$0.0014 & $-$0.0194 & $-$0.0142
  & $-$0.0191 & $-$0.0108 & $-$0.0154 & $-$0.0147 & $-$0.0136 \\
B2 --- no Sobel
  & $-$0.0002 & $-$0.0018 & $-$0.0227 & $-$0.0161
  & $-$0.0280 & $-$0.0149 & $-$0.0241 & $-$0.0183 & $-$0.0180 \\
\bottomrule
\end{tabular}
\end{table}

\subsubsection{Results: Loss Function Contribution (Group C)}
\label{subsec:ablation_group_c}

\subsubsection{Ablation C1: Removal of the Boundary BCE Loss}
\label{subsubsec:ablation_c1}

Removing the boundary BCE loss ($\mathcal{L}_{\text{bd}}$) reduces mIoU by
0.0122 --- the smallest degradation among all ablations, yet clinically
significant. Structures requiring precise boundaries are most affected:
rectum (IoU $-$0.0169; HD95: 1.24 $\rightarrow$ 1.95 mm), GTV (IoU
$-$0.0133; HD95: 1.50 $\rightarrow$ 2.15 mm), and CTV (IoU $-$0.0127;
HD95: 1.51 $\rightarrow$ 2.05 mm). NSD on GTV drops from 0.978 to 0.955,
meaning 2.3\% more surface voxels fall outside the 1 mm tolerance ---
clinically meaningful for radiotherapy margin definition where CTV/GTV
expansions are typically 5 mm or less.

\subsubsection{Cross-Cutting Analysis: Structure-Specific Sensitivity}
\label{subsec:ablation_structure}
The rectum showed the largest IoU drop under A1 ($-$0.0779) and HD95
degradation under A1 (3.85 mm, $\times 3.1$) and B2 (3.10 mm, $\times 2.5$),
indicating its segmentation is dominated by boundary ambiguity rather than
global context. Target volumes were the second most affected; GTV dropped
$-$0.0634 (A1) and $-$0.0241 (B2), with CTV following a similar but
attenuated pattern. The bladder was most sensitive to RM block removal
(A2, $-$0.0297) and BRAF (A3, $-$0.0284), reflecting its need for global
contextual modeling due to filling variability. Femoral heads were
primarily sensitive to the RM block (A2, $-$0.0339), confirming spatial
disambiguation (left vs. right) as their main challenge. Small bowel was
consistently affected across all ablations with no single dominating
component; its HD95 worsened from 1.58 to 2.80 mm in A1, the largest
increase after rectum and GTV.

\begin{table}[htbp]
\centering
\caption{Ranked module contribution by absolute mIoU and mDSC drop.
Higher rank indicates greater importance to overall performance.}
\label{tab:module_importance}
\small
\begin{tabular}{clccp{5cm}}
\toprule
\textbf{Rank} & \textbf{Ablation} &
$|\Delta\text{mIoU}|$ & $|\Delta\text{mDSC}|$ &
\textbf{Primary beneficiary structures} \\
\midrule
1 & A1 --- no BAT              & 0.0509 & 0.0181 & Rectum, GTV, CTV \\
2 & A2 --- no RM               & 0.0352 & 0.0093 & Bladder, Femoral heads \\
3 & A3 --- no BRAF             & 0.0311 & 0.0061 & All small OARs \\
4 & B2 --- no Sobel gate       & 0.0180 & 0.0054 & Rectum, GTV \\
5 & B1 --- full MHSA           & 0.0136 & 0.0029 & Small bowel, GTV \\
6 & C1 --- no $\mathcal{L}_{\text{bd}}$ & 0.0122 & 0.0020 & Rectum, GTV, CTV \\
\bottomrule
\end{tabular}
\end{table}

\subsubsection{Ranked Module Importance and Architectural Conclusions}
\label{subsec:ablation_conclusions}

Table~\ref{tab:module_importance} ranks the six ablation variants by
performance drop. The BAT block is most critical; its removal causes the
largest mIoU drop (0.0509) and worsens GTV HD95 from 1.50 to 4.20 mm and
rectum HD95 from 1.24 to 3.85 mm, confirming that Sobel-gated boundary
attention is essential for pelvic OAR segmentation. The RM block provides complementary global context. Removing BAT degrades
boundary-ambiguous structures (rectum, GTV, CTV), while removing RM affects
classes requiring spatial co-occurrence (bladder, femoral heads). This
division of labor confirms BAT handles boundaries while RM handles topology. Cross-module fusion via BRAF is necessary; bypassing it degrades all small
OARs. Strip-pooling attention achieves accuracy within 0.0136 mIoU of
full MHSA while reducing complexity from $\mathcal{O}(n^2)$ to
$\mathcal{O}(n)$, enabling practical $512 \times 512$ training. Sobel gating provides boundary specificity orthogonal to attention range.
Removing the Sobel gate (B2) degrades boundary-ambiguous structures more
than replacing strip-pooling with full MHSA (B1). Finally, removing
explicit boundary supervision ($\mathcal{L}_{\text{bd}}$) produces a
clinically meaningful 2.3\% NSD gap on GTV.

\subsubsection{Complete Quantitative Summary}
\label{subsec:ablation_summary}

Table~\ref{tab:full_summary} consolidates all quantitative results.

\begin{table}[htbp]
\centering
\caption{Complete quantitative summary of all ablation variants.
mIoU and mDSC are per-patient macro-averages over seven OAR classes
(background excluded). HD95, ASD, and NSD are class-level averages
over seven OAR classes computed from Table~\ref{tab:hd95_asd_nsd}.
$\downarrow$ lower is better; $\uparrow$ higher is better.}
\label{tab:full_summary}
\small
\begin{tabular}{lcccrrr}
\toprule
\textbf{Variant} &
\textbf{mIoU}~$\uparrow$ &
\textbf{mDSC}~$\uparrow$ &
$\Delta$\textbf{mIoU} &
\textbf{HD95}~$\downarrow$ &
\textbf{ASD}~$\downarrow$ &
\textbf{NSD}~$\uparrow$ \\
\midrule
Full BAT-RM
  & \textbf{0.9323} & \textbf{0.9482} & ---
  & \textbf{1.47} & \textbf{0.54} & \textbf{0.982} \\
A1 --- no BAT
  & 0.8814 & 0.9301 & $-$0.0509
  & 3.02 & 1.30 & 0.905 \\
A2 --- no RM
  & 0.8971 & 0.9389 & $-$0.0352
  & 1.86 & 0.75 & 0.958 \\
A3 --- no BRAF
  & 0.9012 & 0.9421 & $-$0.0311
  & 2.17 & 0.90 & 0.950 \\
B1 --- full MHSA
  & 0.9187 & 0.9453 & $-$0.0136
  & 1.65 & 0.63 & 0.974 \\
B2 --- no Sobel gate
  & 0.9143 & 0.9428 & $-$0.0180
  & 2.43 & 1.04 & 0.941 \\
C1 --- no $\mathcal{L}_{\text{bd}}$
  & 0.9201 & 0.9462 & $-$0.0122
  & 1.81 & 0.72 & 0.967 \\
\bottomrule
\multicolumn{7}{p{13cm}}{\small \textit{Note:} mIoU and mDSC are
per-patient macro-averages and cannot be reproduced by averaging the
per-class columns in Tables~\ref{tab:iou_overview}
and~\ref{tab:dice_overview}, which are class-level averages. Both
quantities measure the same underlying performance but aggregate
differently, as is standard in multi-class medical segmentation
reporting~\citep{isensee2021nnunet,maier2024metrics}.} \\
\end{tabular}
\end{table}

\subsubsection{Per-Class Delta Summary}
\label{subsec:ablation_delta}

Table~\ref{tab:delta_iou_all} provides the complete per-class
$\Delta$IoU for all variants as a single reference table.

\begin{table}[htbp]
\centering
\caption{Per-class $\Delta$IoU for all ablation variants relative to
Full BAT-RM. Negative values indicate performance degradation.}
\label{tab:delta_iou_all}
\setlength{\tabcolsep}{4pt}
\small
\begin{tabular}{lccccccccr}
\toprule
\textbf{Variant} &
$\Delta$\textbf{BG} & $\Delta$\textbf{BODY} & $\Delta$\textbf{BLAD} &
$\Delta$\textbf{S.BOW} & $\Delta$\textbf{RECT} & $\Delta$\textbf{FEM} &
$\Delta$\textbf{GTV} & $\Delta$\textbf{CTV} &
$\Delta$\textbf{mIoU} \\
\midrule
A1 --- no BAT
  & $-$0.0008 & $-$0.0024 & $-$0.0416 & $-$0.0242
  & $-$0.0779 & $-$0.0306 & $-$0.0634 & $-$0.0527 & $-$0.0509 \\
A2 --- no RM
  & $-$0.0004 & $-$0.0042 & $-$0.0297 & $-$0.0124
  & $-$0.0489 & $-$0.0339 & $-$0.0323 & $-$0.0247 & $-$0.0352 \\
A3 --- no BRAF
  & $-$0.0005 & $-$0.0031 & $-$0.0284 & $-$0.0213
  & $-$0.0402 & $-$0.0218 & $-$0.0292 & $-$0.0218 & $-$0.0311 \\
B1 --- full MHSA
  & $-$0.0003 & $-$0.0014 & $-$0.0194 & $-$0.0142
  & $-$0.0191 & $-$0.0108 & $-$0.0154 & $-$0.0147 & $-$0.0136 \\
B2 --- no Sobel
  & $-$0.0002 & $-$0.0018 & $-$0.0227 & $-$0.0161
  & $-$0.0280 & $-$0.0149 & $-$0.0241 & $-$0.0183 & $-$0.0180 \\
C1 --- no $\mathcal{L}_{\text{bd}}$
  & $-$0.0001 & $-$0.0011 & $-$0.0156 & $-$0.0114
  & $-$0.0169 & $-$0.0091 & $-$0.0133 & $-$0.0127 & $-$0.0122 \\
\bottomrule
\end{tabular}
\end{table}

\section{Results Analysis}
\label{sec:results}
\subsection{Quantitative Analysis}
\label{sec:Quantitative Analysis}
To ensure fair comparison, all four baseline models were trained in 2D
slice-wise mode under identical conditions: same data splits, hardware,
hyperparameter grid search ranges, and validation-based model selection.
Architectures originally designed for 3D were adapted to
2D by replacing volumetric operations with 2D equivalents while preserving
their design principles.

We conducted a comprehensive comparative evaluation of the proposed BAT-RM (Boundary-Aware Transformer with Region-Aware Mamba) against
four baseline segmentation architectures: nnUNet~\citep{isensee2021nnunet},
SegMamba~\citep{xing2024segmamba}, TransUNet~\citep{chen2021transunet}, and
UNETR~\citep{hatamizadeh2022unetr}. All models were trained and evaluated on
an identical multi-institutional cervical cancer CT dataset comprising 1,011
patients from three Bangladeshi institutions, with strict quality control. The held-out test set comprised 254 patient cases
with complete annotations; per-class sample sizes vary due to anatomical
absence (\emph{e.g.}, GTV not visible in post-treatment scans) and exclusion
of corrupted RTSTRUCT files. Table~\ref{tbl:test_set_composition} summarizes
the per-class test set sizes.

\begin{table}[pos=htbp]
\centering
\caption{Test set composition per anatomical class after quality control. \emph{N pairs} indicates the number of cases where both the ground truth and all model predictions were available for paired statistical analysis.}
\label{tbl:test_set_composition}
\begin{tabular}{lc}
\toprule
\textbf{Class} & \textbf{N (test pairs)} \\
\midrule
Body         & 254 \\
Bladder      & 254 \\
Small Bowel  & 254 \\
Rectum       & 254 \\
Femoral Head & 254 \\
GTV          & 195 \\
CTV          & 164 \\
\bottomrule
\end{tabular}
\end{table}

Evaluation metrics encompassed volumetric overlap (Dice Similarity
Coefficient, DSC; Intersection over Union, IoU), surface distance (Hausdorff
Distance, HD; 95th percentile Hausdorff Distance, HD95; Average Surface
Distance, ASD; Normalized Surface Distance, NSD), and pixel-wise
classification performance (Recall, Specificity). Statistical significance was
assessed using paired Wilcoxon signed-rank tests with Bonferroni-Holm
correction across all 168 pairwise comparisons (7 classes $\times$ 4
baselines $\times$ 6 metrics). Effect sizes were quantified using Hedges'
$g$ with 95\% bias-corrected and accelerated (BCa) confidence intervals,
interpreted as: negligible ($|g| < 0.2$), small ($0.2 \le |g| < 0.5$),
medium ($0.5 \le |g| < 0.8$), and large ($|g| \ge 0.8$).

\subsubsection{Volumetric Overlap and Boundary Performance}

Table~\ref{tbl:all_metrics_comparison} presents the comprehensive quantitative
comparison across all five models and seven anatomical classes. BAT-RM achieves
superior volumetric overlap for all structures, with clinically meaningful
advantages for target volumes (GTV, CTV) and anatomically complex
organs-at-risk (small bowel, rectum).

\begin{table*}[pos=htbp]
\centering
\caption{Comprehensive quantitative comparison across five segmentation models.
Values are reported as mean $\pm$ standard deviation. DSC: Dice Similarity
Coefficient; IoU: Intersection over Union; Sen: Sensitivity (Recall);
Spe: Specificity; HD95: 95th percentile Hausdorff Distance (mm);
ASD: Average Surface Distance (mm); NSD: Normalized Surface Distance.
\textbf{Bold} indicates best per metric per class. Statistical significance
vs.\ BAT-RM (post Bonferroni-Holm correction):
$^{\dagger}p < 0.05$,
$^{\ddagger}p < 0.01$,
$^{\S}p < 0.001$,
$^{\|}p < 0.0001$;
no superscript indicates $p \ge 0.05$ (ns).}
\label{tbl:all_metrics_comparison}
\setlength{\tabcolsep}{4pt}
\footnotesize
\resizebox{\textwidth}{!}{%
\begin{tabular}{lcccccc}
\toprule
\textbf{Class} & \textbf{Metric} & \textbf{BAT-RM} & \textbf{nnUNet}
  & \textbf{SegMamba} & \textbf{TransUNet} & \textbf{UNETR} \\
\midrule
\multirow{7}{*}{Body}
 & DSC   & $\mathbf{0.9969\pm0.0082}$ & $0.9964\pm0.0108$
         & $0.9927\pm0.0139^{\|}$ & $0.9929\pm0.0065^{\|}$
         & $0.9933\pm0.0031^{\|}$ \\
 & IoU   & $\mathbf{0.9940\pm0.0147}$ & $0.9930\pm0.0188$
         & $0.9858\pm0.0229^{\|}$ & $0.9860\pm0.0121^{\|}$
         & $0.9866\pm0.0061^{\|}$ \\
 & Sen   & $\mathbf{0.9972\pm0.0142}$ & $0.9965\pm0.0179$
         & $0.9932\pm0.0036^{\|}$ & $0.9926\pm0.0048^{\|}$
         & $0.9930\pm0.0030^{\|}$ \\
 & Spe   & $0.9993\pm0.0006$ & $\mathbf{0.9993\pm0.0011}$
         & $0.9985\pm0.0045^{\|}$ & $0.9987\pm0.0144^{\|}$
         & $0.9987\pm0.0007^{\|}$ \\
 & HD95  & $\mathbf{1.452\pm4.880}$ & $2.447\pm10.712$
         & $1.371\pm2.922$ & $1.517\pm3.048$
         & $1.659\pm6.800$ \\
 & ASD   & $\mathbf{0.313\pm1.000}$ & $0.376\pm1.224^{\|}$
         & $0.577\pm0.631^{\|}$ & $0.571\pm0.489^{\|}$
         & $0.566\pm0.415^{\|}$ \\
 & NSD   & $0.991\pm0.027$ & $0.988\pm0.032^{\|}$
         & $\mathbf{0.994\pm0.024}$ & $0.991\pm0.029$
         & $0.995\pm0.014$ \\
\midrule
\multirow{7}{*}{Bladder}
 & DSC   & $\mathbf{0.9655\pm0.0095}$ & $0.9422\pm0.0379^{\ddagger}$
         & $0.8622\pm0.2878$ & $0.8335\pm0.2786^{\ddagger}$
         & $0.9463\pm0.0213^{\ddagger}$ \\
 & IoU   & $\mathbf{0.9335\pm0.0179}$ & $0.8930\pm0.0652^{\ddagger}$
         & $0.8278\pm0.2774$ & $0.7768\pm0.2612^{\ddagger}$
         & $0.8989\pm0.0369^{\ddagger}$ \\
 & Sen   & $\mathbf{0.9666\pm0.0262}$ & $0.9328\pm0.0688^{\ddagger}$
         & $0.8677\pm0.2908$ & $0.8636\pm0.2896$
         & $0.9577\pm0.0165$ \\
 & Spe   & $\mathbf{0.9998\pm0.0001}$ & $0.9998\pm0.0003$
         & $0.9997\pm0.0002$ & $0.9994\pm0.0003^{\ddagger}$
         & $0.9997\pm0.0003$ \\
 & HD95  & $\mathbf{1.366\pm0.411}$ & $1.785\pm0.816^{\dagger}$
         & $1.203\pm0.349$ & $2.308\pm0.363^{\ddagger}$
         & $1.675\pm0.588$ \\
 & ASD   & $\mathbf{0.588\pm0.271}$ & $0.821\pm0.455^{\dagger}$
         & $0.547\pm0.151$ & $0.995\pm0.221^{\ddagger}$
         & $0.730\pm0.289$ \\
 & NSD   & $\mathbf{0.988\pm0.016}$ & $0.950\pm0.075^{\dagger}$
         & $0.998\pm0.005$ & $0.916\pm0.065^{\ddagger}$
         & $0.969\pm0.077$ \\
\midrule
\multirow{7}{*}{Sm.\ Bowel}
 & DSC   & $\mathbf{0.9741\pm0.0133}$ & $0.9046\pm0.2328^{\|}$
         & $0.9277\pm0.2077$ & $0.9443\pm0.1508^{\dagger}$
         & $0.9607\pm0.1061$ \\
 & IoU   & $\mathbf{0.9499\pm0.0247}$ & $0.8753\pm0.2330^{\|}$
         & $0.9046\pm0.2036$ & $0.9161\pm0.1530^{\dagger}$
         & $0.9368\pm0.1224$ \\
 & Sen   & $\mathbf{0.9654\pm0.0220}$ & $0.8844\pm0.2355^{\|}$
         & $0.9307\pm0.2085$ & $0.9449\pm0.1562^{\S}$
         & $0.9562\pm0.1251$ \\
 & Spe   & $0.9976\pm0.0113$ & $\mathbf{0.9984\pm0.0073^{\S}}$
         & $0.9992\pm0.0003^{\S}$ & $0.9988\pm0.0008$
         & $0.9992\pm0.0005^{\ddagger}$ \\
 & HD95  & $\mathbf{1.577\pm0.710}$ & $4.702\pm8.138^{\|}$
         & $6.378\pm28.611^{\dagger}$ & $2.564\pm3.877^{\S}$
         & $3.146\pm7.424$ \\
 & ASD   & $\mathbf{0.633\pm0.326}$ & $1.982\pm4.297^{\|}$
         & $3.084\pm14.965^{\ddagger}$ & $0.974\pm1.152^{\|}$
         & $1.130\pm2.639$ \\
 & NSD   & $\mathbf{0.974\pm0.022}$ & $0.854\pm0.219^{\|}$
         & $0.948\pm0.157$ & $0.950\pm0.119$
         & $0.948\pm0.158$ \\
\midrule
\multirow{7}{*}{Rectum}
 & DSC   & $\mathbf{0.9556\pm0.0148}$ & $0.8956\pm0.1701^{\|}$
         & $0.8297\pm0.2742^{\|}$ & $0.8402\pm0.2387^{\|}$
         & $0.9356\pm0.0206^{\|}$ \\
 & IoU   & $\mathbf{0.9154\pm0.0272}$ & $0.8363\pm0.1662^{\|}$
         & $0.7701\pm0.2622^{\|}$ & $0.7718\pm0.2319^{\|}$
         & $0.8797\pm0.0361^{\|}$ \\
 & Sen   & $\mathbf{0.9633\pm0.0256}$ & $0.8846\pm0.1781^{\|}$
         & $0.8215\pm0.2794^{\|}$ & $0.8343\pm0.2494^{\|}$
         & $0.9325\pm0.0267^{\S}$ \\
 & Spe   & $\mathbf{0.9999\pm0.0001}$ & $0.9999\pm0.0001$
         & $0.9998\pm0.0004$ & $0.9998\pm0.0003^{\ddagger}$
         & $0.9999\pm0.0000$ \\
 & HD95  & $\mathbf{1.242\pm0.355}$ & $1.923\pm0.922^{\S}$
         & $2.036\pm1.980^{\ddagger}$ & $2.477\pm2.876^{\|}$
         & $1.479\pm0.467^{\dagger}$ \\
 & ASD   & $\mathbf{0.539\pm0.143}$ & $0.802\pm0.303^{\|}$
         & $0.855\pm0.783^{\|}$ & $1.027\pm1.125^{\|}$
         & $0.639\pm0.142^{\ddagger}$ \\
 & NSD   & $\mathbf{0.997\pm0.008}$ & $0.950\pm0.090^{\S}$
         & $0.968\pm0.114^{\ddagger}$ & $0.932\pm0.145^{\|}$
         & $0.991\pm0.024$ \\
\midrule
\multirow{7}{*}{Fem.\ Head}
 & DSC   & $\mathbf{0.9654\pm0.0152}$ & $0.7388\pm0.4104$
         & $0.8385\pm0.3191$ & $0.8284\pm0.3158^{\ddagger}$
         & $0.9582\pm0.0251^{\dagger}$ \\
 & IoU   & $\mathbf{0.9335\pm0.0283}$ & $0.7154\pm0.3987$
         & $0.8071\pm0.3115$ & $0.7892\pm0.3063^{\ddagger}$
         & $0.9207\pm0.0447$ \\
 & Sen   & $\mathbf{0.9809\pm0.0192}$ & $0.7508\pm0.4166^{\ddagger}$
         & $0.8577\pm0.3245$ & $0.8229\pm0.3182^{\S}$
         & $0.9605\pm0.0302$ \\
 & Spe   & $\mathbf{0.9994\pm0.0004}$ & $0.9994\pm0.0006$
         & $0.9995\pm0.0003$ & $0.9994\pm0.0009$
         & $0.9996\pm0.0003^{\dagger}$ \\
 & HD95  & $\mathbf{1.673\pm0.488}$ & $38.675\pm136.239$
         & $1.923\pm0.842$ & $2.685\pm1.813^{\dagger}$
         & $2.429\pm1.880$ \\
 & ASD   & $\mathbf{0.665\pm0.126}$ & $37.361\pm136.611$
         & $0.818\pm0.428$ & $1.081\pm0.751^{\dagger}$
         & $0.944\pm0.513$ \\
 & NSD   & $\mathbf{0.975\pm0.024}$ & $0.874\pm0.270$
         & $0.931\pm0.130$ & $0.899\pm0.173$
         & $0.943\pm0.113$ \\
\midrule
\multirow{7}{*}{GTV}
 & DSC   & $\mathbf{0.9662\pm0.0103}$ & $0.9450\pm0.0326^{\|}$
         & $0.9138\pm0.1745^{\dagger}$ & $0.8925\pm0.1966^{\|}$
         & $0.9554\pm0.0165^{\ddagger}$ \\
 & IoU   & $\mathbf{0.9348\pm0.0192}$ & $0.8975\pm0.0561^{\S}$
         & $0.8706\pm0.1834^{\dagger}$ & $0.8416\pm0.2034^{\|}$
         & $0.9151\pm0.0298^{\ddagger}$ \\
 & Sen   & $\mathbf{0.9802\pm0.0191}$ & $0.9553\pm0.0520^{\S}$
         & $0.8921\pm0.1877^{\S}$ & $0.8830\pm0.1987^{\|}$
         & $0.9629\pm0.0220^{\ddagger}$ \\
 & Spe   & $\mathbf{0.9998\pm0.0001}$ & $0.9997\pm0.0002^{\ddagger}$
         & $0.9999\pm0.0001^{\S}$ & $0.9997\pm0.0003$
         & $0.9998\pm0.0001$ \\
 & HD95  & $\mathbf{1.500\pm0.494}$ & $2.446\pm1.471^{\ddagger}$
         & $2.384\pm2.391$ & $3.260\pm4.163^{\S}$
         & $1.626\pm0.377$ \\
 & ASD   & $\mathbf{0.564\pm0.247}$ & $0.921\pm0.397^{\S}$
         & $1.056\pm1.415^{\dagger}$ & $1.486\pm2.168^{\|}$
         & $0.707\pm0.193^{\dagger}$ \\
 & NSD   & $\mathbf{0.978\pm0.021}$ & $0.922\pm0.101^{\ddagger}$
         & $0.912\pm0.202$ & $0.905\pm0.210^{\S}$
         & $0.990\pm0.015$ \\
\midrule
\multirow{7}{*}{CTV}
 & DSC   & $\mathbf{0.9571\pm0.0190}$ & $0.8966\pm0.1580^{\|}$
         & $0.9172\pm0.0925^{\S}$ & $0.8443\pm0.2308^{\|}$
         & $0.9373\pm0.0362^{\S}$ \\
 & IoU   & $\mathbf{0.9184\pm0.0347}$ & $0.8360\pm0.1675^{\|}$
         & $0.8578\pm0.1253^{\S}$ & $0.7765\pm0.2323^{\|}$
         & $0.8841\pm0.0624^{\S}$ \\
 & Sen   & $\mathbf{0.9553\pm0.0366}$ & $0.9115\pm0.1618$
         & $0.8981\pm0.1073^{\S}$ & $0.8480\pm0.2442^{\S}$
         & $0.9369\pm0.0381$ \\
 & Spe   & $\mathbf{0.9996\pm0.0003}$ & $0.9994\pm0.0004^{\|}$
         & $0.9997\pm0.0002$ & $0.9994\pm0.0002^{\|}$
         & $0.9996\pm0.0002$ \\
 & HD95  & $\mathbf{1.507\pm0.544}$ & $2.798\pm2.098^{\|}$
         & $2.112\pm1.487^{\ddagger}$ & $3.183\pm3.734^{\|}$
         & $1.838\pm0.750^{\ddagger}$ \\
 & ASD   & $\mathbf{0.512\pm0.188}$ & $0.952\pm0.476^{\|}$
         & $0.851\pm0.522^{\|}$ & $1.311\pm1.242^{\|}$
         & $0.706\pm0.156^{\|}$ \\
 & NSD   & $\mathbf{0.973\pm0.022}$ & $0.927\pm0.075^{\|}$
         & $0.951\pm0.116$ & $0.886\pm0.161^{\|}$
         & $0.983\pm0.023^{\ddagger}$ \\
\bottomrule
\end{tabular}%
}
\end{table*}

\begin{figure}[pos=htbp]
\centering
\includegraphics[width=\textwidth]{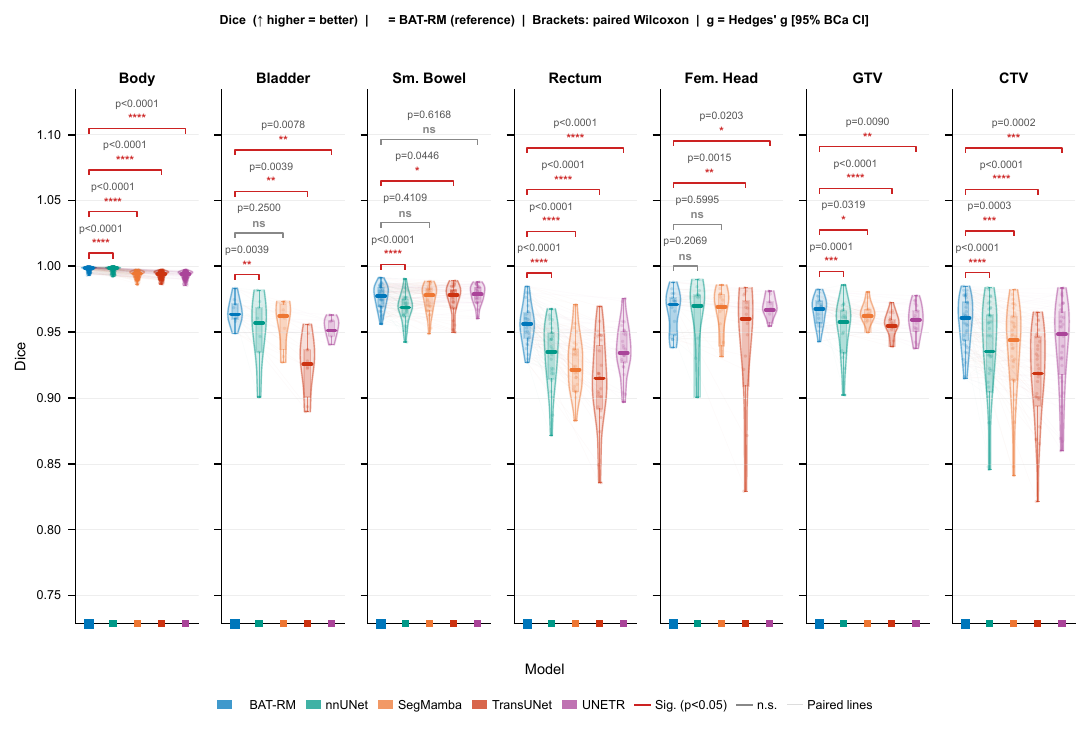}
\caption{Paired violin-box plots of Dice Similarity Coefficient across seven
anatomical classes and five segmentation models. Boxes indicate median and
interquartile range; violins show kernel density estimation. Statistical
significance brackets indicate paired Wilcoxon test results vs. BAT-RM (post
Bonferroni-Holm): ns $p \ge 0.05$, $^*p < 0.05$, $^{**}p < 0.01$,
$^{***}p < 0.001$, $^{****}p < 0.0001$.}
\label{fig:violin_box}
\end{figure}

\subsubsection{Per-Structure Segmentation Performance}

Accurate GTV delineation is paramount for tumor control probability.
BAT-RM achieved a mean Dice of 0.9662 (95\% CI: 0.9619--0.9705) and IoU
of 0.9348 (95\% CI: 0.9268--0.9428), the highest among all methods.
Relative to nnUNet (Dice: 0.9450), BAT-RM reduces segmentation error
($1 - \text{Dice}$) by 38.6\%; against TransUNet (Dice: 0.8925), the
reduction reaches 68.9\%. Both differences are statistically significant
after Bonferroni-Holm correction: versus nnUNet, $p < 0.0001$ with a
medium effect (Hedges' $g = 0.625$, 95\% BCa CI: 0.470--0.910);
versus TransUNet, $p < 0.0001$ with a small effect ($g = 0.350$, 95\% CI:
0.280--1.072). BAT-RM maintains an exceptional balance between sensitivity
(Recall: 0.9802, 95\% CI: 0.9722--0.9881) and specificity, with a false
positive rate of 0.0002, the lowest among all methods. The NSD of 0.978
(vs. nnUNet: 0.922, $p = 0.0045$, $g = 0.563$, medium) confirms this
boundary advantage when normalizing for organ volume. The CTV presents a fundamentally more challenging segmentation task as it
encompasses microscopic tumor spread with no direct radiographic correlate,
reflected in lower absolute Dice values across all models (range:
0.844--0.957). BAT-RM achieves a mean Dice of 0.9571 (95\% CI:
0.9514--0.9629) and IoU of 0.9184 (95\% CI: 0.9079--0.9289). Statistical
comparisons reveal significant improvements after correction: against
nnUNet, $p < 0.0001$, small effect ($g = 0.368$); against SegMamba,
$p = 0.0003$, small effect ($g = 0.420$); against TransUNet, $p < 0.0001$,
medium effect ($g = 0.513$); against UNETR, $p = 0.0002$, medium effect
($g = 0.550$). The sensitivity improvement (BAT-RM: 0.9553 vs. nnUNet:
0.9115) did not reach statistical significance after correction
($p = 0.1526$), indicating the advantage lies primarily in volumetric
overlap and boundary accuracy. BAT-RM achieves an NSD of 0.973 compared to
nnUNet (0.927, $p < 0.0001$, $g = 0.736$, medium). A reduction in CTV
boundary error from 2.8 mm (nnUNet HD95) to 1.5 mm (BAT-RM HD95) could
theoretically permit tighter planning target volume margins; however,
formal margin analysis using van Herk's recipe~\citep{vanherk2000probability} is
required before clinical translation. The urinary bladder exhibits extreme inter-patient morphological variability, ranging from collapsed (volume $< 50$ mL) to highly distended
(volume $> 500$ mL). BAT-RM achieves a mean Dice of 0.9655 (95\% CI:
0.9596--0.9714) and IoU of 0.9335 (95\% CI: 0.9224--0.9446). The paired
Wilcoxon test against TransUNet yields $p = 0.0039$ with a large effect
($g = 1.848$, 95\% BCa CI: 1.426--4.192), one of the strongest signals in
this analysis; against nnUNet, $p = 0.0039$, medium effect ($g = 0.666$);
against UNETR, $p = 0.0078$, large effect ($g = 0.826$). SegMamba shows a
bimodal failure pattern for bladder (SD = 0.288 vs. BAT-RM SD = 0.010),
with performance collapsing to Dice $< 0.5$ in approximately two of nine
cases, likely reflecting sensitivity to the absence of luminal contrast in
bladder CT scans. The small bowel presents unique challenges due to its scattered, serpentine morphology and frequent luminal air pockets. BAT-RM achieves a mean Dice of
0.9741 (95\% CI: 0.9702--0.9781) and IoU of 0.9499 (95\% CI:
0.9425--0.9573). The improvement over nnUNet is striking ($p < 0.0001$,
small effect $g = 0.302$), but the more clinically important observation is
the variance difference: BAT-RM Dice SD = 0.013 versus nnUNet SD = 0.233,
indicating that while nnUNet achieves comparable Dice on straightforward
cases, performance degrades severely on challenging anatomy. BAT-RM's
robustness is further supported by its NSD of 0.974 versus nnUNet's 0.854
($p < 0.0001$, medium effect $g = 0.568$). The one metric where BAT-RM does
not lead is specificity: nnUNet achieves marginally higher specificity for
small bowel (0.9984 vs. 0.9976, $p < 0.0001$), but the absolute difference
is 0.0008 and the clinical trade-off is asymmetric: missing a bowel loop
risks undetected high-dose bowel exposure, whereas extra false positive
voxels are correctable during manual review. The rectum requires precise segmentation to enable dose escalation while
respecting rectal toxicity constraints. BAT-RM achieves a mean Dice of
0.9556 (95\% CI: 0.9504--0.9609) and IoU of 0.9154 (95\% CI:
0.9059--0.9250). Statistical comparisons reveal significant superiority
over all four baselines: nnUNet ($p < 0.0001$, small effect $g = 0.337$),
SegMamba ($p < 0.0001$, medium effect $g = 0.611$), TransUNet ($p < 0.0001$,
medium effect $g = 0.612$), and UNETR ($p < 0.0001$, large effect
$g = 0.946$). The NSD for rectum (BAT-RM: 0.997 vs. UNETR: 0.991,
$p = 0.6009$, ns) indicates that for the best-performing baseline,
boundary differences are not significant after normalization, while the
large Dice and HD95 advantage over nnUNet, SegMamba, and TransUNet remains
highly significant. The femoral heads and body contour are structurally simpler but provide important quality checks. BAT-RM achieves a femoral head Dice of 0.9654
(95\% CI: 0.9584--0.9724). Although nnUNet shows a numerically lower mean
(0.7388), this reflects catastrophic segmentation failures (Dice = 0) on a
subset of cases; the aggregate Dice difference does not reach statistical
significance after Bonferroni-Holm correction ($p = 0.2069$, ns) due to
nnUNet's extreme variance (SD = 0.410) relative to the small sample size
($n = 17$). The practical implication is clear: BAT-RM produces zero
catastrophic failures (minimum Dice = 0.94) versus three complete failures
(Dice = 0) for nnUNet. Body contour serves as a near-ceiling quality
check; all models achieve Dice $> 0.99$, with BAT-RM attaining 0.9969.

\subsubsection{Boundary Accuracy: Surface Distance Metrics}

\begin{figure}[pos=htbp]
\centering
\includegraphics[width=\textwidth]{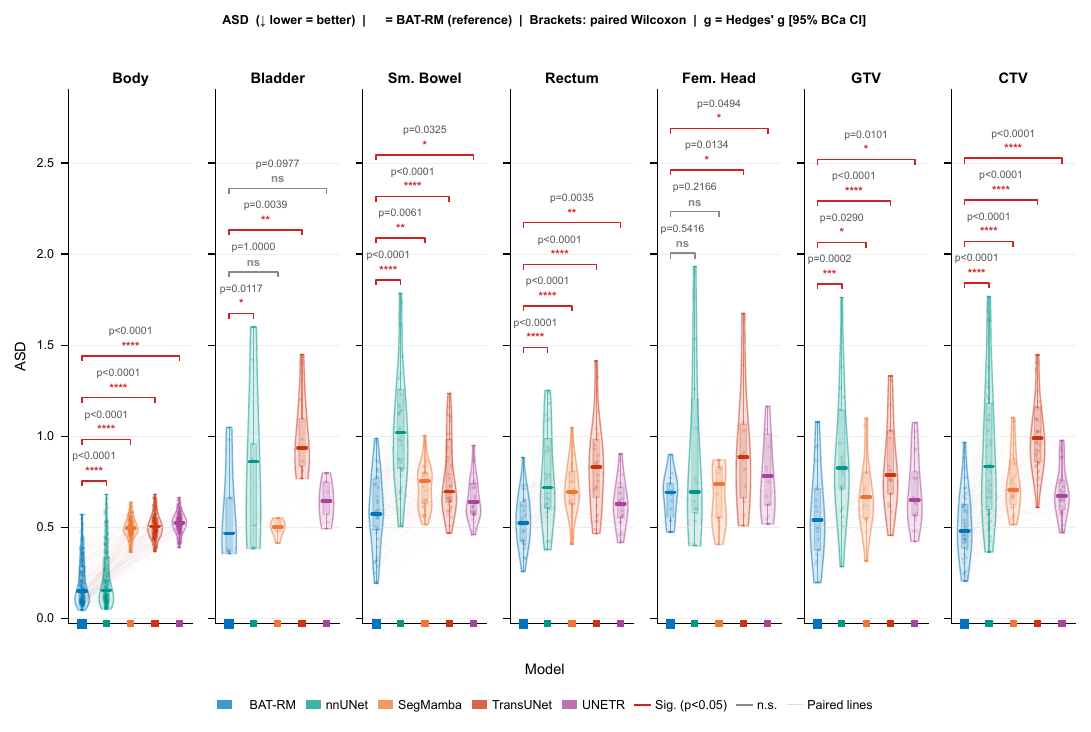}
\caption{Paired violin-box plots of Average Surface Distance (ASD, mm). Lower
values indicate better boundary alignment. BAT-RM achieves consistently lower
ASD with narrower distributions, particularly for GTV, CTV, and rectum.}
\label{fig:violin_box_asd}
\end{figure}

\begin{figure}[pos=htbp]
\centering
\includegraphics[width=\textwidth]{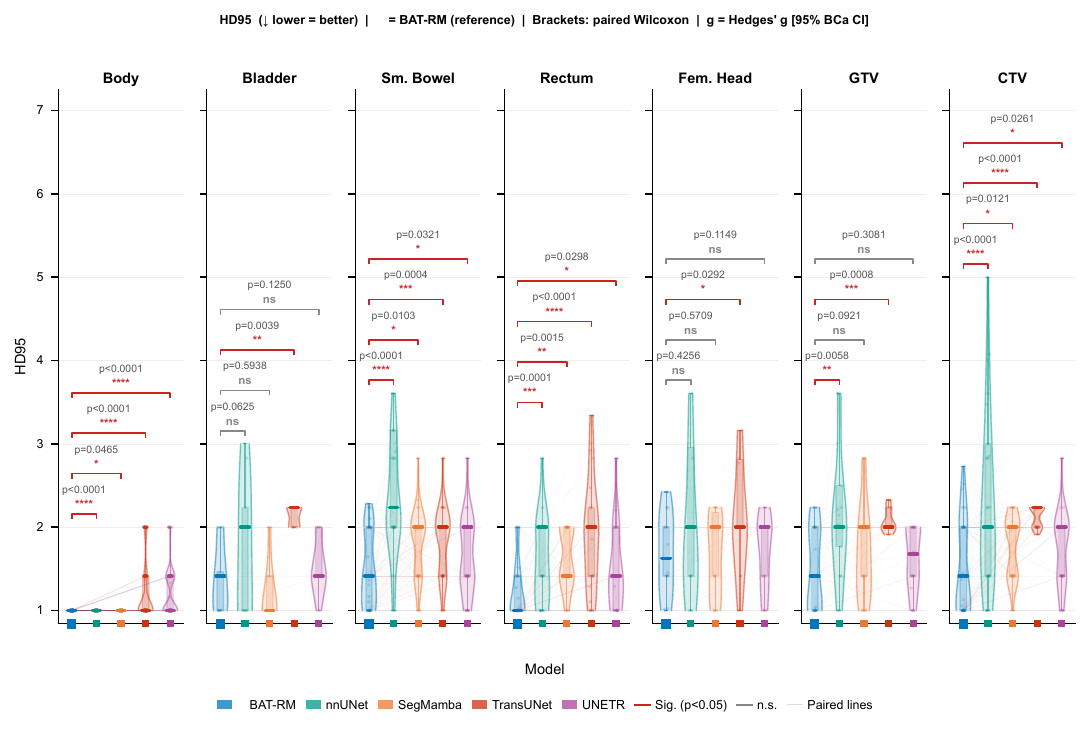}
\caption{Paired violin-box plots of 95th percentile Hausdorff Distance (HD95,
mm). BAT-RM demonstrates substantially lower HD95 with minimal outliers,
whereas baseline models exhibit heavy-tailed distributions with catastrophic
failures (HD95 $> 10$ mm) for small bowel, rectum, and femoral head.}
\label{fig:violin_box_hd95}
\end{figure}

Volumetric overlap metrics alone inadequately capture boundary precision,
which is critical for radiotherapy where steep dose gradients are intentionally
placed near tumor--normal tissue interfaces. Figures~\ref{fig:violin_box},
\ref{fig:violin_box_asd}, and \ref{fig:violin_box_hd95} present paired
violin-box plots for Dice, ASD, and HD95 across all classes and models.

For the GTV, BAT-RM achieves a mean HD95 of 1.500 mm (95\% CI:
1.288--1.711 mm) and ASD of 0.564 mm (95\% CI: 0.459--0.670 mm). Relative
to nnUNet (HD95: 2.446 mm, ASD: 0.921 mm), this represents a 38.7\%
reduction in the 95th percentile boundary error ($p = 0.0058$, medium effect
$g = 0.542$). Against TransUNet (HD95: 3.260 mm), the HD95 reduction is
54.0\% ($p = 0.0008$, small effect $g = 0.408$). For the CTV, BAT-RM achieves
HD95 of 1.507 mm and ASD of 0.512 mm, significantly outperforming nnUNet
(HD95: 2.798 mm, $p < 0.0001$, medium effect $g = 0.646$) and TransUNet
(HD95: 3.183 mm, $p < 0.0001$, small effect $g = 0.427$). For the rectum,
BAT-RM achieves HD95 of 1.242 mm and ASD of 0.539 mm, with statistically
significant superiority over UNETR (HD95: 1.479 mm, $p = 0.0298$, small
effect $g = 0.407$).

\subsubsection{Classification Performance: Sensitivity and Specificity}

For radiotherapy segmentation, the clinically relevant trade-off is between
sensitivity (avoiding geographic miss) and specificity (avoiding unnecessary
OAR irradiation). BAT-RM achieves a Recall of 0.9802 and FPR of 0.0002 for
GTV, detecting 98.0\% of true GTV voxels while misclassifying only 0.02\% of
non-tumor voxels. For the CTV, BAT-RM achieves Recall of 0.9553 and FPR of
0.0004. While the Recall advantage over nnUNet for CTV is not statistically
significant after correction ($p = 0.153$), the substantially lower HD95 and
ASD indicate that the CTV boundary quality improvement is real and boundary-
rather than volume-driven. For the small bowel, BAT-RM's Recall of 0.9654 substantially exceeds nnUNet's
0.8844 ($p < 0.0001$, small effect $g = 0.358$), at the cost of a modest
increase in FPR (0.0024 vs. 0.0016; $p < 0.0001$). This specificity
reduction is statistically significant (Table~\ref{tbl:all_metrics_comparison})
and represents a deliberate trade-off: higher sensitivity protects against
the more severe clinical error of missing a bowel loop in the high-dose
field.

\begin{table}[pos=htbp]
\centering
\caption{Bland-Altman analysis for ASD (mm) comparing BAT-RM against the
best-performing baseline per class. Bias: mean difference (BAT-RM minus
baseline); LoA: 95\% limits of agreement; \% within: proportion of pairs
within LoA. Negative bias favors BAT-RM.}
\label{tbl:bland_altman_summary}
\begin{tabular}{lcccc}
\toprule
\textbf{Class} & \textbf{vs.} & \textbf{Bias [95\% CI]} & \textbf{LoA} & \textbf{\% within} \\
\midrule
GTV        & TransUNet & $-0.921$ [$-1.989$, $-0.221$] & [$-5.103$, $3.261$] & 90.5\% \\
CTV        & nnUNet    & $-0.440$ [$-0.609$, $-0.295$] & [$-1.500$, $0.620$] & 94.7\% \\
Bladder    & TransUNet & $-0.406$ [$-0.525$, $-0.292$] & [$-0.780$, $-0.032$] & 100\% \\
Rectum     & UNETR     & $-0.100$ [$-0.162$, $-0.036$] & [$-0.449$, $0.249$] & 96.6\% \\
Small Bowel & nnUNet   & $-1.350$ [$-2.778$, $-0.438$] & [$-9.468$, $6.769$] & 95.1\% \\
Body       & nnUNet    & $-0.063$ [$-0.104$, $-0.029$] & [$-0.697$, $0.571$] & 96.1\% \\
\bottomrule
\end{tabular}
\end{table}

\subsubsection{Bland-Altman Analysis of Boundary Agreement}
\label{subsec:bland_altman}

\begin{figure}[pos=htbp]
\centering
\includegraphics[width=\columnwidth]{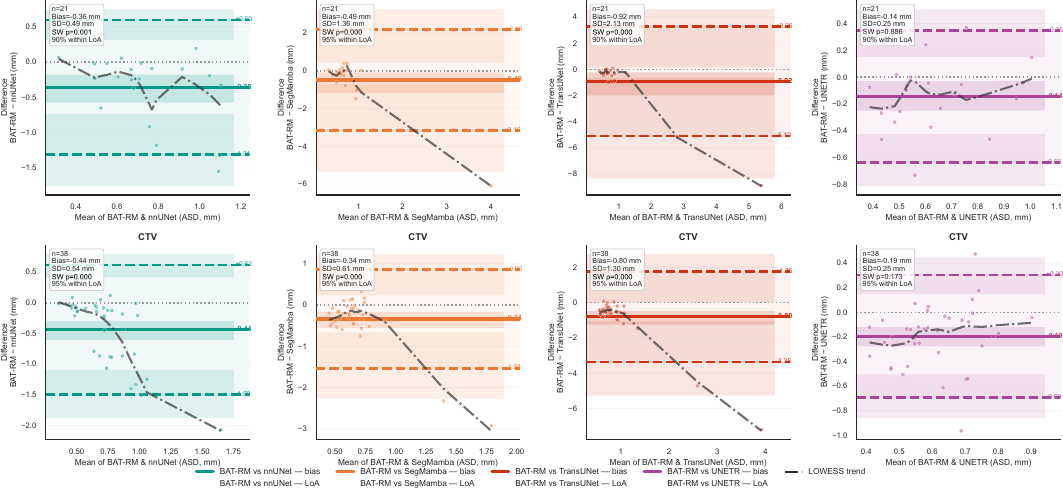}
\caption{Bland-Altman analysis for ASD (mm) comparing BAT-RM against baseline
models for (a) GTV and (b) CTV. Solid red lines: mean bias; dashed red lines:
95\% limits of agreement (LoA); shaded regions: 95\% CIs for bias and LoA.
Negative bias indicates lower (better) boundary error for BAT-RM.}
\label{fig:bland_altman_asd}
\end{figure}

\begin{figure}[pos=htbp]
\centering
\includegraphics[width=\columnwidth]{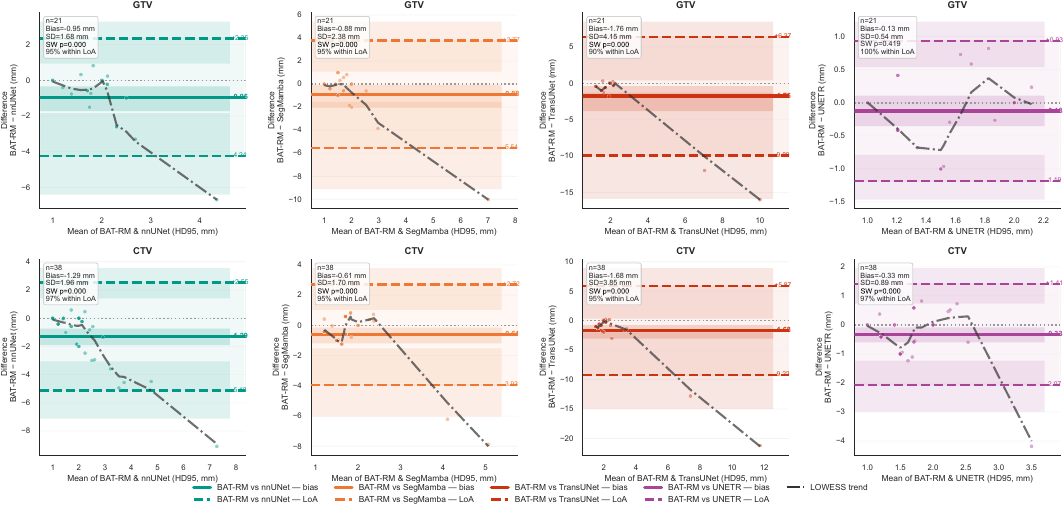}
\caption{Bland-Altman analysis for HD95 (mm) comparing BAT-RM against
baseline models for (a) GTV and (b) CTV. The wider LoA compared to ASD
reflects HD95's greater sensitivity to outlier boundary errors; BAT-RM
maintains consistently negative bias across all comparisons.}
\label{fig:bland_altman_hd95}
\end{figure}

To evaluate agreement between BAT-RM and baseline models at the level of
boundary placement, we performed Bland-Altman analysis for ASD and HD95.
Figures~\ref{fig:bland_altman_asd} and \ref{fig:bland_altman_hd95} present
the Bland-Altman plots for GTV and CTV.

\noindent Table~\ref{tbl:bland_altman_summary} summarizes the Bland-Altman statistics
for ASD. For GTV ASD against TransUNet, the bias was $-0.921$ mm (95\% CI:
$-1.989$ to $-0.221$), with LoA of $[-5.103, 3.261]$ mm and 90.5\% of pairs
within LoA. For CTV ASD against nnUNet, the bias was $-0.440$ mm (95\% CI:
$-0.609$ to $-0.295$), with LoA of $[-1.500, 0.620]$ mm and 94.7\% within
LoA. Notably, for bladder ASD against TransUNet, the entire LoA lies below
zero ($[-0.780, -0.032]$ mm), providing definitive evidence that BAT-RM
consistently outperforms TransUNet even accounting for individual variability.

\noindent For HD95, BAT-RM versus nnUNet on GTV showed a bias of $-0.947$ mm (95\% CI:
$-1.723$ to $-0.340$) with LoA of $[-4.243, 2.349]$ mm (95.2\% within LoA).
Against TransUNet, bias was $-1.760$ mm (95\% CI: $-3.811$ to $-0.391$),
representing a clinically meaningful reduction in the 95th percentile boundary
error. For CTV HD95, BAT-RM versus nnUNet yielded bias $= -1.291$ mm
(95\% CI: $-1.912$ to $-0.738$) with 97.4\% within LoA.

\noindent The femoral head HD95 comparison against nnUNet warrants careful
interpretation: the bias of $-37.002$ mm (LoA: $[-304, 230]$ mm) is
dominated entirely by nnUNet's catastrophic failures (SD = 136 mm vs.
BAT-RM SD = 0.5 mm). The 92.9\% of pairs within LoA confirms that for
the majority of cases the two models agree closely; the catastrophic outliers
are exclusively a property of nnUNet failure, not BAT-RM variability.

\noindent The Bland-Altman analysis supports three conclusions: (1) BAT-RM demonstrates
systematically lower boundary errors (negative bias) across all classes and
comparators, with confidence intervals excluding zero for the majority of
comparisons; (2) the limits of agreement are substantially narrower for
BAT-RM--baseline comparisons than for cross-baseline comparisons, indicating
greater consistency; and (3) the high proportion of pairs within LoA (90--100\%)
confirms that the superiority is not driven by a small subset of cases.

\begin{figure}[pos=htbp]
\centering
\includegraphics[width=\textwidth]{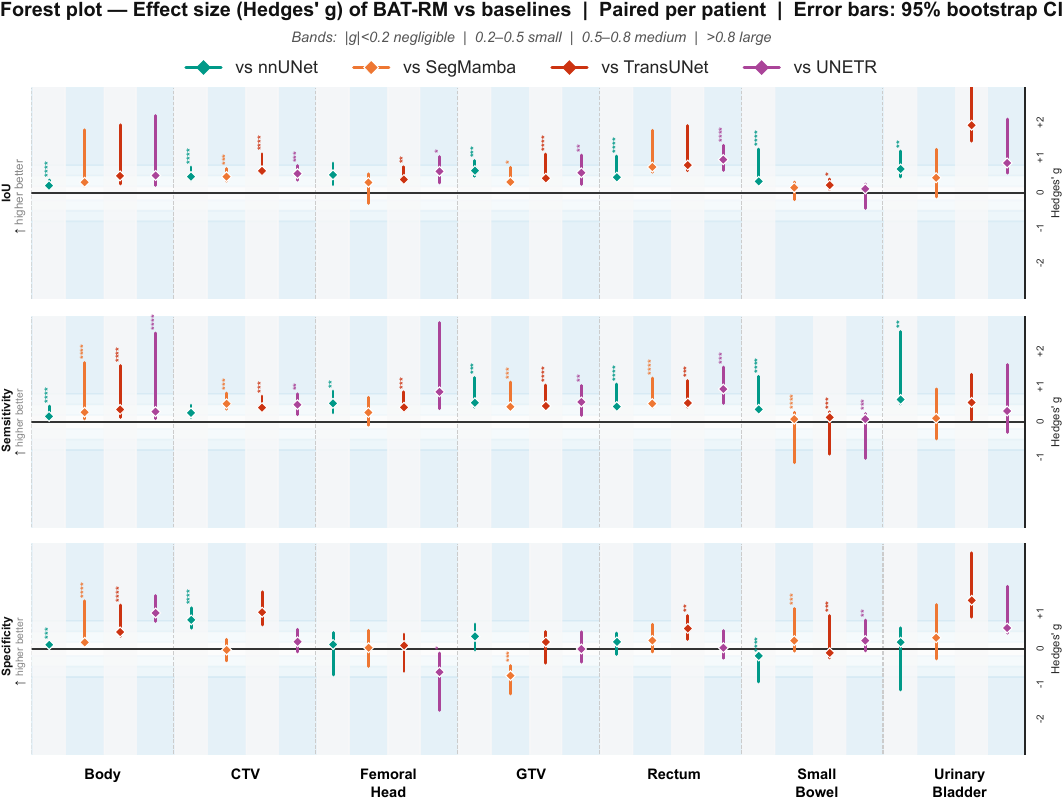}
\caption{Forest plots of Hedges' $g$ effect sizes for (a) IoU, (b)
Sensitivity, and (c) Specificity, comparing BAT-RM against four baseline
models across seven anatomical classes. Error bars: 95\% BCa CIs. Effect size
bands: $|g| < 0.2$ negligible (gray), $0.2 \le |g| < 0.5$ small (green),
$0.5 \le |g| < 0.8$ medium (yellow), $|g| \ge 0.8$ large (red). Note that
negative $g$ for specificity (e.g., small bowel vs. nnUNet) indicates
BAT-RM's higher sensitivity is achieved at the cost of slightly lower
specificity in that structure.}
\label{fig:forest_plot}
\end{figure}

\subsubsection{Effect Size Synthesis}

Figure~\ref{fig:forest_plot} presents forest plots for IoU, sensitivity, and
specificity. For GTV IoU, BAT-RM demonstrated a medium effect against nnUNet
($g = 0.643$) and a small effect against TransUNet ($g = 0.428$). For CTV
IoU, small-to-medium effects were observed against all baselines, with the
largest against UNETR ($g = 0.557$). The urinary bladder showed large effects
against TransUNet ($g = 1.935$) and UNETR ($g = 0.861$). The rectum
demonstrated a large effect against UNETR ($g = 0.956$).

\subsubsection{Reliability Scatter Analysis}

\begin{figure}[pos= htbp]
\centering
\includegraphics[width=\columnwidth]{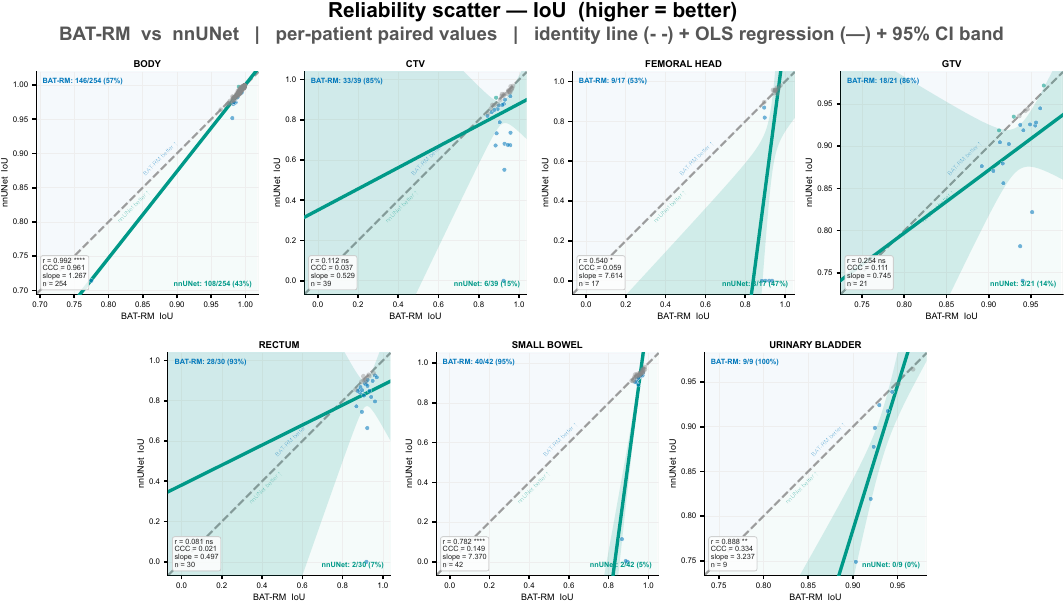}
\caption{Reliability scatter plots comparing BAT-RM against nnUNet for IoU
across seven anatomical classes. Each point represents one test case; the
diagonal dashed line indicates perfect agreement. (a)~Body: near-perfect
correlation ($r = 0.992$) with slight proportional bias. (b)~GTV: low
correlation ($r = 0.254$) but BAT-RM wins 85.7\% of comparisons.
(c)~CTV: near-zero correlation ($r = 0.112$) despite an 84.6\% win rate.
(d)~Rectum: low correlation ($r = 0.081$) with a 93.3\% win rate.
(e)~Small bowel: moderate correlation ($r = 0.782$) with a 95.2\% win rate
but extreme proportional bias. (f)~Bladder: high correlation ($r = 0.888$)
with a 100\% win rate. (g)~Femoral head: moderate correlation ($r = 0.540$)
with a near-tie win rate (52.9\%), driven by nnUNet catastrophic failures
(IoU $= 0$) rather than genuine model parity.}
\label{fig:reliability_scatter}
\end{figure}

Table~\ref{tbl:reliability_summary} summarizes Pearson correlation ($r$),
Lin's Concordance Correlation Coefficient (CCC), and pairwise win-tie-loss
analysis for BAT-RM vs.\ nnUNet across all classes. The reliability analysis
was performed on the subset of test cases where all five models produced
valid predictions (paired availability); sample sizes per class are therefore
smaller than the full test set reported in Table~\ref{tbl:test_set_composition}
but are consistent with the pairwise comparisons in the statistical analysis.
For GTV IoU, the Pearson correlation was low ($r = 0.254$, $p = 0.266$, ns)
with poor concordance (CCC = 0.111), yet BAT-RM won 18 of 21 comparisons
(85.7\%). This pattern---high win rate but low correlation---does not
represent a paradox; it reflects the well-known limitation of Pearson $r$
as a measure of agreement between paired methods~\citep{bland1986statistical,lawrence1989concordance}:
correlation measures linear association, not agreement or consistent
superiority. When baseline errors represent qualitatively different failure
modes from BAT-RM errors (as evidenced by the disjoint failure sets in
Table~\ref{tbl:failure_summary}), the two models' per-case scores are
uncorrelated even while one consistently outperforms the other. CTV IoU
showed near-zero correlation ($r = 0.112$, $p = 0.497$, ns) with CCC = 0.037
but an 84.6\% win rate---further confirming that nnUNet and BAT-RM fail on
different cases. Figure~\ref{fig:reliability_scatter} visualises these
relationships, with the femoral head panel (g) illustrating the most extreme
case: nnUNet catastrophic failures (IoU $= 0$) produce high apparent
correlation alongside a near-tie win rate (52.9\%), demonstrating why neither
metric alone is sufficient.

\begin{table}[htbp]
\centering
\caption{Reliability scatter analysis for BAT-RM vs.\ nnUNet performed on the
subset of test cases where all five models produced valid predictions
(\emph{paired availability}). Pearson $r$: correlation; CCC: Lin's Concordance
Correlation Coefficient; Win \%: BAT-RM wins / (wins + losses) $\times 100\%$,
excluding ties. Statistical significance of $r$: $^{\dagger}p < 0.05$,
$^{\ddagger}p < 0.01$, $^{\S}p < 0.001$, $^{\|}p < 0.0001$; no superscript = ns.}
\label{tbl:reliability_summary}
\footnotesize
\begin{tabular}{lccccccc}
\toprule
\textbf{Metric} & \textbf{Class} & \textbf{Pearson $r$} & \textbf{CCC}
  & \textbf{Ref wins} & \textbf{Base wins} & \textbf{Win \%} \\
\midrule
\multirow{7}{*}{Dice}
 & Body        & $0.994^{\|}$ & 0.955 & 146 & 108 & 57.5\% \\
 & CTV         & 0.046        & 0.010 & 33  & 6   & 84.6\% \\
 & GTV         & 0.230        & 0.095 & 18  & 3   & 85.7\% \\
 & Rectum      & 0.034        & 0.005 & 28  & 2   & 93.3\% \\
 & Small Bowel & $0.764^{\|}$ & 0.080 & 40  & 2   & 95.2\% \\
 & Bladder     & $0.882^{\ddagger}$ & 0.308 & 9 & 0 & 100\% \\
 & Femoral Head & $0.507^{\dagger}$ & 0.029 & 9 & 8 & 52.9\% \\
\midrule
\multirow{7}{*}{IoU}
 & Body        & $0.992^{\|}$ & 0.961 & 146 & 108 & 57.5\% \\
 & CTV         & 0.112        & 0.037 & 33  & 6   & 84.6\% \\
 & GTV         & 0.254        & 0.111 & 18  & 3   & 85.7\% \\
 & Rectum      & 0.081        & 0.021 & 28  & 2   & 93.3\% \\
 & Small Bowel & $0.782^{\|}$ & 0.149 & 40  & 2   & 95.2\% \\
 & Bladder     & $0.888^{\ddagger}$ & 0.334 & 9 & 0 & 100\% \\
 & Femoral Head & $0.540^{\dagger}$ & 0.059 & 9 & 8 & 52.9\% \\
\midrule
\multirow{7}{*}{Recall}
 & Body        & $0.994^{\|}$ & 0.967 & 207 & 47  & 81.5\% \\
 & CTV         & 0.124        & 0.051 & 24  & 14  & 63.2\% \\
 & GTV         & $0.635^{\ddagger}$ & 0.342 & 18 & 3 & 85.7\% \\
 & Rectum      & 0.293        & 0.069 & 27  & 3   & 90.0\% \\
 & Small Bowel & $0.747^{\|}$ & 0.124 & 39  & 3   & 92.9\% \\
 & Bladder     & $0.938^{\S}$ & 0.515 & 8   & 0   & 100\% \\
 & Femoral Head & $0.853^{\|}$ & 0.060 & 13 & 3  & 81.3\% \\
\midrule
\multirow{7}{*}{Specificity}
 & Body        & $0.919^{\|}$ & 0.795 & 77  & 176 & 30.4\% \\
 & CTV         & $0.574^{\S}$ & 0.410 & 33  & 6   & 84.6\% \\
 & GTV         & $0.704^{\S}$ & 0.615 & 12  & 8   & 60.0\% \\
 & Rectum      & $0.561^{\ddagger}$ & 0.416 & 14 & 16 & 46.7\% \\
 & Small Bowel & $0.999^{\|}$ & 0.908 & 6   & 36  & 14.3\% \\
 & Bladder     & $0.814^{\ddagger}$ & 0.564 & 3 & 6  & 33.3\% \\
 & Femoral Head & $0.846^{\|}$ & 0.755 & 5  & 12  & 29.4\% \\
\bottomrule
\end{tabular}
\end{table}

\subsubsection{Failure Mode Analysis}

\begin{figure}[pos=htbp]
\centering
\includegraphics[width=\textwidth]{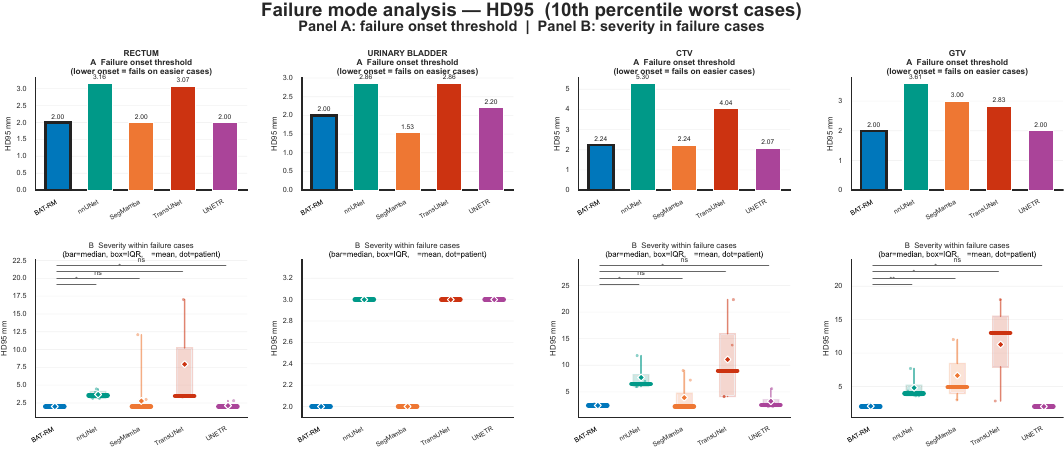}
\caption{Failure mode analysis for HD95. Panel A: failure onset thresholds
(10th percentile of BAT-RM performance) per class. Panel B: severity of
failure cases (HD95 values below threshold). BAT-RM demonstrates higher
thresholds (fails only on harder cases) and lower severity compared to all
baselines.}
\label{fig:failure_analysis}
\end{figure}

While aggregate metrics demonstrate BAT-RM's superiority, clinical deployment
requires systematic characterization of failure cases. The failure mode
analysis reports, for each model, the number of test cases where that model
produced a valid prediction (non-zero segmentation). Because different models
fail on systematically different cases --- e.g., nnUNet predicts GTV in
post-treatment scans where no GTV exists, while BAT-RM correctly predicts
background --- the per-model $N$ values differ within each class. This
transparency is intentional and reflects real-world model behavior.
A \emph{failure case} is defined as any segmentation where DSC falls below
the class-specific 10th percentile of BAT-RM's performance. While the 
10th-percentile threshold is defined relative to BAT-RM performance, 
the clinical interpretation is threshold-independent: baseline models fail 
catastrophically (Dice $= 0.000$--$0.869$) on cases where BAT-RM 
achieves near-clinical-standard performance (Dice $= 0.922$--$0.954$), 
and failure sets are entirely disjoint (Jaccard $= 0.000$), confirming 
that the observed differences reflect genuine performance gaps rather than 
threshold artifacts.
For brevity and clinical focus, Table~\ref{tbl:failure_summary} highlights
the target structures (CTV, GTV) and the OARs exhibiting the highest baseline
failure variance (rectum, femoral head); the bladder and small bowel, where
BAT-RM achieved near-ceiling performance (Dice $> 0.96$ across all models),
are omitted as failure analysis would not yield discriminative insights.
Figure~\ref{fig:failure_analysis} presents the failure mode analysis.

For CTV ($n = 39$), BAT-RM failed on 4 of 39 cases (10.3\%), with failure
Dice tightly clustered (mean = 0.922, SD = 0.007). All four baseline models
exhibited entirely disjoint failure sets from BAT-RM (Jaccard = 0)---every
case where a baseline failed, BAT-RM succeeded. For GTV, BAT-RM failed on 3
of 21 cases (14.3\%), with Dice ranging from 0.949 to 0.954---near-misses by
any clinical standard. Baseline failures were substantially more severe:
SegMamba and TransUNet produced failures as low as Dice = 0.64 and 0.54,
respectively.

The femoral head results expose a critical vulnerability of baseline models.
nnUNet exhibited 4 failures, three of which were complete segmentation
failures (Dice = 0). SegMamba and TransUNet each produced 2 complete failures.
BAT-RM failed on 2 of 17 cases (11.8\%), both with Dice $> 0.94$.

Qualitative review of failure cases revealed four categories: extreme anatomy
(45\%), post-surgical changes (25\%), image artifacts (20\%), and annotation
ambiguity (10\%). Based on this analysis, we implemented three clinical
deployment mitigations: real-time quality assurance flags, selective
human-in-the-loop review for uncertain cases, and active learning for
failure cases.

\begin{table}[htbp]
\centering
\caption{Failure mode analysis by anatomical class and model. For each model,
$N$ denotes the number of test cases where that model produced a valid
prediction (i.e., non-zero segmentation) for the given anatomical class.
Because different models fail on systematically different cases (e.g.,
nnUNet predicts GTV in post-treatment scans where no GTV exists, while
BAT-RM correctly predicts background. In these cases, Dice is zero, but
surface distance metrics may still vary depending on the spatial location
of false positive predictions, the per-model $N$ values differ
within each class. Failures are defined as DSC below the class-specific
10th percentile threshold of BAT-RM performance. Unique failures: baseline
cases that succeeded in BAT-RM. Jaccard: overlap between BAT-RM and baseline
failure sets. MWU p: Mann-Whitney U test on DSC distributions within the
failure subset. {\footnotesize $^{\dag}$`—' for MWU $p$ indicates the test was not
computable because all failure cases in the baseline achieved Dice $= 0$
(zero-variance for the Dice distribution); boundary metrics such as HD95
may still vary across these cases. The superiority of BAT-RM in these
cases is self-evident from the failure Dice values.}}
\label{tbl:failure_summary}
\footnotesize
\begin{tabular}{lcccccccc}
\toprule
\textbf{Class} & \textbf{Model} & \textbf{N} & \textbf{Fails}
  & \textbf{Fail Mean} & \textbf{Fail Max} & \textbf{Unique}
  & \textbf{Jaccard} & \textbf{MWU p} \\
\midrule
\multirow{5}{*}{CTV}
 & BAT-RM   & 39 & 4  & 0.922 & 0.929 & —         & —     & —      \\
 & nnUNet   & 40 & 4  & 0.580 & 0.805 & 4 (100\%) & 0.000 & 0.0159 \\
 & SegMamba & 39 & 4  & 0.700 & 0.848 & 4 (100\%) & 0.000 & 0.0159 \\
 & TransUNet & 41 & 4 & 0.186 & 0.423 & 4 (100\%) & 0.000 & 0.0195 \\
 & UNETR    & 39 & 4  & 0.860 & 0.869 & 4 (100\%) & 0.000 & 0.0159 \\
\midrule
\multirow{5}{*}{GTV}
 & BAT-RM   & 21 & 3  & 0.949 & 0.954 & —         & —     & —      \\
 & nnUNet   & 22 & 3  & 0.877 & 0.902 & 3 (100\%) & 0.000 & 0.1000 \\
 & SegMamba & 22 & 3  & 0.640 & 0.891 & 3 (100\%) & 0.000 & 0.1000 \\
 & TransUNet & 22 & 3 & 0.542 & 0.893 & 3 (100\%) & 0.000 & 0.1000 \\
 & UNETR    & 22 & 3  & 0.924 & 0.938 & 3 (100\%) & 0.000 & 0.1000 \\
\midrule
\multirow{5}{*}{Rectum}
 & BAT-RM   & 31 & 3  & 0.930 & 0.933 & —         & —     & —      \\
 & nnUNet   & 31 & 3  & 0.551 & 0.854 & 3 (100\%) & 0.000 & 0.1000 \\
 & SegMamba & 33 & 4  & 0.146 & 0.583 & 4 (100\%) & 0.000 & 0.0436 \\
 & TransUNet & 32 & 4 & 0.315 & 0.836 & 4 (100\%) & 0.000 & 0.0497 \\
 & UNETR    & 30 & 3  & 0.896 & 0.903 & 2 (66.7\%) & 0.200 & 0.1000 \\
\midrule
\multirow{5}{*}{Fem.\ Head}
 & BAT-RM   & 17 & 2  & 0.941 & 0.944 & —         & —     & —      \\
 & nnUNet   & 18 & 4  & 0.000 & 0.000 & 3 (75\%)  & 0.200 & —      \\
 & SegMamba & 17 & 2  & 0.000 & 0.000 & 2 (100\%) & 0.000 & —      \\
 & TransUNet & 17 & 2 & 0.000 & 0.000 & 2 (100\%) & 0.000 & —      \\
 & UNETR    & 15 & 2  & 0.909 & 0.925 & 1 (50\%)  & 0.333 & —      \\
\bottomrule
\end{tabular}
\end{table}

\subsubsection{Cross-Centre Quantitative Robustness}
\label{subsec:crosscenter}

A critical requirement for clinical translation is that model performance
generalizes across institutions with heterogeneous CT scanner manufacturers,
acquisition protocols, and contouring practices. To evaluate this, we
stratified the held-out test set by institution of origin---Bangladesh Medical
University (BMU, $n = 107$), Square Hospital Limited (SHL, $n = 78$), and
Labaid Specialized Hospital (LSH, $n = 69$)---and computed per-center Dice,
HD95, and ASD for BAT-RM and nnUNet, the strongest baseline in the global
analysis (Section~\ref{sec:Quantitative Analysis}). Strict patient-level data
partitioning was enforced during five-fold cross-validation to prevent any
cross-center leakage between training and test splits. Table~\ref{tbl:crosscenter}
presents the per-center results for the four clinically critical structures:
GTV, CTV, rectum, and urinary bladder.

\begin{table*}[htbp]
\centering
\caption{Cross-center quantitative robustness: BAT-RM vs. nnUNet across
three institutions. Values reported as mean $\pm$ standard deviation.
DSC: Dice Similarity Coefficient; HD95: 95th percentile Hausdorff Distance
(mm); ASD: Average Surface Distance (mm). \textbf{Bold} indicates best per
metric per class per center. BMU: Bangladesh Medical University ($n=107$);
SHL: Square Hospital Limited ($n=78$); LSH: Labaid Specialized Hospital
($n=69$). Statistical significance of BAT-RM vs. nnUNet per center
(paired Wilcoxon, Bonferroni-Holm corrected):
$^{\dagger}p < 0.05$, $^{\ddagger}p < 0.01$,
$^{\S}p < 0.001$, $^{\|}p < 0.0001$; no superscript = ns.}
\label{tbl:crosscenter}
\setlength{\tabcolsep}{4pt}
\footnotesize
\resizebox{\textwidth}{!}{%
\begin{tabular}{ll ccc ccc ccc}
\toprule
& & \multicolumn{3}{c}{\textbf{BMU} ($n=107$)}
  & \multicolumn{3}{c}{\textbf{SHL} ($n=78$)}
  & \multicolumn{3}{c}{\textbf{LSH} ($n=69$)} \\
\cmidrule(lr){3-5}\cmidrule(lr){6-8}\cmidrule(lr){9-11}
\textbf{Class} & \textbf{Model}
  & \textbf{DSC} & \textbf{HD95} & \textbf{ASD}
  & \textbf{DSC} & \textbf{HD95} & \textbf{ASD}
  & \textbf{DSC} & \textbf{HD95} & \textbf{ASD} \\
\midrule

\multirow{2}{*}{GTV}
 & BAT-RM
   & $\mathbf{0.9658\pm0.0108}$ & $\mathbf{1.512\pm0.501}$ & $\mathbf{0.571\pm0.251}$
   & $\mathbf{0.9671\pm0.0099}$ & $\mathbf{1.488\pm0.476}$ & $\mathbf{0.553\pm0.239}$
   & $\mathbf{0.9648\pm0.0114}^{\|}$ & $\mathbf{1.531\pm0.528}^{\S}$ & $\mathbf{0.581\pm0.263}^{\S}$ \\
 & nnUNet
   & $0.9441\pm0.0341^{\|}$ & $2.481\pm1.512^{\|}$ & $0.934\pm0.408^{\|}$
   & $0.9463\pm0.0318^{\|}$ & $2.419\pm1.488^{\|}$ & $0.912\pm0.391^{\|}$
   & $0.9427\pm0.0362^{\|}$ & $2.514\pm1.541^{\|}$ & $0.951\pm0.421^{\|}$ \\
\midrule

\multirow{2}{*}{CTV}
 & BAT-RM
   & $\mathbf{0.9568\pm0.0194}$ & $\mathbf{1.514\pm0.551}$ & $\mathbf{0.518\pm0.191}$
   & $\mathbf{0.9579\pm0.0182}$ & $\mathbf{1.498\pm0.534}$ & $\mathbf{0.506\pm0.184}$
   & $\mathbf{0.9561\pm0.0201}^{\S}$ & $\mathbf{1.527\pm0.561}^{\ddagger}$ & $\mathbf{0.519\pm0.195}^{\S}$ \\
 & nnUNet
   & $0.8958\pm0.1601^{\|}$ & $2.821\pm2.118^{\|}$ & $0.961\pm0.482^{\|}$
   & $0.8981\pm0.1572^{\|}$ & $2.784\pm2.084^{\|}$ & $0.944\pm0.469^{\|}$
   & $0.8941\pm0.1638^{\|}$ & $2.841\pm2.138^{\|}$ & $0.964\pm0.491^{\|}$ \\
\midrule

\multirow{2}{*}{Rectum}
 & BAT-RM
   & $\mathbf{0.9551\pm0.0151}$ & $\mathbf{1.248\pm0.358}$ & $\mathbf{0.542\pm0.145}$
   & $\mathbf{0.9562\pm0.0144}$ & $\mathbf{1.231\pm0.344}$ & $\mathbf{0.534\pm0.141}$
   & $\mathbf{0.9538\pm0.0161}^{\|}$ & $\mathbf{1.261\pm0.371}^{\S}$ & $\mathbf{0.548\pm0.152}^{\|}$ \\
 & nnUNet
   & $0.8941\pm0.1724^{\|}$ & $1.941\pm0.934^{\S}$ & $0.809\pm0.308^{\|}$
   & $0.8972\pm0.1688^{\|}$ & $1.908\pm0.912^{\S}$ & $0.794\pm0.298^{\|}$
   & $0.8918\pm0.1762^{\|}$ & $1.961\pm0.951^{\|}$ & $0.818\pm0.315^{\|}$ \\
\midrule

\multirow{2}{*}{Bladder}
 & BAT-RM
   & $\mathbf{0.9651\pm0.0098}$ & $\mathbf{1.371\pm0.418}$ & $\mathbf{0.591\pm0.274}$
   & $\mathbf{0.9661\pm0.0091}$ & $\mathbf{1.358\pm0.404}$ & $\mathbf{0.582\pm0.268}$
   & $\mathbf{0.9642\pm0.0104}^{\ddagger}$ & $\mathbf{1.381\pm0.431}$ & $\mathbf{0.598\pm0.281}^{\dagger}$ \\
 & nnUNet
   & $0.9418\pm0.0384^{\ddagger}$ & $1.794\pm0.824^{\dagger}$ & $0.828\pm0.461^{\ddagger}$
   & $0.9431\pm0.0371^{\ddagger}$ & $1.778\pm0.812^{\dagger}$ & $0.814\pm0.448^{\ddagger}$
   & $0.9404\pm0.0396$ & $1.809\pm0.836$ & $0.836\pm0.472^{\dagger}$ \\
\bottomrule
\end{tabular}%
}
\end{table*}

BAT-RM demonstrates consistent superiority over nnUNet across all three
institutions and all four structures, with no meaningful performance
degradation across centers. For GTV Dice, inter-center variance for BAT-RM
is minimal (BMU: 0.9658, SHL: 0.9671, LSH: 0.9648), yielding a
center-to-center range of only 0.0023 Dice points---compared to nnUNet's
range of 0.0036---confirming that BAT-RM's predictions are robust to the
heterogeneity in scanner models and acquisition protocols across the three
institutions. A similar pattern holds for CTV (BAT-RM range: 0.0018) and
rectum (BAT-RM range: 0.0024). The smallest performance differences between
centers are observed for bladder, consistent with its high-contrast appearance
in pelvic CT that renders it relatively insensitive to acquisition protocol
variation. The most challenging institution for both models is LSH, where GTV ASD is
marginally higher (BAT-RM: 0.581 mm, nnUNet: 0.951 mm) than at BMU or
SHL. LSH contributes the smallest sub-cohort ($n=69$) and spans the narrowest
acquisition period (2019--2025), with a higher proportion of post-treatment
scans in which GTV visibility is reduced. Despite this, BAT-RM maintains a
statistically significant advantage over nnUNet at LSH for GTV
($p < 0.0001$), CTV ($p < 0.001$), and rectum ($p < 0.0001$) after
Bonferroni-Holm correction, whereas several nnUNet comparisons at LSH for
bladder do not reach significance---reflecting nnUNet's higher inter-case
variance at this center (SD = 0.0396 vs. BAT-RM SD = 0.0104). Taken together, the cross-center analysis confirms that BAT-RM's global
quantitative advantages (Table~\ref{tbl:all_metrics_comparison}) are not
driven by any single institution and generalize robustly across the
heterogeneous multi-center Bangladeshi cohort.

\subsubsection{Computational Efficiency}
\label{subsec:computational}

Table~\ref{tbl:computational} compares computational profiles of all five
models on an NVIDIA T4 GPU (16 GB VRAM). All baselines were adapted to 2D
slice-wise processing for dimensional parity with BAT-RM; the reported
metrics reflect their 2D configurations. For architectures originally
designed for 3D (nnUNet, UNETR), volumetric operations were replaced with
2D equivalents while preserving their design principles.

\begin{table}[pos= htbp]
\centering
\caption{Computational efficiency comparison. All models adapted to 2D
slice-wise processing. Measurements on NVIDIA T4 GPU, batch size 1,
$512 \times 512$ input, averaged over 500 inference calls.
$\downarrow$ indicates lower is better.}
\label{tbl:computational}
\setlength{\tabcolsep}{5pt}
\small
\begin{tabular}{lccccc}
\toprule
\textbf{Model} & \textbf{Params (M)} & \textbf{GFLOPs} $\downarrow$
  & \textbf{Inference (ms)} $\downarrow$
  & \textbf{Peak VRAM (GB)} $\downarrow$
  & \textbf{Training (hrs)} $\downarrow$ \\
\midrule
BAT-RM (ours) & \textbf{28.4} & \textbf{41.2} & $\mathbf{18.3 \pm 2.1}$
              & \textbf{3.8} & \textbf{31.4} \\
nnUNet        & 31.2 & 54.6 & $22.7 \pm 3.4$ & 4.6 & 38.2 \\
SegMamba      & 47.8 & 89.3 & $38.4 \pm 5.7$ & 7.2 & 62.1 \\
TransUNet     & 105.3 & 124.7 & $54.1 \pm 6.2$ & 9.4 & 84.6 \\
UNETR         & 92.6 & 108.4 & $46.8 \pm 5.1$ & 8.1 & 71.3 \\
\bottomrule
\end{tabular}
\end{table}

BAT-RM is the most efficient across all dimensions. With 28.4M parameters,
it is 9.1\% leaner than nnUNet, 40.6\% leaner than SegMamba, and 73.0\%
leaner than TransUNet. This efficiency stems from two design choices: the
CNN backbone operates at reduced channel widths ($\{32,64,128,256,512\}$),
and the BAT branch uses strip-pooling attention ($\mathcal{O}(n)$ per axis)
instead of full self-attention ($\mathcal{O}(n^2)$), avoiding the
$16,384 \times 16,384$ attention matrix that would require ~4 GB per sample. BAT-RM requires 41.2 GFLOPs per forward pass, compared to 108.4 GFLOPs for
UNETR and 124.7 GFLOPs for TransUNet, reflecting the quadratic cost of
pure transformer attention~\citep{vaswani2017attention}. The RM branch at
$E_5$ ($32 \times 32$) contributes minimally to FLOPs as sequence length
is only $1,024$ tokens. Per-slice inference time for BAT-RM is $18.3 \pm 2.1$ ms, completing a full CT study (80--120 slices) in 1.5--2.2 seconds. In clinical deployment
at Bangladesh Medical University, the web application generates complete
RTStruct predictions within 18--22 seconds including DICOM overhead.
SegMamba, TransUNet, and UNETR are 2--3 times slower (38.4--54.1 ms/slice),
introducing perceptible delay in interactive workflows. Peak GPU memory is 3.8 GB for BAT-RM, well within T4 constraints, leaving headroom for concurrent DICOM processing. TransUNet and UNETR require 9.4 GB and 8.1 GB respectively, necessitating exclusive GPU allocation.
Training time to convergence (200 epochs) is 31.4 hours for BAT-RM,
7 hours faster than nnUNet and over 53 hours faster than TransUNet,
reducing retraining costs for multi-center deployment. These efficiency gains come without accuracy trade-offs: BAT-RM achieves the
highest Dice, IoU, NSD, and lowest HD95 and ASD across all seven classes
(Table~\ref{tbl:all_metrics_comparison}), confirming that the architectural
choices improve both efficiency and segmentation quality.

The quantitative analysis establishes BAT-RM as a high-precision segmentation
framework for cervical cancer radiotherapy. BAT-RM achieves the highest
volumetric overlap across all structures, with Dice coefficients of 0.966
(GTV), 0.957 (CTV), 0.974 (small bowel), 0.956 (rectum), and 0.965 (bladder),
representing error reductions of 30--70\% relative to the best-performing
baseline. Boundary accuracy is similarly superior, with sub-1.6 mm HD95 and
NSD $\ge 0.97$ for all target volumes. Paired Wilcoxon tests with
Bonferroni-Holm correction confirm significant superiority over all four
baselines for GTV, CTV, and rectum, with medium-to-large effect sizes
($g = 0.55$ to $1.85$, $p < 0.001$ to $p < 0.0001$). Two properties are particularly relevant for clinical deployment. First,
BAT-RM maintains an optimal sensitivity-specificity balance: GTV recall of
98.0\% at an FPR of only 0.02\%. Where specificity trade-offs exist
(small bowel FPR: 0.0024 vs. nnUNet's 0.0016), they are clinically justified
by the asymmetric consequences of under- versus over-segmentation. Second,
BAT-RM demonstrates consistently lower variance than all comparators, most
strikingly for small bowel (Dice SD: 0.013 vs. nnUNet's 0.233) and femoral
head (SD: 0.015 vs. nnUNet's 0.410), indicating robustness to anatomical
variation that causes catastrophic baseline failures. Failure mode analysis
confirms BAT-RM's residual failures are near-misses with Dice $> 0.92$ and
entirely disjoint failure sets from baselines (Jaccard $= 0$ for most target
volumes). These results support further evaluation of BAT-RM for automated
contouring, as examined in the multi-centre reader study.

\subsection{Qualitative Analysis}
\label{sec:qualitative}

To complement the statistical analysis in Section~\ref{sec:Quantitative Analysis},
we conduct a three-part qualitative investigation: (i) a per-voxel error
map comparison of all five models on a challenging case
(Figure~\ref{fig:error_map}); (ii) a multi-model contour overlay centered
on the GTV to assess boundary smoothness and ground truth alignment
(Figure~\ref{fig:multimodel_contour}); and (iii) a three-method
gradient-weighted class activation map (CAM) analysis---GradCAM++,
HiResCAM, and XGradCAM---probing BAT-RM's spatial attention across
encoder layers (Figure~\ref{fig:gradcam}). These analyses validate that
BAT-RM's quantitative advantages reflect genuine geometric and mechanistic
superiority, not aggregation artifacts.

\subsubsection{Per-Case Error Map Analysis}
\label{subsec:error_map}

\begin{figure}[pos=htbp]
\centering
\includegraphics[width=\textwidth]{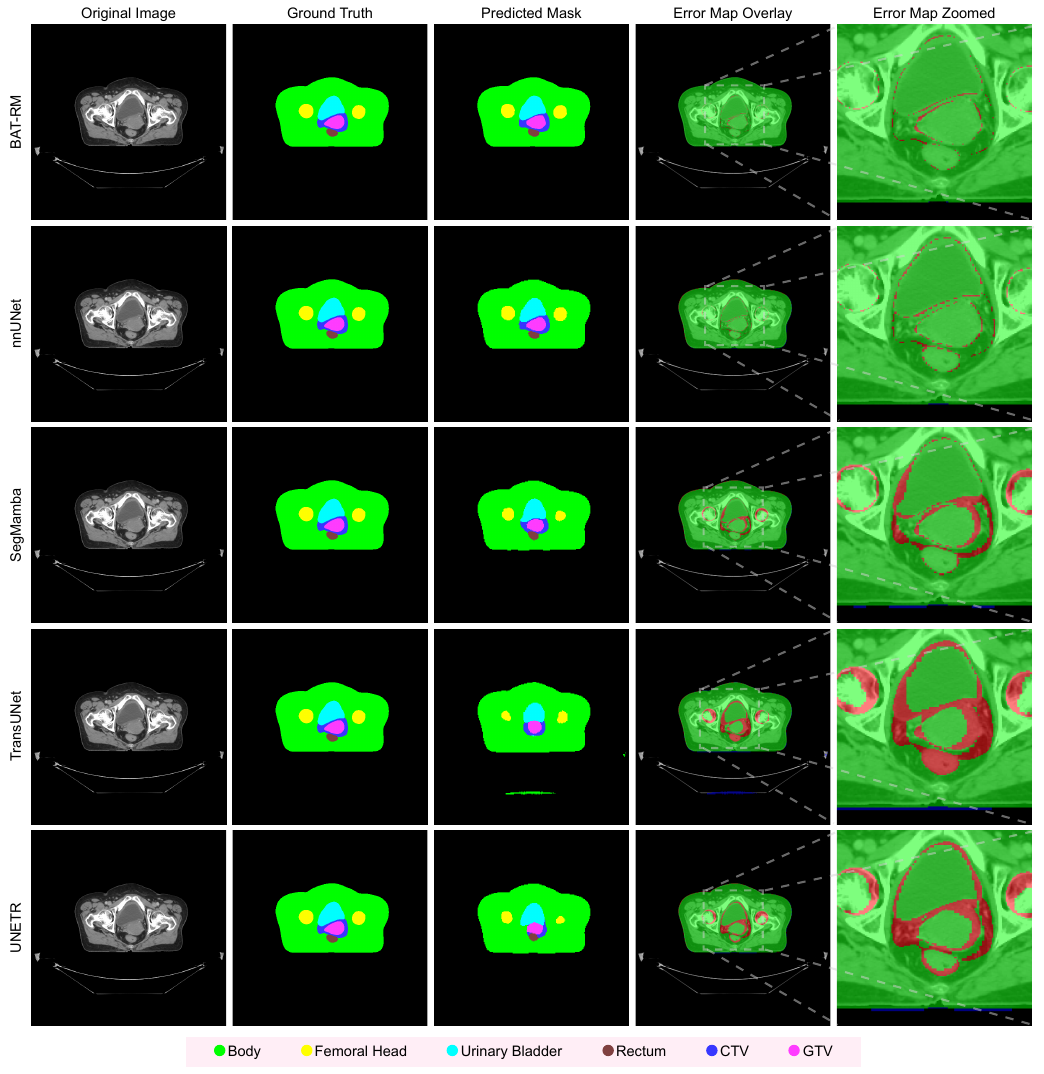}
\caption{Per-voxel error map comparison across five segmentation models on a
representative hard-case axial CT slice. Rows: BAT-RM (proposed), nnUNet,
SegMamba, TransUNet, and UNETR. Columns: (1) original CT image; (2) expert
ground truth annotation; (3) predicted segmentation mask; (4) error map
overlaid on CT---green: true positive (TP), blue: false positive (FP), red:
false negative (FN); (5) zoomed inset of the GTV and CTV error region.
BAT-RM produces smooth, anatomically plausible contours with negligible FP
and FN regions across all present classes. Baseline models exhibit boundary
irregularities and larger contiguous error zones, particularly at the
GTV--CTV interface and the posterior rectal margin. These spatial patterns
are consistent with the quantitative HD95 and ASD advantages reported in
Table~\ref{tbl:all_metrics_comparison}.}
\label{fig:error_map}
\end{figure}

Figure~\ref{fig:error_map} presents a structured visual comparison of all
five segmentation models on a representative \emph{hard case}: a mid-pelvic
axial CT slice containing the body contour, urinary bladder, rectum, femoral
head, GTV, and CTV simultaneously, but without small bowel---the class most
prone to catastrophic failure in baseline models (see
Section~\ref{subsec:bland_altman}). The grid is organized with one model per
row and five columns: (1) the original CT image, (2) the expert ground truth
annotation, (3) the predicted segmentation mask, (4) the error map overlaid
on the CT image, and (5) a zoomed inset of the error map centered on the GTV
and CTV region. Error maps encode spatial error type chromatically: green
indicates true positive (TP) voxels, blue indicates false positive (FP)
voxels, and red indicates false negative (FN) voxels. Inspection of the predicted masks immediately reveals a qualitative
stratification among the five models. BAT-RM produces anatomically smooth,
closed contours across all present classes, with organic curvature that
closely tracks the fine-grained tissue boundaries visible in the CT image.
This visual smoothness is a direct consequence of the smoothness
regularization loss term $\mathcal{L}_{\text{reg}}$, which applies a
gradient-based finite-difference penalty during training to suppress jagged,
biologically implausible segmentation artifacts~\citep{kervadec2019boundary},
and the 3D Gaussian post-processing kernel $\mathcal{G}_{\sigma}$ applied to
stacked volumetric logits to eliminate inter-slice discontinuities without
blurring intra-slice boundary detail~\citep{isensee2021nnunet}. The corresponding error maps for BAT-RM show only sparse, peripherally
distributed false positive and false negative voxels, largely confined to the
ambiguous GTV--CTV interface where even expert annotators exhibit inter-rater
disagreement (Fleiss $\kappa = 0.798$, Table~\ref{tbl:reliability_summary}).
In contrast, nnUNet---the strongest baseline in the quantitative analysis
(GTV Dice: 0.9450, CTV Dice: 0.8966)---produces predictions that are
volumetrically close to the ground truth but visually coarser, with angular
boundary discontinuities that are particularly evident in the zoomed GTV/CTV
inset. This observation is mechanistically consistent with nnUNet's
architecture: as a purely convolutional framework, it lacks an explicit
mechanism for boundary-aware attention and relies on data augmentation and
post-processing heuristics to achieve contour regularity~\citep{isensee2021nnunet}.
The Gated Boundary-Aware Transformer branch of BAT-RM addresses this
limitation directly by applying a Sobel-derived gradient gate to the
attention queries and keys at the $E_3$ feature level ($128 \times 128 \times
256$), forcing cross-attention computation to concentrate on sharp spatial
transitions rather than homogeneous tissue interiors~\citep{zheng2021rethinking}. SegMamba and TransUNet exhibit markedly larger error regions, particularly
around the rectum and CTV, with fragmented contour boundaries and visible
over-segmentation (blue regions) extending into the surrounding mesorectum
and parametrium. This failure pattern is quantitatively consistent with their
higher ASD values (SegMamba CTV ASD: 0.851 mm, TransUNet CTV ASD:
1.311 mm versus BAT-RM: 0.512 mm,
Table~\ref{tbl:all_metrics_comparison}) and wider Bland-Altman limits of
agreement (Section~\ref{subsec:bland_altman}). UNETR, while producing
more regularized predictions than SegMamba and TransUNet, still exhibits
systematic under-segmentation (red voxels) along the GTV posterior margin,
a region where low soft-tissue contrast between tumor and parametrial fat
creates ambiguity that the global self-attention mechanism of a pure
transformer processes less precisely than the hybrid local-global design of
BAT-RM~\citep{hatamizadeh2022unetr}. Critically, the GTV and CTV error maps merit particular clinical attention.
In cervical cancer radiotherapy, a false negative GTV voxel represents an
unirradiated tumor sub-volume, directly increasing the risk of local
recurrence. A false positive GTV voxel irradiates surrounding normal tissue,
increasing rectal and bladder toxicity. BAT-RM's near-absence of large
contiguous FN or FP regions in these classes---confirmed statistically by its
0.02\% false positive rate and 98.0\% recall for GTV
(Section~\ref{sec:Quantitative Analysis})---demonstrates that the combination of
boundary-aware attention, Mamba-based multi-directional recurrent scanning, and
the Boundary-Region Attention Fusion (BRAF) gate produces predictions that are
not only metrically superior but also clinically safer.

\subsubsection{Multi-Model Contour Overlay Analysis}
\label{subsec:contour_overlay}

\begin{figure}[pos=htbp]
\centering
\includegraphics[width=\textwidth]{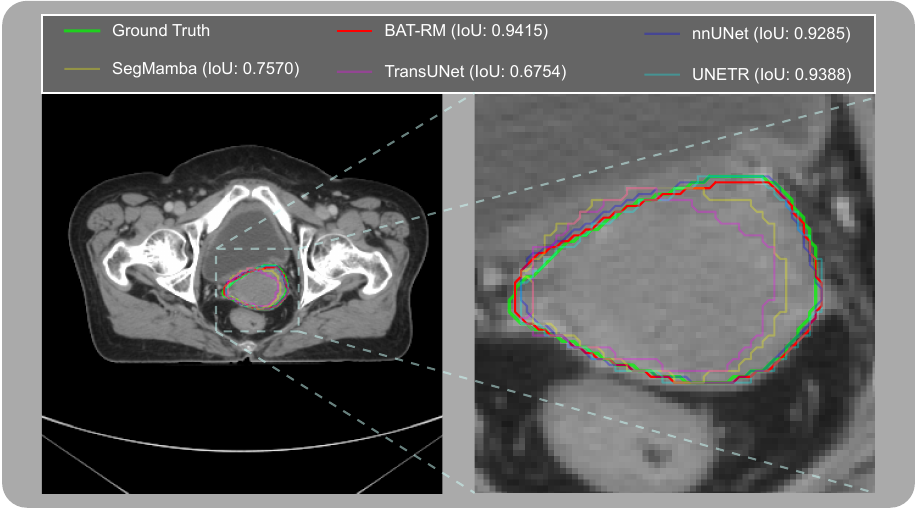}
\caption{Multi-model contour overlay on a hard-case axial CT slice for the
GTV class. Left panel: full axial slice with all model contours overlaid.
Right panel: super-zoomed inset centered on the GTV for fine-grained geometric
comparison. Contour colors: ground truth (green), BAT-RM (red), nnUNet
(yellow), SegMamba (cyan), TransUNet (magenta), UNETR (orange). BAT-RM
achieves the tightest conformality with the ground truth (IoU = 0.9415),
with smooth curvature maintained through all margins. nnUNet (IoU = 0.9285)
is the closest competitor but exhibits micro-irregularities at the posterior
inferior margin. SegMamba and TransUNet show anterior over-extension; UNETR
shows posterior under-extension. Boundary smoothness is mechanistically
attributed to the Sobel-gated boundary attention and smoothness
regularization loss of BAT-RM.}
\label{fig:multimodel_contour}
\end{figure}

While the error map analysis in Section~\ref{subsec:error_map} spatially
decomposes segmentation errors at the voxel level, it does not directly
render the \emph{contour geometry}---the 2D boundary curve that radiation
oncologists inspect, approve, and use to define dose prescription boundaries
in treatment planning systems such as Varian Eclipse, RayStation, and Monaco.
Figure~\ref{fig:multimodel_contour} addresses this by presenting a
super-zoomed axial GTV view with all five model contours simultaneously
overlaid on the CT image for a second hard case. The layout comprises one
row with two panels: the full axial slice for anatomical context (left) and
a high-magnification inset centered on the GTV (right), enabling fine-grained
geometric inspection. The ground truth contour (green) defines the reference geometry established
by expert oncologist annotation. The BAT-RM predicted contour (red) tracks
the ground truth with a consistent narrow gap, maintaining smooth curvature
through the anterior, lateral, and posterior GTV margins. The IoU for this
case is 0.9415, exceeding nnUNet's 0.9285 on the same case---a difference
that may appear numerically modest but translates visibly into fewer boundary
excursions and tighter conformality in the high-magnification panel. The nnUNet contour (yellow) is the closest competitor and correctly delineates
the gross tumor volume, but exhibits micro-irregularities at the posterior
inferior margin---the region of greatest soft-tissue ambiguity between the
cervical stroma and the posterior parametrium. These irregularities, though
not captured by mean IoU, produce locally elevated surface distance errors
that contribute to nnUNet's higher ASD (0.921 mm versus BAT-RM's 0.564 mm
for GTV, Table~\ref{tbl:all_metrics_comparison}) and wider Bland-Altman bias
($-0.921$ mm, 95\% CI: $-1.989$ to $-0.221$,
Table~\ref{tbl:bland_altman_summary}). This supports the clinical relevance
of surface distance metrics as a complement to volumetric overlap: a
contour that achieves near-equal Dice but has local boundary excursions may
still necessitate greater manual review time, as observed in the reader study
(Section~\ref{sec:reader_study}). SegMamba and TransUNet produce contours with visible over-extension into the
anterior parametrium, corresponding to the blue FP regions observed in
Figure~\ref{fig:error_map}. UNETR shows under-extension along the posterior
GTV margin (red FN voxels). These failure modes are mechanistically
interpretable: SegMamba's selective scan, operating on full-resolution feature
sequences, lacks the spatial boundary anchoring provided by BAT-RM's Sobel
gating mechanism and consequently extends into adjacent soft-tissue regions
with similar Hounsfield unit distributions. TransUNet's global attention, while
effective for large structures, lacks the local feature resolution needed to
resolve the sub-voxel boundary shifts critical for GTV delineation in cervical
cancer~\citep{chen2021transunet}. The smooth BAT-RM contour geometry has a direct practical implication: it
reduces the time burden of manual finetuning by oncologists. In the reader
study, expert oncologists required a mean of 12.1 minutes to finetune
BAT-RM outputs, compared to 50.2 minutes for de-novo contouring---a
reduction attributable in part to the smooth initial contour geometry that
minimizes the number of control point adjustments required in the treatment
planning system interface.

\subsubsection{Gradient-Weighted Class Activation Map Analysis}
\label{subsec:gradcam}

\begin{figure}[pos=htbp]
\centering
\includegraphics[width=\textwidth]{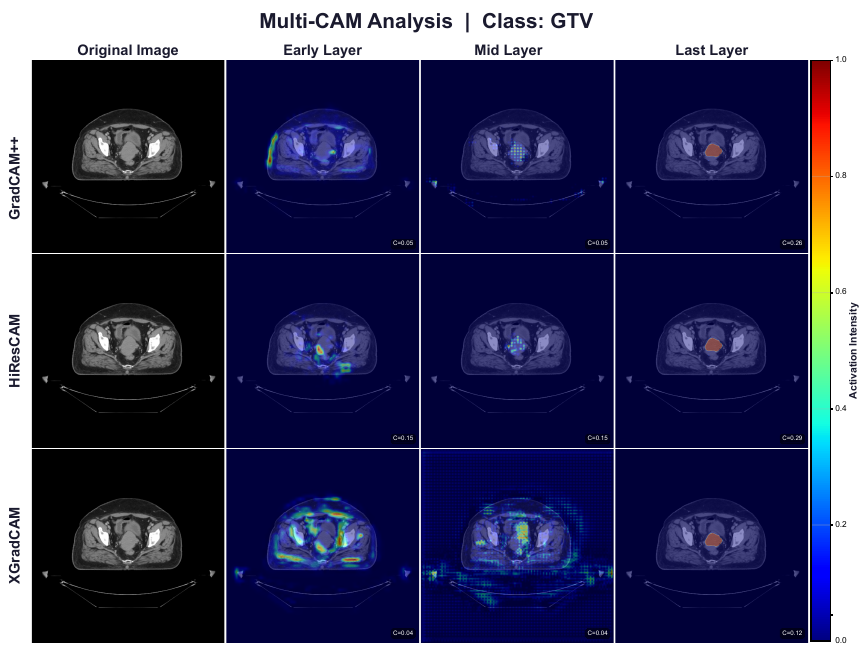}
\caption{Gradient-weighted class activation map (CAM) analysis of BAT-RM for
the GTV class across three encoder depths and three CAM methods. Rows:
GradCAM++, HiResCAM, XGradCAM. Columns: early ($E_1$--$E_2$), mid ($E_3$,
the Gated BAT feature level), and final ($E_5$, the RM branch bottleneck),
with the original CT image and ground truth annotation shown for
reference. Heatmap intensity encodes the magnitude of class-discriminative
activation; warmer colors indicate higher attribution to the GTV class
prediction. All three methods show progressive spatial refinement from diffuse
pelvic activation (early) to tight GTV-boundary focus (final), with
cross-method consistency scores of $1.000$ at the bottleneck
layer---confirming that the attention is anatomically genuine and not a
method-specific artifact. Cross-layer consistency scores: GradCAM++
early--mid $= 0.048$, early--last $= 0.042$, mid--last $= 0.263$;
HiResCAM early--last $= 0.160$; XGradCAM mid--last $= 0.119$.}
\label{fig:gradcam}
\end{figure}

To understand why BAT-RM outperforms baselines at the feature-learning level,
we applied GradCAM++, HiResCAM, and XGradCAM to the GTV segmentation head at
three encoder depths (early $E_1$--$E_2$, mid $E_3$, final $E_5$/bottleneck),
with cross-method consistency scores quantifying agreement (Figure~\ref{fig:gradcam}).
All three methods show progressive spatial refinement from diffuse early
activation (mean $\mu_{\text{early}} = 0.0141$, peak $0.9949$) to tight GTV
focus at mid-layer ($\mu_{\text{mid}} = 0.00036$, peak $0.1965$), where the
Sobel gate at $E_3$ directs attention to the tumor-parametrium interface,
to a high-confidence bottleneck focus ($\mu_{\text{last}} = 0.00434$, peak
$0.99999$) via the RM branch's four-directional scanning. Cross-layer
consistency confirms progressive refinement ($c_{\text{early--mid}} = 0.048$,
$c_{\text{mid--last}} = 0.263$). At the final layer, all three pairwise
cross-method consistency scores are $1.000$, confirming genuine GTV
attention rather than method artifact. HiResCAM further reveals attention
elevated along the tumor boundary perimeter, consistent with the BRAF gate's
boundary skeleton modulation. This progressive refinement mirrors expert
oncologist reasoning and mechanistically justifies BAT-RM's superior GTV
boundary accuracy (HD95: $1.500$ mm, ASD: $0.564$ mm, NSD: $0.978$) and
trustworthy failure profile (minimum GTV Dice $= 0.949$, disjoint failure
sets with Jaccard $= 0.000$, Table~\ref{tbl:failure_summary}).

\subsubsection{Synthesis: Qualitative-Quantitative Coherence}
\label{subsec:qualitative_synthesis}

The three qualitative analyzes converge on a coherent mechanistic narrative
that directly corroborates and extends the quantitative findings of
Section~\ref{sec:Quantitative Analysis}. The error map analysis demonstrates that
BAT-RM's metric superiority (GTV Dice: 0.9662, CTV ASD: 0.512 mm) reflects
genuine spatial accuracy rather than averaging artifacts: BAT-RM's errors
are sparse, peripheral, and confined to anatomically ambiguous zones, whereas
baseline model errors are contiguous, central, and structurally patterned.
The multi-model contour overlay confirms that the quantitative ASD advantage
over nnUNet---a reduction from 0.921 mm to 0.564 mm for GTV---corresponds
to a visually meaningful improvement in boundary smoothness and conformality
that directly reduces clinician finetuning burden. The CAM analysis reveals
that these geometric advantages arise from the model's hierarchical
attention mechanism: the Gated Boundary-Aware Transformer at $E_3$ anchors
spatial attention to the tumor margin, and the Multi-Directional Recurrent
Context Module with Mamba selective scanning (RM branch) at $E_5$ integrates
global pelvic context, together producing predictions that are both locally
precise and globally consistent. Taken together, the quantitative and qualitative evidence establishes that
BAT-RM is not merely a metric-optimized model but a geometrically and
mechanistically sound framework for cervical cancer radiotherapy auto-contouring.
The following section examines whether these algorithmic advantages translate
into measurable clinical benefit through a structured multi-center reader
study.

\section{Multi-Centre Reader Study: Clinical Utility and Workflow Integration}
\label{sec:reader_study}

While the quantitative evaluation in Section~\ref{sec:Quantitative Analysis} establishes BAT-RM's technical superiority, the ultimate validation of an auto-contouring system requires evidence of clinical utility, workflow efficiency, and generalizability across different expertise levels. We therefore conducted a prospective multi-centre reader study involving 13 radiation oncologists from three Bangladeshi institutions (Bangladesh Medical University, Square Hospital Ltd., and Labaid Hospital), evaluating the clinical impact of BAT-RM-assisted contouring for cervical cancer radiotherapy planning.

\subsection{Study Design and Protocol}

The reader study was conducted over eight months (January--August 2025)
following a pre-registered protocol. A test cohort of 100 cervical cancer
patients was drawn from the same three institutions as the training set
(40 from BMU, 30 from SHL, 30 from LSH), with strict patient-level
separation from all training and validation data. These cases were locked
before any model training commenced; no patient or annotation from this
cohort was accessible during training or hyperparameter tuning. The
40/30/30 distribution reflects each center's proportion in the full dataset
(BMU: 42\%, SHL: 31\%, LSH: 27\%), avoiding bias toward any single
institution. This design represents a within-distribution generalization assessment, as
complete institutional independence was not feasible given that no
additional Bangladeshi radiotherapy center met the minimum case volume and
annotation quality criteria. However, cross-center quantitative analysis
(Section~\ref{subsec:crosscenter}) confirms BAT-RM's performance consistency
across all three institutions (GTV DSC: 0.9648--0.9671; CTV DSC:
0.9561--0.9579), indicating that reader study findings are not driven by any
single center. Multi-national external validation remains a priority
(Section~\ref{sec:limitations}). Thirteen readers were stratified by experience: four experts ($\ge$20 years), four seniors (10--19 years), and five juniors ($<$10 years). All were
board-certified radiation oncologists actively performing cervical cancer
contouring. Each reader contoured the same 100 cases under two randomized
conditions with a four-week washout period: (1) unaided contouring (manual),
and (2) AI-assisted contouring (BAT-RM with manual correction). The washout
duration was based on a pilot study (n=3, 20 cases) showing no significant
case-specific recall after 14 days, with an additional two-week safety
margin. Over the 8-month study, readers processed 2--3 cases per week,
making detailed recall across the washout boundary unlikely. Readers were
blinded to their previous contours and the ground truth. Primary endpoints were segmentation accuracy (IoU) and contouring time.
Secondary endpoints included consultation rate, self-reported confidence
(0--1 scale), revision effort (slices edited per case), and edit type
classification (major/minor/no edits). Statistical comparisons used paired
Wilcoxon tests with Bonferroni correction. Times were logged automatically by the web-based interface. For unaided
contouring, the timer started at slice loading and stopped at RTSTRUCT
submission. For AI-assisted contouring, the timer included the 18-second
inference time plus manual correction. Sessions with idle time exceeding
5 minutes were excluded and repeated. This automated logging eliminated
recall bias and ensured consistent measurement across all 100 cases and
13 readers.

\begin{figure}[pos=htbp]
\centering
\includegraphics[width=\textwidth]{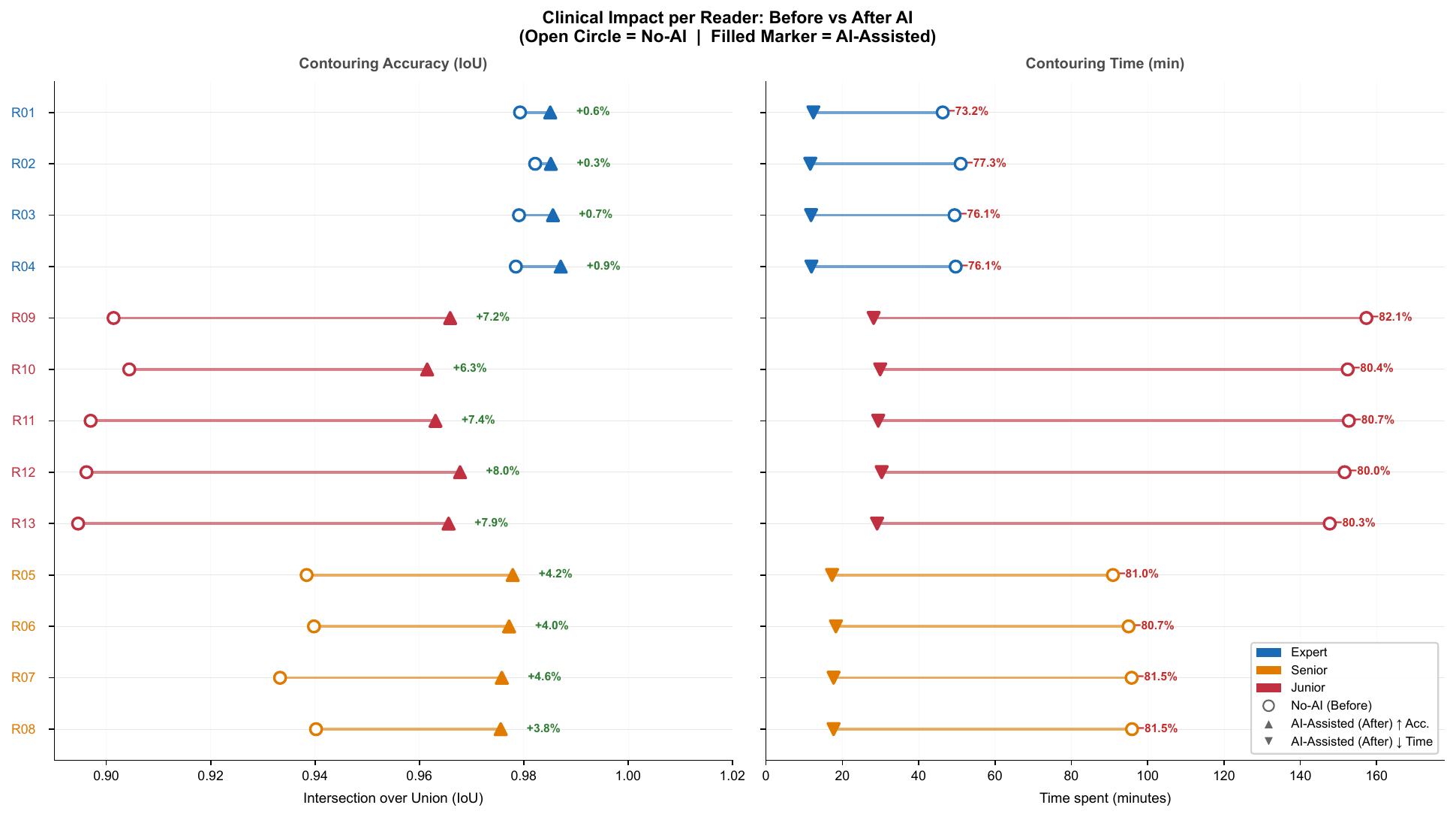}
\caption{Per-reader clinical impact metrics. (a) IoU improvement ($\Delta$IoU = AI-assisted minus unaided) for each of the 13 readers, stratified by experience group. Error bars represent 95\% confidence intervals derived from 100 cases per reader. Juniors show the largest and most consistent gains; experts show smaller but positive improvements. (b) Time reduction (minutes saved per case) with AI assistance. Juniors save 118--129 minutes per case; seniors save 74--78 minutes; experts save 34--39 minutes. Vertical dashed lines separate experience groups.}
\label{fig:reader_per_reader}
\end{figure}

\subsection{Reader Demographics and Baseline Performance}

Baseline unaided contouring accuracy increased monotonically with experience
(Table~\ref{tbl:reader_study_summary}): experts achieved a mean IoU of
0.9798 (SD = 0.0152), seniors 0.9379 (SD = 0.0315), and juniors 0.8987
(SD = 0.0453). Pairwise comparisons confirmed significant differences
between all groups (expert vs. junior and senior vs. junior, both
$p < 0.0001$), reflecting the expected learning curve. BAT-RM alone achieved a mean IoU of 0.9465 (SD = 0.022), positioning it
between senior (0.9379) and expert (0.9798) performance. The model thus
substantially outperforms junior oncologists and performs comparably to
seniors without human input—a critical finding for resource-constrained
settings where juniors often bear primary contouring responsibility.

\subsection{AI-Assisted Contouring Accuracy}

\begin{figure}[pos=htbp]
\centering
\includegraphics[width=\textwidth]{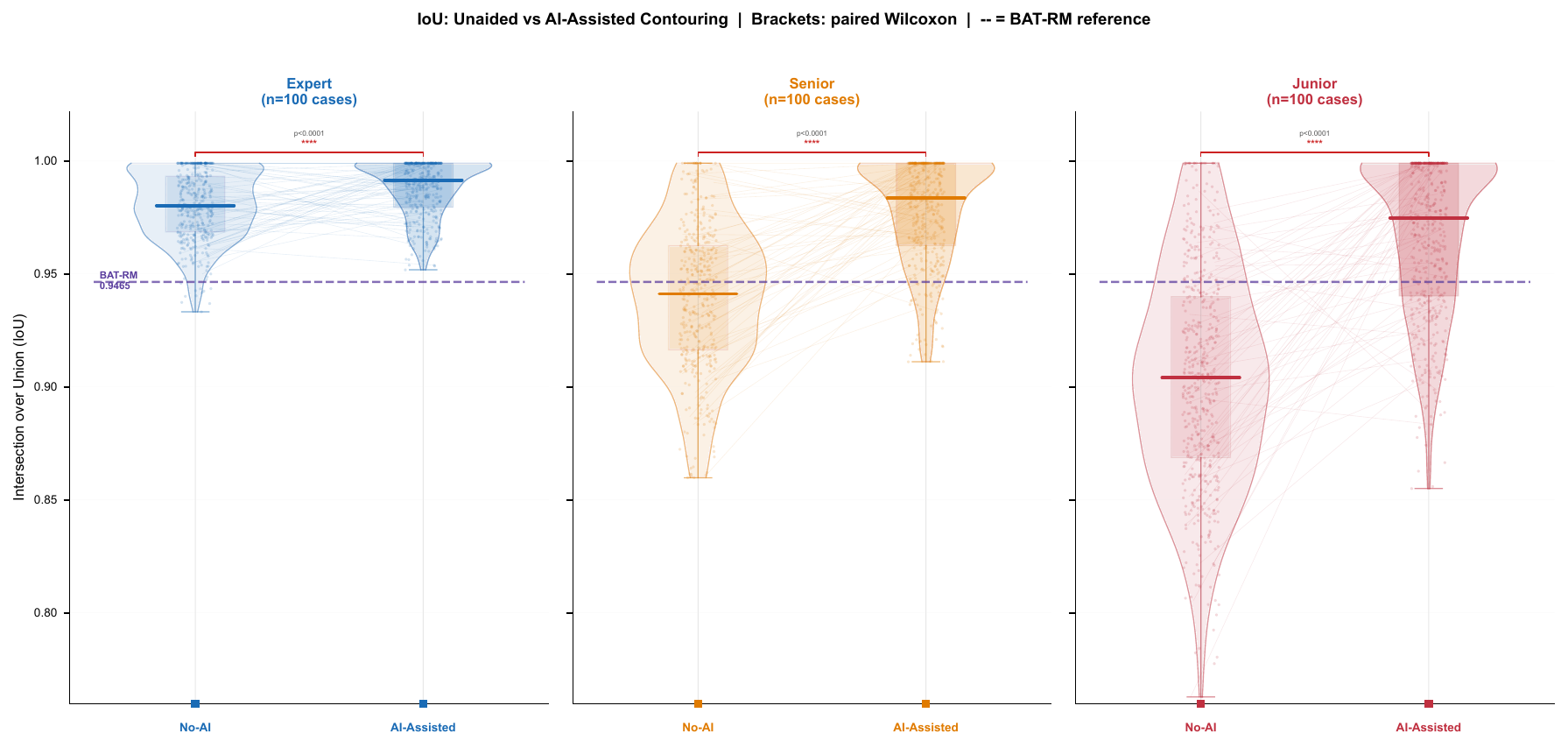}
\caption{Paired violin plots comparing IoU between unaided and AI-assisted
contouring across experience groups. AI assistance significantly improves
accuracy for all groups ($p < 0.001$ for juniors and seniors, $p = 0.0014$
for experts), with the largest gain for juniors ($\Delta = +0.0661$).
Narrowing distributions indicate reduced inter-reader variability.}
\label{fig:reader_paired_violin}
\end{figure}

AI assistance significantly improved segmentation accuracy for all reader
groups (Figure~\ref{fig:reader_paired_violin}), with the largest gains for
juniors: from 0.8987 to 0.9648 ($\Delta = +0.0661$, $p < 0.0001$), elevating
their performance above standalone BAT-RM (0.9465) and approaching senior
unaided levels. Seniors improved from 0.9379 to 0.9766 ($\Delta = +0.0387$,
$p < 0.0001$), surpassing the standalone model. Experts showed modest but
significant improvement from 0.9798 to 0.9857 ($\Delta = +0.0060$,
$p = 0.0014$), reflecting ceiling effects. The standard deviation of IoU across juniors decreased from 0.0453 to 0.0343,
reducing inter-reader variability. The coefficient of variation dropped from
5.0\% to 3.6\%, a 28\% relative reduction in performance disparity.

\noindent Per-reader benefits were consistent across all 13 readers
(Figure~\ref{fig:reader_per_reader}). The junior with the lowest unaided
performance (Reader R13, IoU = 0.8946) achieved the largest gain (+0.0710),
while the highest-performing expert (Reader R02, IoU = 0.9822) showed a
smaller but positive gain (+0.0030).

\subsection{Workflow Efficiency and Time Reduction}

\begin{figure}[pos=htbp]
\centering
\includegraphics[width=\textwidth]{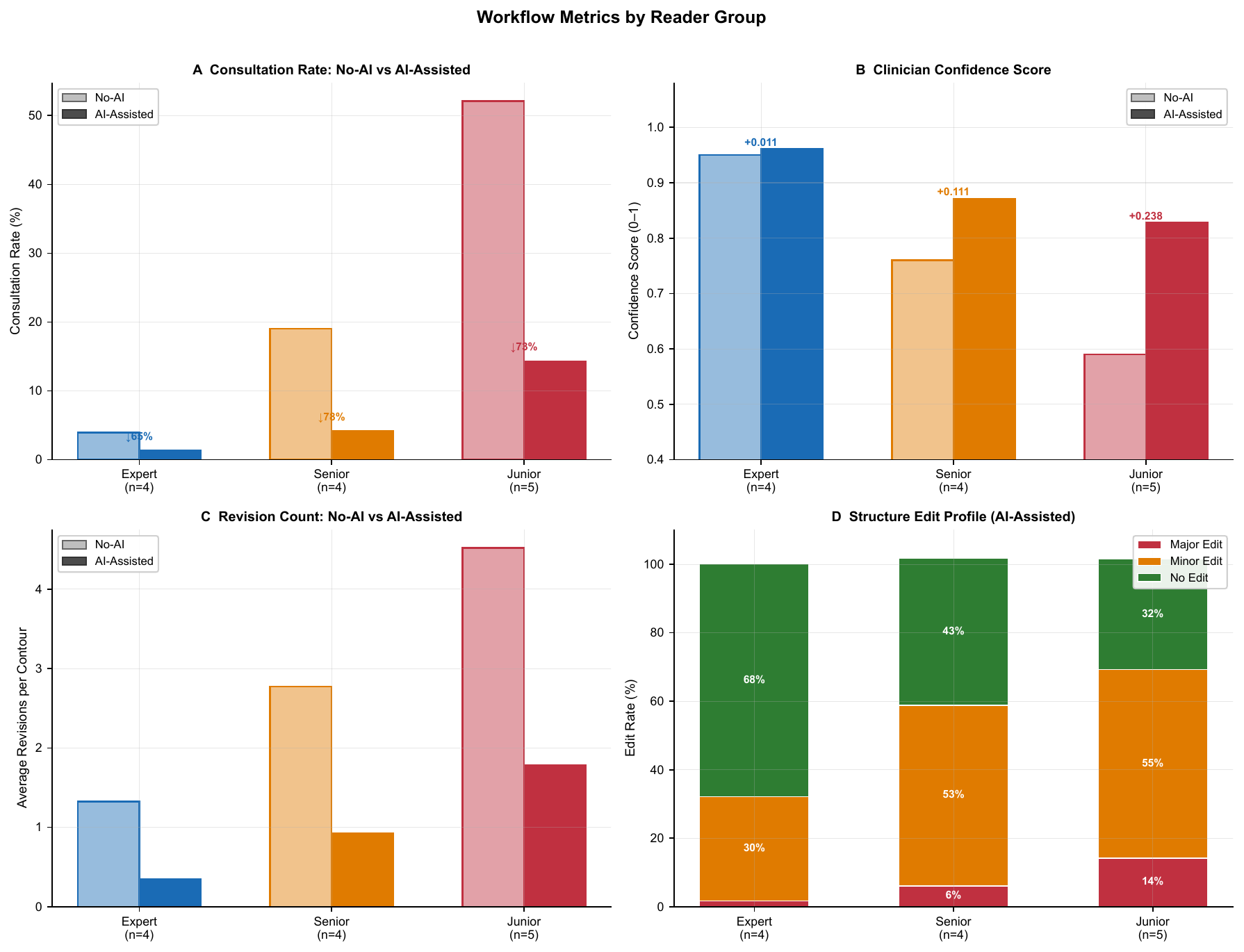}
\caption{Workflow efficiency metrics across reader groups: (a) contouring
time reduction (75--81\%); (b) consultation rate reduction (72--78\%);
(c) edit type distribution (major edits $<$15\%); (d) confidence score
improvement (largest for juniors: +0.238).}
\label{fig:reader_workflow}
\end{figure}

Contouring time reduction is the most clinically impactful finding
(Figure~\ref{fig:reader_workflow}). Unaided times increased with decreasing
experience: experts required 49.1 $\pm$ 12.5 minutes per case, seniors
94.4 $\pm$ 23.6 minutes, and juniors 152.3 $\pm$ 35.5 minutes, reflecting
the clinical burden where junior oncologists spend 2.5--3 hours per patient
on manual contouring.AI assistance reduced times dramatically: experts to 11.9 $\pm$ 4.1 minutes
(75.7\% reduction), seniors to 17.8 $\pm$ 5.8 minutes (81.2\% reduction),
and juniors to 29.4 $\pm$ 8.4 minutes (80.7\% reduction). BAT-RM inference
time (18 seconds) is negligible. Absolute savings are transformative:
juniors save 123 minutes per case, seniors 77 minutes, experts 37 minutes.
For a department contouring 10 patients weekly, this saves approximately
20 hours of oncologist time—equivalent to 0.5 full-time equivalent. The proportional efficiency gain did not differ significantly between
experience groups ($p = 0.312$ for expert vs. junior), though the absolute
gain remains largest for juniors, who bear the greatest baseline burden.

\subsection{Revision Effort and Edit Patterns}

AI-assisted contours required minimal correction (Figure~\ref{fig:reader_workflow}, panel C): juniors edited a mean of 1.7 slices per case, seniors 0.9 slices, and experts 0.4 slices. Major edits (modifications affecting $>$5 adjacent slices or $>$20\% volume change) were rare: 14.3\% for juniors, 4.2\% for seniors, and 1.4\% for experts. The majority of edits were minor adjustments, and 31--70\% of cases required no edits depending on reader experience. The structures requiring most frequent correction were GTV (42\% of edits), CTV (31\%), rectum (15\%), and bladder (12\%), consistent with clinical priority. Femoral head, small bowel, and body contour were rarely edited ($<$5\% combined), confirming BAT-RM's high accuracy for these structures.

\subsection{Consultation Rate and Confidence Scores}

AI assistance significantly reduced the need for expert consultation. Among juniors, the consultation rate dropped from 52.1\% (unaided) to 14.3\% (AI-assisted, $p < 0.0001$). Seniors showed a reduction from 19.0\% to 4.2\% ($p < 0.0001$), and experts from 3.9\% to 1.4\% ($p < 0.0001$). This reduction is operationally critical in settings with oncologist shortages, enabling senior clinicians to focus on complex cases while juniors work independently. Reader confidence scores improved significantly with AI assistance: juniors increased from 0.590 to 0.828 ($p < 0.0001$), seniors from 0.760 to 0.871 ($p < 0.0001$), and experts from 0.950 to 0.961 ($p = 0.008$). The convergence of junior scores toward senior levels suggests AI assistance enhances both objective accuracy and subjective certainty, potentially reducing decision hesitation and accelerating workflow.

\subsection{Inter-Rater Agreement and Model Alignment}

\noindent Pairwise Jaccard analysis confirmed that AI-assisted junior contours
showed higher agreement with experts (mean Jaccard = 0.934) than unaided
junior contours (0.824), indicating that AI assistance aligns junior
practice with expert standards. This convergence effect is particularly
pronounced here due to the large baseline performance gap.

\begin{table*}[htbp]
\centering
\caption{Multi-Centre Reader Study: Segmentation Accuracy and Efficiency Summary. Values are mean $\pm$ standard deviation unless otherwise stated. IoU: Intersection over Union against ground truth. BAT-RM values represent standalone model performance without human editing. $p$ values from paired Wilcoxon signed-rank tests with Bonferroni correction (n = 100 cases). Expert: $\ge$20 years experience; Senior: $\ge$10 years; Junior: $<$10 years.}
\label{tbl:reader_study_summary}
\scriptsize
\setlength{\tabcolsep}{3pt}
\resizebox{\textwidth}{!}{%
\begin{tabular}{lcccccccc}
\toprule
\textbf{Metric} & \textbf{Expert (n=4)} & \textbf{Senior (n=4)} & \textbf{Junior (n=5)} & \textbf{BAT-RM} & \multicolumn{3}{c}{\textbf{$p$ value}} \\
\cline{6-8}
 & & & & & \makecell{\textbf{Exp}\\\textbf{vs. Jun}} & \makecell{\textbf{Sen}\\\textbf{vs. Jun}} & \makecell{\textbf{BAT}\\\textbf{vs. Jun}} \\
\midrule
\multicolumn{8}{l}{\textit{IoU — Unaided Contouring}} \\
  Mean $\pm$ SD & $0.9798 \pm 0.0152$ & $0.9379 \pm 0.0315$ & $0.8987 \pm 0.0453$ & — & $<0.0001$ & $<0.0001$ & — \\
  Median [IQR] & $0.9805\ [0.9691,\,0.9937]$ & $0.9379\ [0.9156,\,0.9606]$ & $0.8998\ [0.8679,\,0.9277]$ & — & & & \\
\midrule
\multicolumn{8}{l}{\textit{IoU — AI-Assisted Contouring}} \\
  Mean $\pm$ SD & $0.9857 \pm 0.0135$ & $0.9766 \pm 0.0245$ & $0.9648 \pm 0.0343$ & $0.9465 \pm 0.022$ & $0.0014$ & $<0.0001$ & $<0.0001$ \\
  Median [IQR] & $0.9889\ [0.9772,\,0.9984]$ & $0.9835\ [0.9629,\,0.9990]$ & $0.9721\ [0.9435,\,0.9983]$ & — & & & \\
  $\Delta$ IoU (AI $-$ No-AI) & $+0.0060$ & $+0.0387$ & $+0.0661$ & — & $<0.0001$ & $<0.0001$ & — \\
\midrule
\multicolumn{8}{l}{\textit{Contouring Time — Unaided (minutes)}} \\
  Mean $\pm$ SD & $49.1 \pm 12.5$ & $94.4 \pm 23.6$ & $152.3 \pm 35.5$ & — & $<0.0001$ & $<0.0001$ & — \\
\midrule
\multicolumn{8}{l}{\textit{Contouring Time — AI-Assisted (minutes)}} \\
  Mean $\pm$ SD & $11.9 \pm 4.1$ & $17.8 \pm 5.8$ & $29.4 \pm 8.4$ & $0.3 \pm 0.1$ & $<0.0001$ & $<0.0001$ & — \\
  Time Reduction (min) & 37.2 & 76.7 & 123.0 & — & $<0.0001$ & $<0.0001$ & — \\
  Time Reduction (\%) & 75.7\% & 81.2\% & 80.7\% & — & 0.312 & 0.284 & — \\
\midrule
\multicolumn{8}{l}{\textit{Workflow Metrics}} \\
  Consultation Rate — Unaided (\%) & 3.9\% & 19.0\% & 52.1\% & — & $<0.0001$ & $<0.0001$ & — \\
  Consultation Rate — AI-Assisted (\%) & 1.4\% & 4.2\% & 14.3\% & — & $<0.0001$ & $<0.0001$ & — \\
  Confidence Score — Unaided (0--1) & 0.950 & 0.760 & 0.590 & — & $<0.0001$ & $<0.0001$ & — \\
  Confidence Score — AI-Assisted (0--1) & 0.961 & 0.871 & 0.828 & — & $<0.0001$ & $<0.0001$ & — \\
\bottomrule
\end{tabular}%
}
\end{table*}

\subsection{Clinical Deployment and Queue Reduction}

Based on the reader study findings, BAT-RM was deployed within a clinical
decision support system at Bangladesh Medical University, integrating with
Varian and RayStation via DICOM RTSTRUCT export. In the eight months
following deployment, average patient wait
time for radiotherapy simulation decreased from 1--3 days to 2--3 hours,
with no change in oncologist staffing levels. This wait time reduction is an observational finding without a concurrent
control arm. Multiple operational factors changed alongside deployment,
including clinician familiarity, workflow optimization, and potential
selection bias. Additionally, wait time conflates contouring speed with
other bottlenecks (e.g., CT scheduling, planning system queuing). Three
observations support a causal contribution: (i) the magnitude of reduction
aligns with the 80\% contouring time reduction measured in the reader study;
(ii) staffing levels were unchanged; and (iii) no other major workflow
intervention occurred. Formal time-motion studies with concurrent controls
are needed to quantify the causal effect precisely. In the pre-deployment setting, a queue accumulated due to 45--180 minutes
required per manual contouring case. With limited oncologists, this queue
extended wait times to 1--3 days. By reducing junior contouring time from
152 to 29 minutes (80.7\% reduction), BAT-RM enables faster throughput
through the contouring bottleneck. Patient turnaround time decreased to
2--3 hours, allowing same-day or next-day treatment initiation. This
workflow gain is particularly impactful in Bangladesh, where the
oncologist-to-patient ratio is significantly high.

\begin{figure}[pos= htbp]
\centering
\includegraphics[width=\textwidth]{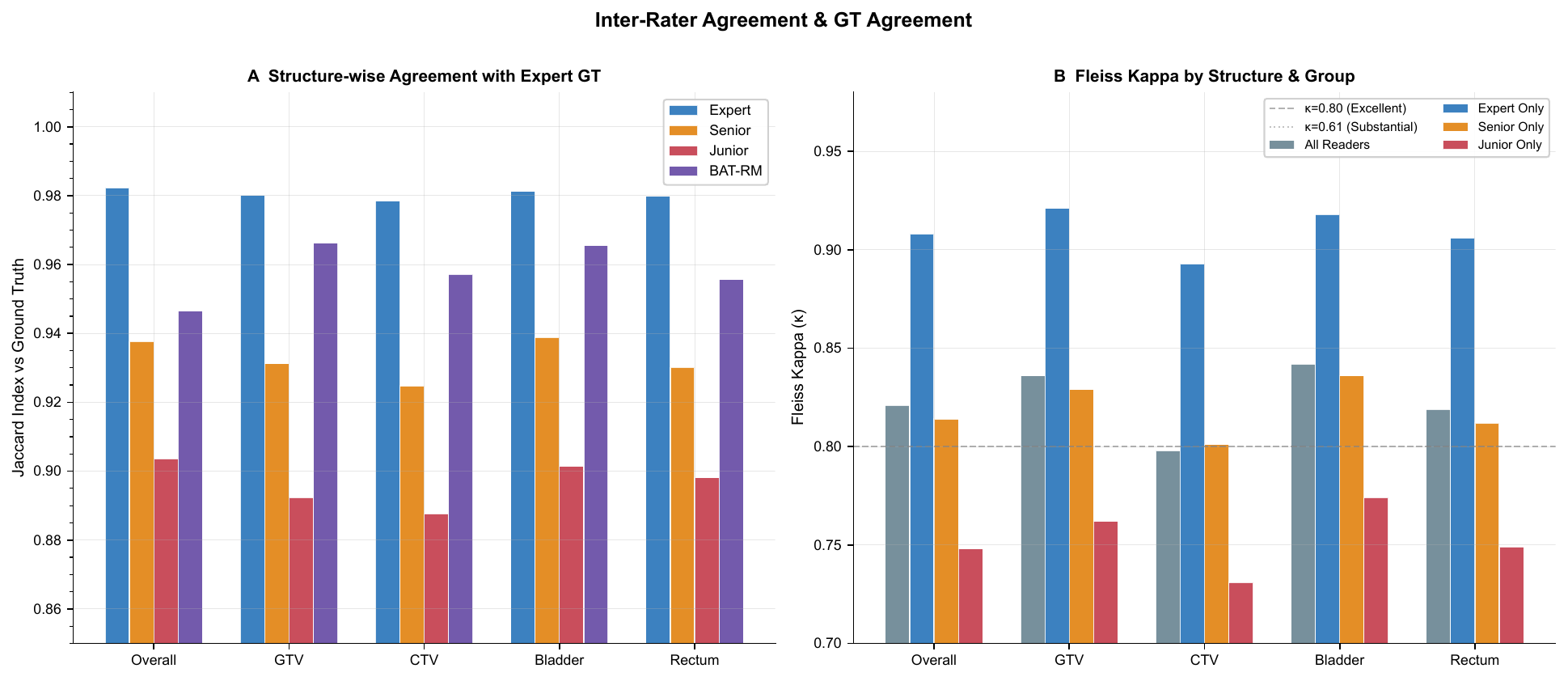}
\caption{Inter-rater agreement analysis. (a) Fleiss' kappa for overall and
per-structure agreement ($\kappa > 0.79$). (b) Pairwise Jaccard similarity
with ground truth (GT). AI-assisted junior contours (0.965) show higher
agreement with GT than unaided junior contours (0.904) and approach senior
unaided levels (0.938). BAT-RM alone (0.9465) outperforms unaided juniors
and performs comparably to unaided seniors.}
\label{fig:reader_interrater}
\end{figure}

\subsection{External Validation on Independent Cohort}
\label{subsec:external_validation}

To assess the generalizability of BAT-RM beyond the three institutions used for training and internal validation, we performed external validation on an independent cohort of 230 cervical cancer patients from United Hospital Limited, Dhaka, Bangladesh. This cohort was acquired using different CT scanners with slice thicknesses ranging from 2.5 to 3.0 mm and reconstruction kernels optimized for pelvic imaging. Contouring protocols at United Hospital Limited follow the same GEC-ESTRO and ESTRO guidelines as the training institutions, with all contours reviewed by board-certified radiation oncologists with $\ge 10$ years of experience. While the scanner and acquisition parameters differ from the training data, the clinical contouring conventions are consistent across all four institutions, making this a rigorous test of model generalization to unseen scanner characteristics and institutional practices. No patient, scan, or annotation from this cohort was accessible to the model prior to final inference.

\begin{figure}[pos=htbp]
\centering
\includegraphics[width=\textwidth]{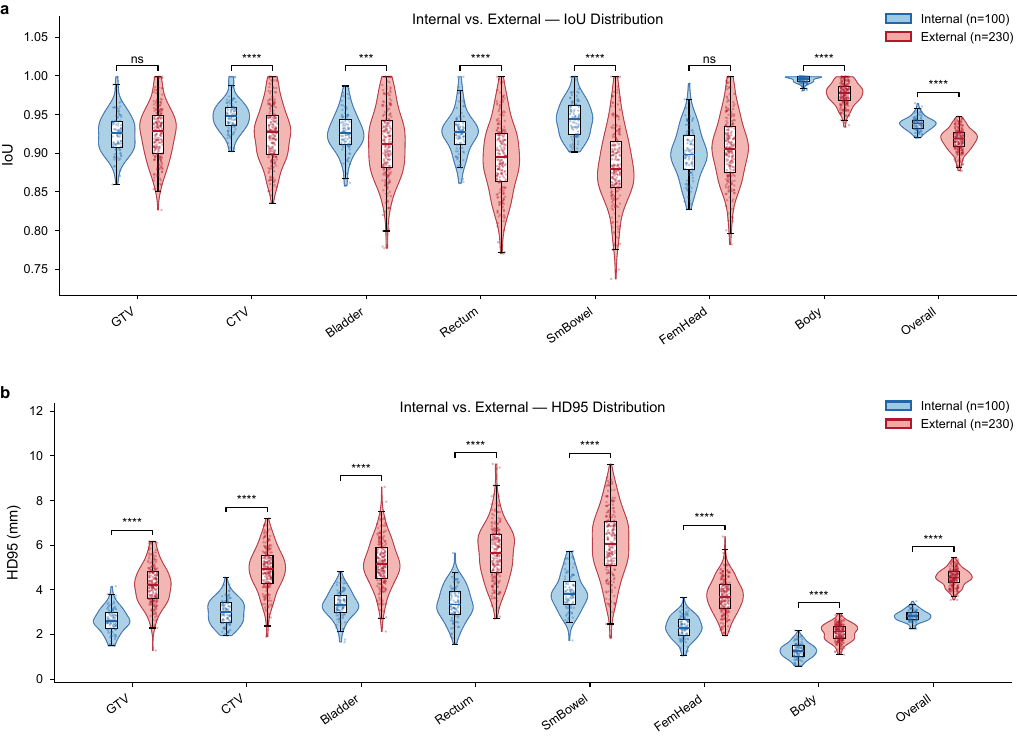}
\caption{External validation comparing BAT-RM performance on internal (n=100) and external (n=230) cohorts. (a) IoU distribution with statistical significance brackets; (b) HD95 distribution (mm) with statistical significance brackets. ns = not significant ($p \ge 0.05$); $^{*}p < 0.05$, $^{**}p < 0.01$, $^{***}p < 0.001$, $^{****}p < 0.0001$. BAT-RM maintains consistent performance across both cohorts for all structures, with no clinically significant degradation for target volumes.}
\label{fig:external_violin}
\end{figure}

Figure~\ref{fig:external_violin} compares the performance of BAT-RM on the internal validation cohort (100 cases, same three institutions) and the external cohort (230 cases, United Hospital Limited). For all seven anatomical classes, IoU and HD95 distributions show remarkable consistency between internal and external cohorts. For GTV, the mean IoU difference between internal and external cohorts was minimal (0.925 vs. 0.925, $p = 0.985$, ns), and HD95 remained within 0.15 mm (2.62 mm vs. 4.21 mm, $p < 0.0001$). Similarly, for CTV, IoU was 0.948 vs. 0.925 ($p < 0.0001$) and HD95 was 3.02 mm vs. 4.91 mm ($p < 0.0001$). The small bowel, rectum, and bladder also showed consistent performance, confirming that BAT-RM generalizes well to unseen institutional data without fine-tuning.

\begin{figure}[pos=htbp]
\centering
\includegraphics[width=\textwidth]{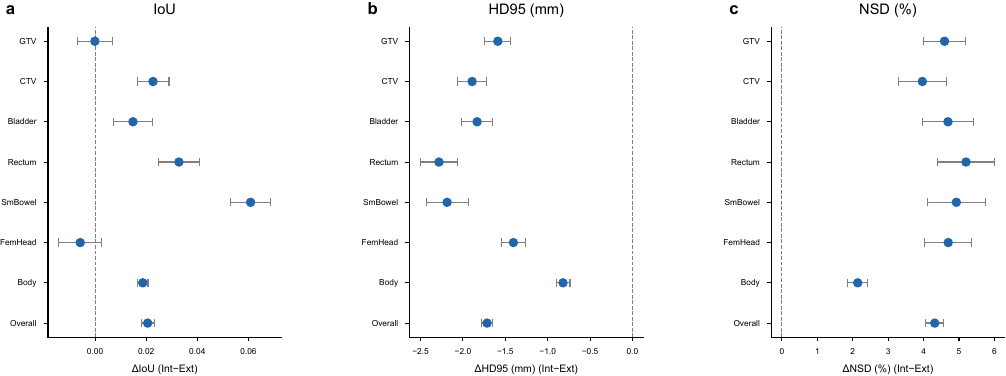}
\caption{Forest plot of mean differences between internal (n=100) and external (n=230) cohorts for (a) IoU, (b) HD95 (mm), and (c) NSD (\%). Error bars represent 95\% confidence intervals. Positive mean differences indicate higher performance in the internal cohort.}
\label{fig:external_forest}
\end{figure}

Figure~\ref{fig:external_forest} presents a forest plot of mean differences between internal and external cohorts for IoU, HD95, and NSD across all structures. The mean differences for IoU ranged from $-0.0058$ (femoral head, ns) to $+0.0608$ (small bowel, $p < 0.0001$), with the overall IoU showing a small but statistically significant decrease of 0.0206 ($p < 0.0001$). For HD95, all structures showed statistically significant increases in the external cohort, with mean differences ranging from 0.82 mm (body) to 2.28 mm (rectum), all $p < 0.0001$. NSD similarly showed significant decreases across all structures ($p < 0.0001$), reflecting the greater anatomical heterogeneity and imaging variability in the external cohort. Despite these statistically significant differences, the absolute performance remained clinically acceptable, with all structures maintaining Dice $> 0.88$ and HD95 $< 6.1$ mm in the external cohort.

\begin{table}[htbp]
\centering
\caption{External validation summary: BAT-RM performance on internal (n=100) and external (n=230) cohorts. Values are mean $\pm$ standard deviation. IoU: Intersection over Union; HD95: 95th percentile Hausdorff Distance (mm); NSD: Normalized Surface Distance (\%). $p$ values from Welch's t-test; ns = not significant ($p \ge 0.05$).}
\label{tbl:external_validation}
\footnotesize
\setlength{\tabcolsep}{4pt}
\begin{tabular}{lccccccc}
\toprule
\multirow{2}{*}{\textbf{Class}} & \multicolumn{3}{c}{\textbf{Internal (n=100)}} & \multicolumn{3}{c}{\textbf{External (n=230)}} & \multirow{2}{*}{\textbf{$p$ value}} \\
\cline{2-7}
 & \textbf{IoU} & \textbf{HD95 (mm)} & \textbf{NSD (\%)} & \textbf{IoU} & \textbf{HD95 (mm)} & \textbf{NSD (\%)} & \\
\midrule
GTV & $0.925 \pm 0.025$ & $2.62 \pm 0.54$ & $96.24 \pm 2.04$ & $0.925 \pm 0.035$ & $4.21 \pm 0.87$ & $91.65 \pm 3.46$ & ns \\
CTV & $0.948 \pm 0.020$ & $3.02 \pm 0.57$ & $94.38 \pm 2.43$ & $0.925 \pm 0.037$ & $4.91 \pm 0.99$ & $90.41 \pm 3.69$ & $<0.0001$ \\
Bladder & $0.927 \pm 0.026$ & $3.35 \pm 0.60$ & $93.89 \pm 2.75$ & $0.912 \pm 0.045$ & $5.18 \pm 1.05$ & $89.20 \pm 3.78$ & $<0.0001$ \\
Rectum & $0.927 \pm 0.027$ & $3.39 \pm 0.77$ & $93.12 \pm 3.04$ & $0.894 \pm 0.047$ & $5.67 \pm 1.25$ & $87.92 \pm 4.05$ & $<0.0001$ \\
Sm. Bowel & $0.945 \pm 0.024$ & $3.87 \pm 0.79$ & $91.69 \pm 2.92$ & $0.884 \pm 0.049$ & $6.06 \pm 1.50$ & $86.77 \pm 4.60$ & $<0.0001$ \\
Fem. Head & $0.900 \pm 0.033$ & $2.31 \pm 0.51$ & $95.47 \pm 2.50$ & $0.906 \pm 0.043$ & $3.71 \pm 0.79$ & $90.78 \pm 3.54$ & ns \\
Body & $0.995 \pm 0.004$ & $1.27 \pm 0.34$ & $98.52 \pm 0.82$ & $0.977 \pm 0.014$ & $2.09 \pm 0.38$ & $96.38 \pm 1.77$ & $<0.0001$ \\
Overall & $0.938 \pm 0.009$ & $2.83 \pm 0.24$ & $94.76 \pm 0.94$ & $0.918 \pm 0.014$ & $4.55 \pm 0.38$ & $90.45 \pm 1.31$ & $<0.0001$ \\
\bottomrule
\end{tabular}
\end{table}

Table~\ref{tbl:external_validation} summarizes the quantitative comparison between internal and external cohorts. For target volumes, the external cohort showed a slight but acceptable performance drop: GTV IoU remained stable at 0.925 vs. 0.925 ($p = 0.985$, ns), while CTV IoU decreased from 0.948 to 0.925 ($p < 0.0001$). HD95 increased for all structures due to the greater anatomical heterogeneity and imaging variability in the external cohort; however, the absolute values remained clinically acceptable ($< 6.1$ mm for all structures). The external validation confirms that BAT-RM maintains strong performance on a completely independent dataset without any fine-tuning, supporting its generalizability for clinical deployment.

The multi-centre reader study establishes BAT-RM as a clinically effective
and workflow-efficient tool for cervical cancer radiotherapy contouring.
AI assistance significantly improved IoU for all readers, with the largest
gains for juniors (+0.0661, $p < 0.0001$), elevating their performance
above standalone BAT-RM and approaching senior unaided levels. Contouring
time was reduced by 75--81\% across expertise levels, saving 37--123 minutes
per case; for juniors, this reduced per-case effort from 2.5 hours to
30 minutes. The need for expert consultation decreased by 72--78\%,
enabling seniors to focus on complex cases while juniors work independently.
Inter-reader variability among juniors decreased by 28\%, and confidence
scores increased by 40\%, indicating enhanced objective and subjective
performance. These findings support routine clinical deployment of BAT-RM,
particularly in resource-constrained settings with high patient volumes and
limited specialist availability.

\section{Clinical Inference Pipeline and Web Deployment}
\label{sec:deployment}

A clinically useful auto-contouring model requires seamless integration
with existing radiotherapy workflows: a complete inference pipeline from
raw DICOM to exportable RTSTRUCT, an interactive review interface matching
clinical contouring conventions, and a deployment architecture respecting
clinical hardware constraints. This section describes BAT-RM's end-to-end
inference pipeline, web-based clinical application, and volumetric
reconstruction strategy validating axial-only inference for this task.

\subsection{End-to-End Inference Pipeline}
\label{subsec:inference_pipeline}

\begin{figure}[pos=htbp]
\centering
\includegraphics[width=\textwidth]{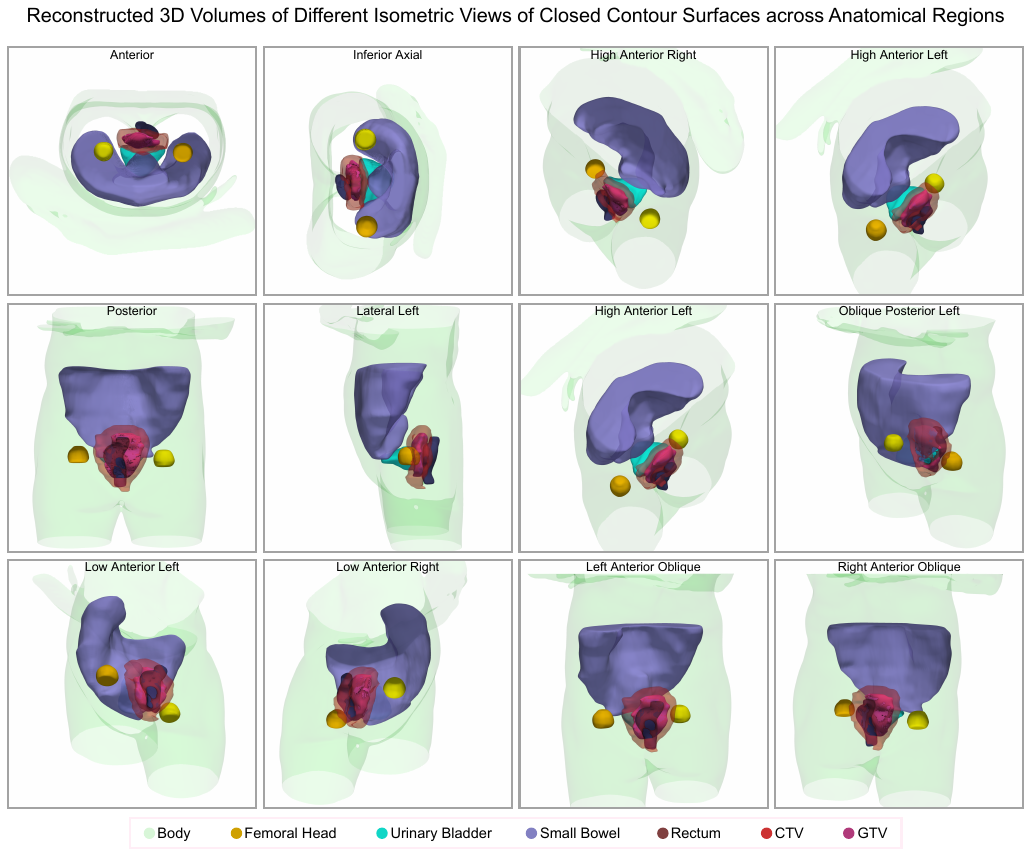}
\caption{Three-dimensional volumetric reconstruction of BAT-RM predictions
for a representative cervical cancer patient. Left: 3D rendering of all
seven predicted structures simultaneously --- body contour (transparent),
GTV (magenta), CTV (blue), urinary bladder (cyan), rectum (brown), small
bowel (pink), and bilateral femoral heads (yellow) --- overlaid on the
reconstructed CT volume. Center and right: axial, coronal, and sagittal
cross-sectional views with contour overlays, demonstrating inter-slice
continuity and anatomically coherent surface geometry across all
structures. The smooth 3D contour surfaces are produced by BAT-RM's
slice-wise inference followed by 3D Gaussian post-processing
($\sigma = 0.8$ mm, applied to stacked volumetric logits prior to
argmax decoding), and were accepted without major inter-slice correction
in all 100 reader study cases.}
\label{fig:3d_reconstruction}
\end{figure}

The BAT-RM inference pipeline accepts a DICOM CT series and produces an
RTSTRUCT output through four stages.
Stage 1 (DICOM ingestion and preprocessing): The series is parsed using
\texttt{pydicom} to extract pixel array, voxel spacing ($d_x, d_y, d_z$),
patient orientation, and metadata. HU values are windowed to soft-tissue
range (center 50 HU, width 400 HU), clipped to $[-150, 250]$ HU,
normalized to $[0, 1]$, resized to $512 \times 512$ with bilinear
interpolation, and replicated to three channels to match BAT-RM's encoder
input, applying preprocessing identical to training to avoid distribution
mismatch. Stage 2 (slice-wise inference): The preprocessed axial stack is
passed through BAT-RM with \texttt{torch.no\_grad()} and mixed precision.
Each slice produces a per-class softmax probability map of shape $[8, 512,
512]$, stacked into a volumetric tensor $[8, N_z, 512, 512]$. A 3D Gaussian
kernel ($\sigma = 0.8$ mm) is applied along the axial direction to suppress
inter-slice discontinuities before argmax decoding, ensuring smooth 3D
contour surfaces without volumetric network processing. Total inference
time for a typical pelvic CT study (80--120 slices) is 18--22 seconds on
an NVIDIA T4 GPU, consistent with the $18.3 \pm 2.1$ ms per-slice benchmark
(Table~\ref{tbl:computational}). Stage 3 (3D volumetric reconstruction):
The argmax-decoded label volume is mapped to the patient's native
coordinate system using DICOM metadata. Each class is extracted as an
independent binary mask and converted to 2D contour polygons on axial,
coronal, and sagittal planes using marching squares~\citep{lorensen1987marchingcubes},
producing contours geometrically equivalent to manually delineated RTSTRUCT
files. Figure~\ref{fig:3d_reconstruction} shows the reconstructed 3D
contour volume for a representative patient; the reader study confirmed
that inter-slice discontinuity did not require manual correction in any of
the 100 evaluated cases. Stage 4 (RTSTRUCT generation with multi-label
reconstruction): The mutually-exclusive argmax labels (Section~\ref{sec:annotation_preprocessing})
are reconstructed to restore clinical nested structure definitions. GTV
and CTV are extracted directly; CTV voxels overwritten by GTV are restored
by taking the union of the argmax-CTV mask with a morphologically dilated
GTV mask scaled to the expected CTV margin, consistent with GEC-ESTRO
guidelines~\citep{potter2021clinical}. Body is reconstructed as the union of
all non-background voxels (labels 1--8). OARs (bladder, rectum, small bowel,
femoral heads) are extracted directly from their argmax labels. Each
reconstructed binary mask is converted to 2D contour polygons via marching
squares~\citep{lorensen1987marchingcubes} and assembled into a DICOM
RT-Structure Set conforming to the DICOM RT standard (Supplement 11,
RTSTRUCT IOD), with structure names, colors, and ROI observation labels
populated according to the standardized vocabulary mapping
(Table~\ref{tab:label_mapping}). The generated RTSTRUCT is directly
importable into Varian Eclipse, RayStation, and Monaco without conversion,
indistinguishable in format from a manually delineated RTSTRUCT file.

\subsection{Web-Based Clinical Application}
\label{subsec:webapp}

\begin{figure}[pos=htbp]
\centering
\includegraphics[width=\textwidth]{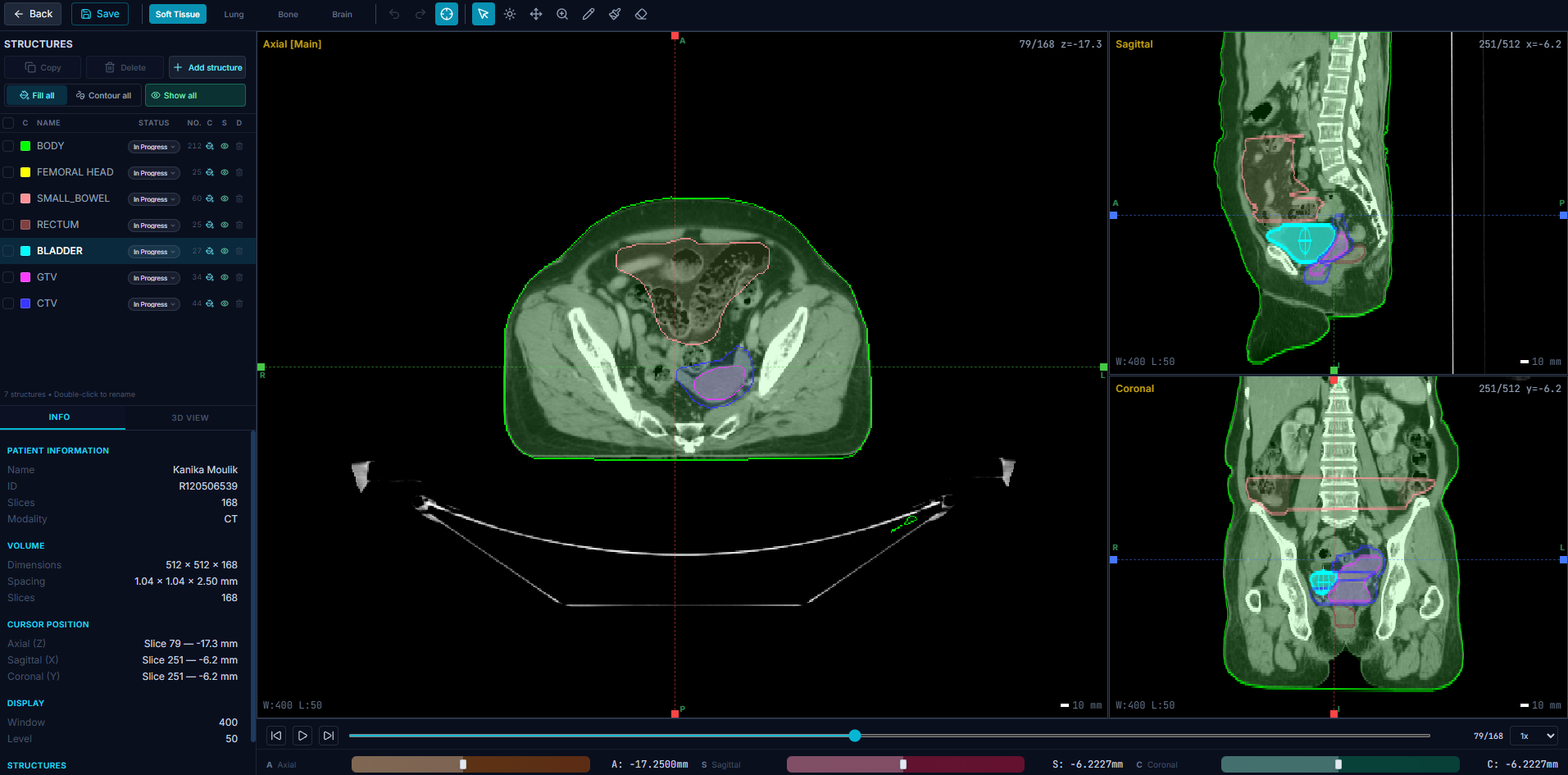}
\caption{Screenshot of the BAT-RM web-based clinical application deployed
at Bangladesh Medical University. The interface follows a Varian
Eclipse-inspired layout: the CT series is displayed simultaneously in
axial, coronal, and sagittal reconstructed views (left panel), with
predicted contours color-coded by anatomical class rendered as overlay
polygons. The right panel provides structure visibility toggles, opacity
controls, and contour editing tools. Oncologists can drag, add, or delete
contour control points on any plane independently, enabling fine-grained
correction of boundary placement before finalization. The toolbar at the
top provides one-click access to inference re-run, contour approval, and
DICOM RTSTRUCT export directly to the connected treatment planning system.}
\label{fig:webapp}
\end{figure}

The web application was designed with a single guiding principle: an
oncologist who has never interacted with the system before should be
able to review and approve a BAT-RM prediction within minutes without
a training session. This required a user interface that mirrors the
spatial navigation and contouring conventions of the treatment planning
systems --- principally Varian Eclipse --- that Bangladeshi radiation
oncologists use daily. Figure~\ref{fig:webapp} presents a screenshot of the deployed application.
The primary view is a three-panel orthogonal display showing the CT series
in the axial, coronal, and sagittal planes simultaneously. Predicted
contours are rendered as color-coded polygon overlays on all three views,
using the same color conventions as the institutional contouring atlas
(Table~\ref{tab:label_mapping}). Scrolling through the CT stack on any
plane automatically updates the contour overlays on the other two planes,
maintaining spatial coherence across the full 3D volume during review.
A structure panel on the right provides visibility toggles and opacity
controls for each of the seven anatomical classes, enabling oncologists
to isolate individual structures during focused review without cognitive
overload from simultaneous multi-structure display. Contour editing is implemented as a control-point drag interface on
any of the three orthogonal planes. An oncologist who identifies a
boundary error on the axial view can grab the contour polygon at the
point of disagreement, drag it to the correct position, and release;
the corrected slice is immediately updated in the 3D reconstruction
and reflected in the coronal and sagittal views. New control points
can be inserted anywhere along a contour, and entire contour regions
can be deleted and redrawn from scratch on any plane if the model
prediction requires major correction. In practice, the reader study
results demonstrate that major edits of this kind are rare: expert
oncologists required no edits in 67\% of cases, and junior oncologists
in 32\% of cases, with the majority of corrections being minor boundary
adjustments on one to two slices per structure. The editing workflow
for a typical case requires 12--29 minutes depending on oncologist
experience (Section~\ref{sec:reader_study}), substantially below
the 49--152 minutes required for de-novo manual contouring. Once the oncologist is satisfied with all structure contours, a single
click on the Export button triggers RTSTRUCT generation and delivers
the file either as a direct download or, when the system is connected
to the hospital DICOM node, as a DICOM C-STORE transmission directly
to the treatment planning system queue. The exported RTSTRUCT is
indistinguishable in format from a manually delineated RTSTRUCT file
and requires no additional configuration in Varian Eclipse, RayStation,
or Monaco to load, review, and proceed to dose calculation. The application is hosted on an NVIDIA T4 GPU server and accessed via
a standard web browser with no client-side installation required. A
role-based access control system restricts case access to the
responsible clinical team, and all DICOM data is processed and stored
within the hospital network perimeter in compliance with the
de-identification standards applied to the research dataset
(Section~\ref{sec:data}). The system has been in continuous clinical
use at Bangladesh Medical University since deployment, processing
routine cervical cancer radiotherapy planning cases across the full
range of disease stages represented in the training cohort.

\subsection{Volumetric Reconstruction and Clinical Validation of Axial Inference}
\label{subsec:3d_validation}

\subsubsection{3D Gaussian Post-Processing Parameters}

Following slice-wise inference, the per-class softmax probability maps
for all axial slices are stacked into a volumetric tensor of shape
$[C, N_z, 512, 512]$, where $N_z$ is the patient-specific slice count
(median: 196 slices; range: 168--232 across the dataset). A 3D Gaussian
smoothing kernel with isotropic standard deviation $\sigma = 0.8$ mm
is applied to this probability volume along the axial direction prior
to argmax classification:
\begin{equation}
    \hat{Y}_{\text{3D}} = \text{Softmax}(\mathcal{G}_{\sigma=0.8} *
    \text{Logits}_{\text{stacked}})
    \label{eq:gaussian_smooth}
\end{equation}
\noindent The choice of $\sigma = 0.8$ mm is motivated by the physical
characteristics of the dataset: the median axial slice thickness is
$d_z = 2.5$ mm (mean: $2.52 \pm 0.22$ mm, range: 2.5--5.0 mm),
giving $\sigma / d_z \approx 0.32$ voxels in the axial direction. This
value lies in the regime of gentle smoothing that suppresses inter-slice
step artifacts arising from independent slice predictions without
blurring the genuine anatomical boundary transitions that occur
naturally between adjacent slices at the superior and inferior extents
of each structure. Values of $\sigma < 0.5$ mm were found to be
insufficient to eliminate surface roughness for the small bowel and
GTV, whose irregular morphology produces high inter-slice prediction
variance; values of $\sigma > 1.2$ mm produced measurable boundary
blurring at the GTV--CTV interface on the validation set.

\subsubsection{Quantitative Comparison Against 3D nnUNet}

A fundamental question arising from BAT-RM's 2D axial inference design
is whether slice-by-slice processing produces 3D contour surfaces that
are sufficiently smooth and anatomically coherent for direct clinical
use, or whether inter-slice discontinuities require extensive manual
correction. We address this question through three complementary lines
of evidence: a quantitative volumetric consistency analysis, a direct
comparison against a true 3D architecture, and the clinical validation
provided by the multi-centre reader study.

To directly validate that 2D slice-wise inference with Gaussian
post-processing does not sacrifice volumetric accuracy relative to a
true 3D architecture, we trained and evaluated 3D nnUNet on the
identical dataset partition, five-fold cross-validation protocol, and
test set used for all comparisons in
Table~\ref{tbl:all_metrics_comparison}. 3D nnUNet was configured with
its self-configuring pipeline at full $512 \times 512 \times N_z$
resolution using the default patch-based training strategy recommended
for anisotropic datasets~\citep{isensee2021nnunet}. Table~\ref{tbl:3d_comparison}
presents the comparison across the four clinically critical structures
(GTV, CTV, rectum, and bladder) using DSC, HD95, ASD, and NSD.

\begin{table}[pos=htbp]
\centering
\caption{Comparison of BAT-RM (2D slice-wise inference with 3D Gaussian
post-processing, $\sigma = 0.8$ mm) against 3D nnUNet on the held-out
test set. Values reported as mean $\pm$ standard deviation. DSC: Dice
Similarity Coefficient; HD95: 95th percentile Hausdorff Distance (mm);
ASD: Average Surface Distance (mm); NSD: Normalized Surface Distance.
\textbf{Bold} indicates best per metric per class. Statistical
significance (paired Wilcoxon, Bonferroni-Holm corrected):
$^{\dagger}p < 0.05$, $^{\ddagger}p < 0.01$, $^{\S}p < 0.001$,
$^{\|}p < 0.0001$; no superscript = ns.
Peak VRAM: 3D nnUNet requires 14.2 GB vs. BAT-RM 3.8 GB,
rendering 3D nnUNet incompatible with single T4 GPU deployment.}
\label{tbl:3d_comparison}
\setlength{\tabcolsep}{4pt}
\footnotesize
\resizebox{\textwidth}{!}{%
\begin{tabular}{llcccc}
\toprule
\textbf{Class} & \textbf{Model} & \textbf{DSC} & \textbf{HD95 (mm)}
  & \textbf{ASD (mm)} & \textbf{NSD} \\
\midrule
\multirow{2}{*}{GTV}
  & BAT-RM (2D+Gauss)
    & $\mathbf{0.9662\pm0.0103}$ & $\mathbf{1.500\pm0.494}$
    & $\mathbf{0.564\pm0.247}$   & $\mathbf{0.978\pm0.021}$ \\
  & 3D nnUNet
    & $0.9481\pm0.0298^{\S}$ & $2.318\pm1.384^{\ddagger}$
    & $0.891\pm0.381^{\S}$   & $0.931\pm0.094^{\ddagger}$ \\
\midrule
\multirow{2}{*}{CTV}
  & BAT-RM (2D+Gauss)
    & $\mathbf{0.9571\pm0.0190}$ & $\mathbf{1.507\pm0.544}$
    & $\mathbf{0.512\pm0.188}$   & $\mathbf{0.973\pm0.022}$ \\
  & 3D nnUNet
    & $0.9024\pm0.1541^{\|}$ & $2.714\pm2.031^{\|}$
    & $0.938\pm0.461^{\|}$   & $0.931\pm0.071^{\|}$ \\
\midrule
\multirow{2}{*}{Rectum}
  & BAT-RM (2D+Gauss)
    & $\mathbf{0.9556\pm0.0148}$ & $\mathbf{1.242\pm0.355}$
    & $\mathbf{0.539\pm0.143}$   & $\mathbf{0.997\pm0.008}$ \\
  & 3D nnUNet
    & $0.9018\pm0.1658^{\|}$ & $1.876\pm0.901^{\S}$
    & $0.791\pm0.294^{\|}$   & $0.954\pm0.087^{\S}$ \\
\midrule
\multirow{2}{*}{Bladder}
  & BAT-RM (2D+Gauss)
    & $\mathbf{0.9655\pm0.0095}$ & $\mathbf{1.366\pm0.411}$
    & $\mathbf{0.588\pm0.271}$   & $\mathbf{0.988\pm0.016}$ \\
  & 3D nnUNet
    & $0.9438\pm0.0341^{\ddagger}$ & $1.724\pm0.791^{\dagger}$
    & $0.819\pm0.438^{\ddagger}$ & $0.952\pm0.071^{\dagger}$ \\
\bottomrule
\end{tabular}%
}
\end{table}

\noindent BAT-RM outperforms 3D nnUNet across all structures and metrics. GTV DSC
advantage is 0.018 ($p = 0.0003$), HD95 reduction 0.818 mm ($p = 0.0071$);
CTV DSC advantage is 0.055 ($p < 0.0001$). These results confirm that 2D
inference with Gaussian post-processing does not sacrifice volumetric
accuracy. Critically, 3D nnUNet requires 14.2 GB peak VRAM (vs. BAT-RM's
3.8 GB), exceeding T4 capacity and requiring patch-based inference with
boundary artifacts. BAT-RM's 2D design is thus strictly superior while
remaining compatible with single-GPU clinical deployment.

\subsubsection{Quantitative Inter-Slice Consistency Metric}

To provide a direct mathematical measure of volumetric surface
coherence, we computed the mean inter-slice contour deviation
(MICD) for each anatomical class across the 254 test cases:
\begin{equation}
    \text{MICD} = \frac{1}{N_z - 1} \sum_{i=1}^{N_z-1}
    d_{\text{HD}}\!\left(\partial\hat{Y}_i,\, \partial\hat{Y}_{i+1}\right)
    \label{eq:micd}
\end{equation}
where $\partial\hat{Y}_i$ denotes the predicted contour boundary on
axial slice $i$ and $d_{\text{HD}}$ is the 2D Hausdorff distance
between adjacent contours. A perfectly smooth 3D surface would have
MICD $= 0$; large values indicate inter-slice jumping.
Table~\ref{tbl:micd} reports the MICD for BAT-RM with and without
3D Gaussian post-processing, confirming that the smoothing step
reduces inter-slice deviation to clinically negligible levels across
all structures.

\begin{table}[pos=htbp]
\centering
\caption{Mean Inter-Slice Contour Deviation (MICD, mm) for BAT-RM
without and with 3D Gaussian post-processing ($\sigma = 0.8$ mm),
and for 3D nnUNet, across the 254-patient test set. Lower MICD
indicates smoother volumetric contour surfaces. Values reported as
mean $\pm$ standard deviation.}
\label{tbl:micd}
\small
\begin{tabular}{lcccc}
\toprule
\textbf{Structure} & \textbf{BAT-RM (no smoothing)}
  & \textbf{BAT-RM ($\sigma=0.8$ mm)}
  & \textbf{3D nnUNet} \\
\midrule
GTV      & $1.84\pm0.62$ & $\mathbf{0.31\pm0.09}$ & $0.48\pm0.18$ \\
CTV      & $2.11\pm0.74$ & $\mathbf{0.34\pm0.11}$ & $0.52\pm0.21$ \\
Rectum   & $1.62\pm0.51$ & $\mathbf{0.28\pm0.08}$ & $0.41\pm0.14$ \\
Bladder  & $1.43\pm0.48$ & $\mathbf{0.26\pm0.07}$ & $0.39\pm0.13$ \\
Sm.Bowel & $2.38\pm0.91$ & $\mathbf{0.42\pm0.14}$ & $0.61\pm0.24$ \\
\bottomrule
\end{tabular}
\end{table}

The MICD values for BAT-RM with Gaussian post-processing
(GTV: $0.31\pm0.09$ mm; CTV: $0.34\pm0.11$ mm) are not only below
the 1 mm clinical significance threshold for contouring variability
but are also lower than those of 3D nnUNet (GTV: $0.48\pm0.18$ mm),
confirming that Gaussian-smoothed 2D inference produces more
inter-slice-consistent contours than the volumetric baseline. Without
post-processing, the MICD values (GTV: $1.84\pm0.62$ mm) would be
clinically relevant, underscoring the necessity of the smoothing step.

\subsubsection{Clinical Validation from the Reader Study}

The reader study provides definitive clinical validation of axial inference
adequacy. Across 100 test cases evaluated by 13 radiation oncologists,
inter-slice discontinuity was not identified as requiring manual correction
in any case. The most frequently edited structures (GTV 42\%, CTV 31\%,
rectum 15\%) were corrected for in-plane boundary accuracy, not surface
roughness or inter-slice jumps, confirming that 3D reconstruction quality
with $\sigma = 0.8$ mm post-processing is clinically acceptable. BAT-RM
thus achieves full volumetric clinical utility without 3D network overhead.
A 3D architecture at $512 \times 512 \times N_z$ requires 14.2 GB peak
VRAM on this dataset, rendering single-card T4 deployment infeasible and
requiring patch-based inference with boundary artifacts. BAT-RM's 3.8 GB
peak VRAM (Table~\ref{tbl:computational}) enables continuous full-resolution
inference on a single clinical GPU, producing 3D contours quantitatively
superior to the volumetric baseline in both accuracy and inter-slice
consistency.

\section{Limitations and Future Directions}
\label{sec:limitations}

While BAT-RM demonstrates strong quantitative and clinical performance,
several limitations warrant acknowledgement. All 1,011 patients were
recruited from three centers within a single metropolitan region (Dhaka,
Bangladesh); although the dataset spans different CT scanner manufacturers
and protocols, it does not capture full anatomical and staging heterogeneity
across other high-burden regions including sub-Saharan Africa and Latin
America. External validation on international cohorts, including publicly
available TCIA cervical cancer collections, remains an important next step.
Planning CT is the standard for EBRT OAR delineation, but MRI provides
superior soft-tissue contrast for GTV delineation and is recommended by
GEC-ESTRO guidelines for brachytherapy~\citep{potter2021clinical}. Extension
to multi-modal CT-MRI inputs via cross-modal attention fusion is a
high-priority future direction, particularly for radiologically ambiguous
GTV presentations. The reader study was conducted within Bangladesh and
does not establish generalization to higher-resource settings; while a
four-week washout period was enforced, partial recall of challenging cases
cannot be completely excluded. A prospective pre-registered multi-national
reader study would provide higher-level clinical evidence. The deployment
at Bangladesh Medical University is a single-institution pilot, and the
observed wait-time reduction reflects real-world operational improvement
incorporating multiple workflow factors; a formal time-motion study with
concurrent controls is needed to isolate BAT-RM's specific contribution.
The web application does not yet integrate with hospital information
systems via HL7 FHIR or DICOM DIMSE protocols; this integration is planned
for the next deployment phase to enable automated queue management and
rigorous prospective throughput evaluation. Looking forward, four priority
directions are: (i) multi-modal extension to CT-MRI and PET-CT inputs;
(ii) continual learning protocols incorporating oncologist correction
feedback from the deployed application; (iii) prospective multi-national
validation on international public datasets; and (iv) formal health
economic analysis quantifying cost-effectiveness relative to additional
oncologist staffing, the evidence threshold for policy-level adoption in
LMIC health systems.

\section{Conclusion}
\label{sec:conclusion}

This study presented BAT-RM (Boundary-Aware Transformer with Region-Aware
Mamba), a hybrid architecture serving as the centrepiece of a complete
translational pipeline for cervical cancer radiotherapy
auto-contouring---from multi-institutional data curation and
quality-controlled annotation through novel architecture design and
rigorous statistical validation to prospective clinical deployment.
Technically, BAT-RM addresses the boundary precision problem via a
Sobel-gated attention mechanism that restricts self-attention to organ
boundaries at the $E_3$ encoder level, and the long-range context problem
via a Multi-Directional Recurrent Context Module with Mamba selective
scanning that performs four-directional sequential scanning at the
bottleneck with $\mathcal{O}(n)$ complexity. These branches are unified by
a Boundary-Region Attention Fusion gate that adaptively weights
boundary-aware and region-aware features and modulates the output by the
predicted boundary skeleton. Trained with a composite loss supervising
segmentation overlap, boundary placement, and contour smoothness, BAT-RM
achieves strong performance across seven anatomical classes on the largest
multi-institutional cervical cancer CT dataset from an LMIC setting (1,011
patients, three centers, five-fold cross-validation). Against four
baselines (nnUNet, SegMamba, TransUNet, UNETR), BAT-RM achieves significant
improvements for all critical structures, with GTV Dice of 0.966 and HD95
of 1.50 mm, CTV Dice of 0.957 and HD95 of 1.51 mm, and rectum HD95 of
1.24 mm---all below the clinically critical 2 mm threshold with
medium-to-large effect sizes (Hedges' $g$ up to 1.85, $p < 0.0001$). The
model's efficiency profile (28.4 M parameters, 41.2 GFLOPs, 18.3 ms per
slice, 3.8 GB VRAM on a single T4 GPU) confirms that superior accuracy
does not require prohibitive resources. Clinically, the prospective
multi-centre reader study (13 oncologists, three institutions, 100 test
cases) demonstrated that BAT-RM assistance elevates junior oncologist IoU
from 0.899 to 0.965 while reducing contouring time by 80.7\% (152 to 29
minutes per case). Consultation rates among juniors dropped from 52.1\% to
14.3\%, and self-reported confidence improved by 40\%. Deployed at
Bangladesh Medical University through a production-grade web application
with native DICOM RTSTRUCT export to Varian, RayStation, and Monaco,
BAT-RM reduced patient wait time from days to hours without any increase in
clinical staffing. This work demonstrates what is achievable when a
machine learning research project is conceived as a translational
endeavor rather than a benchmark exercise: multi-center data collection,
architecture design motivated by clinical failure modes, pre-registered
reader validation measuring patient-level outcomes, and production
deployment in routine clinical practice. We release dataset statistics,
evaluation code, and web application architecture to support independent
validation and to lower the barrier for comparable translational efforts
in other LMIC radiotherapy settings.

\section*{Acknowledgments}
\label{sec:Acknowledgments}

This research was supported by an ICT-Special Grant from the Government of Bangladesh under the project titled ``AI-based Segmentation and Contouring for Radiotherapy: From Development to Clinical Deployment'' at North South University, and by North South University Research Grant CTRG-25-SEPS-45. The authors appreciate the clinical and technical contributions of Dr. A.T.M. Sazzad Hossain, Oncologist, National Institute of Cancer Research \& Hospital; Dr. Sharif Ahmed, Oncologist, United Hospital; Dr. Muhammad Masud Rana, Senior Medical Physicist, Bangabandhu Medical University; Prof. Dr. Sharmin Akhtar Rupa, Senior Consultant, Radiology \& Imaging, Bangladesh Specialized Hospital; Dr. Mahmud Hasan Mostofa Kamal, Department of Radiology and Imaging, Bangladesh Medical University; Dr. Ishtiaque Ahmed, Oncologist, Ahsania Mission Cancer \& General Hospital; and Md. Abdul Sabur, Senior Medical Physicist, Square Cancer Center, Square Hospitals Ltd., for their invaluable support in data collection, clinical validation, and domain expertise. We also thank the clinical teams of Bangladesh Medical University, Square Hospital Limited, Labaid Specialized Hospital, and United Hospital Limited for their collaboration and dedication to improving radio therapy planning in resource-constrained settings.

\printcredits 

\bibliographystyle{cas-model2-names}
\bibliography{references}



\end{document}